\documentclass[CRGEOS,Unicode,manuscript]{cedram}
\usepackage{amsmath,amssymb}
\usepackage{multirow}
\hypersetup{urlcolor=purple, linkcolor=blue, citecolor=blue}

\title{Did lunar tides sustain the early Earth's dynamo?}

\author[J. Vidal]{\firstname{J\'er\'emie} \lastname{Vidal}\IsCorresp\CDRorcid{0000-0002-3654-6633}}
\address{CNRS, ENS de Lyon, Univ. Lyon 1, LGL-TPE, France}
\email[J. Vidal]{jeremie.vidal@ens-lyon.fr}

\author[D. C\'ebron]{\firstname{David} \lastname{C\'ebron}\CDRorcid{0000-0002-3579-8281}}
\address{Universit\'e Grenoble Alpes, CNRS, ISTerre, 38000 Grenoble, France}
\thanks{JV received funding from \textsc{ens} de Lyon under the programme "Terre \& Plan\`etes". DC received funding from the European Research Council (\textsc{erc}) under the European Union's Horizon $2020$ research and innovation programme (grant agreement No $847433$, \textsc{theia} project). DC also acknowledges the French Academy of Sciences \& Electricit\'e de France.}

\DOI{https://doi.org/10.5802/crgeos.324}


\begin{abstract} 
Geological data show that, early in its history, the Earth had a large-scale magnetic field with an amplitude comparable to the one of the present geomagnetic field.
However, its origin remains enigmatic and various mechanisms have been proposed to explain the Earth's field over geological time scales.
Here, we critically evaluate whether tidal forcing could explain the ancient geodynamo, by combining constraints from geophysical models of the Earth-Moon system and predictions from turbulence studies. 
Our analysis shows that lunar tidal forcing could have been sufficiently strong before $-3.25$~Gy to trigger turbulence within the Earth’s core, and potentially to sustain dynamo action during that interval.
Then, we propose new scaling laws for the magnetic field amplitude $B$.
We expect the latter to scale as $B \propto \beta^{4/3}$, where $\beta$ is the equatorial ellipticity of the liquid core, if the turbulence involves weak interactions of three-dimensional inertial waves. 
Alternatively, in the regime of strong tidal forcing, the expected scaling becomes $B \propto \beta$. 
When extrapolated to the Earth's core, it suggests that tidal forcing alone was too weak to possibly explain the ancient geomagnetic field.
Therefore, our study indirectly favours another origin for the early Earth's dynamo on long time scales (e.g. exsolution of light elements atop the core, or thermal convection due to secular cooling). 
\end{abstract}

\begin{document}

% Use the \maketitle command after the abstract
\maketitle

%% Beginning of text
%% Uncomment the following command when preparing the final version of
%% an accepted manuscript. You can also check whether your figures fit
%% in the narrow columns.
\twocolumngrid

%-----------------------------------------------------------------------
%-----------------------------------------------------------------------
\section{Introduction}
%-----------------------------------------------------------------------
This research article concludes a series published in the \emph{Comptes Rendus} \citep{deguen2024fluid,nataf2024dynamic,vidal2024geophysical,plunian2025three}, sparked by the Amp\`ere Prize awarded in $2021$ by the French Academy of Sciences to the \textsc{g\'eodynamo} research team\footnote{ISTerre, Universit\'e Grenoble Alpes, France, \url{https://www.isterre.fr/english/research/research-teams/geodynamo/}}.
The latter was originally formed to study convective flows within the Earth's core and the geomagnetic field. 
Indeed, more than a century after the seminal idea of \citet{larmor1919could}, it is well accepted that the geomagnetic field is driven by dynamo action due to turbulent motions in the Earth's liquid core \citep[e.g.][]{roberts2013genesis,dormy2025rapidly}. 
Moreover, paleomagnetism indubitably shows that the Earth has hosted a large-scale magnetic field for billions of years \citep[e.g.][]{macouin2004long}. 
Therefore, the dynamics of the liquid core over geological time scales can be indirectly probed by studying the geomagnetic field.
For example, geomagnetic data allows reconstructing surface core flows \citep[e.g.][]{istas2023transient,rogers2025effects}, which may enhance future predictions of geomagnetic variations \citep[e.g.][]{madsen2025modelling}, or anchoring models aimed at understanding the origin of magnetic reversals \citep[e.g.][]{driscoll2009effects,frasson2025geomagnetic}. 
In addition, understanding the emergence of the geodynamo in the ancient Earth is also of paramount importance, since it would provide invaluable insights into the long-term evolution of the Earth since its accretion \citep{halliday2023accretion}.

%-----------------------------------------------------------------------
\subsection{Geologic constraints for dynamo models}
%-----------------------------------------------------------------------
\begin{figure}
    \centering
    \includegraphics[width=0.49\textwidth]{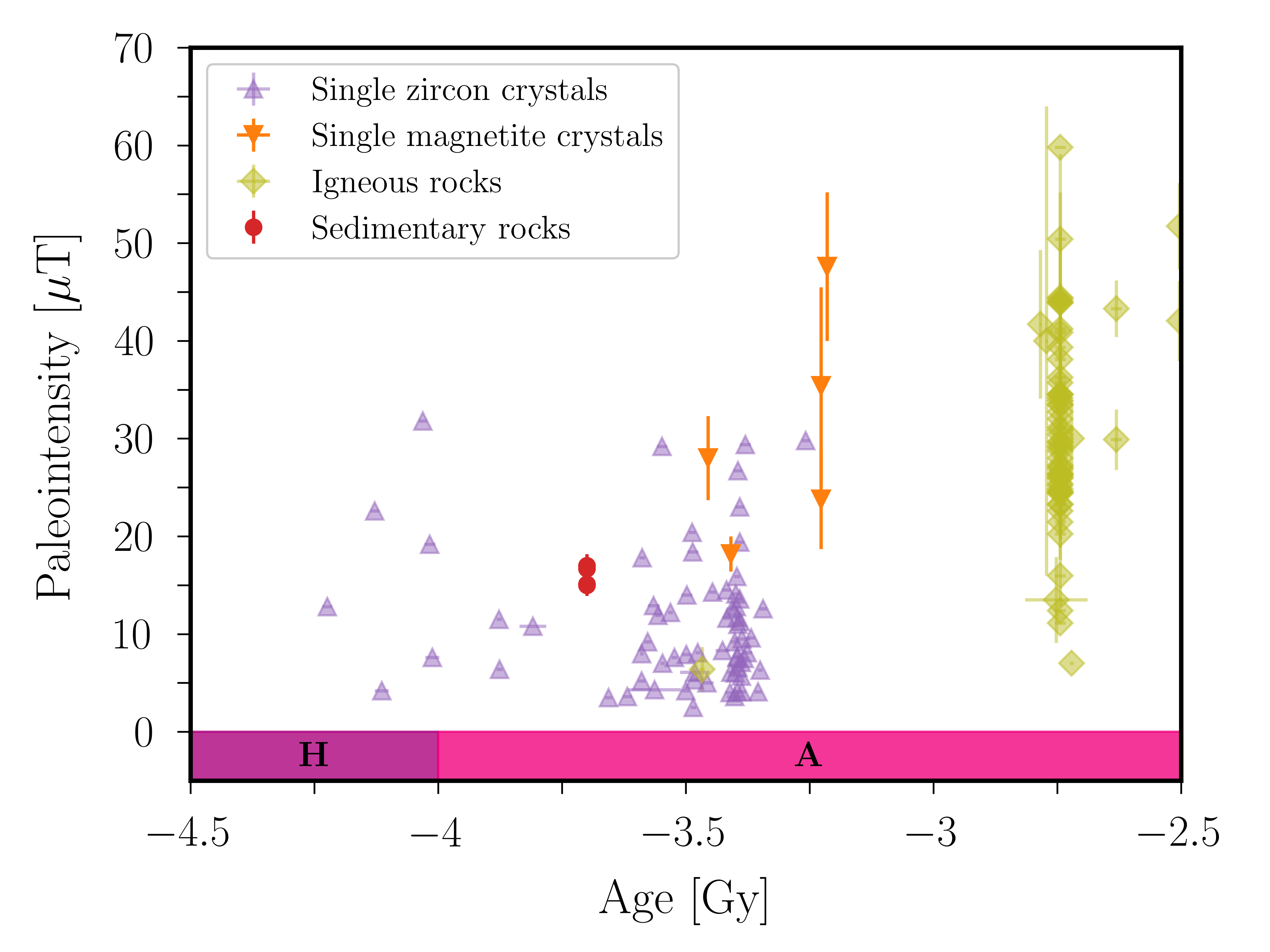}
    \caption{Paleointensity at the Earth's surface during the Hadean and Archean periods. Measurements have been performed on either single-silicate crystals (e.g. zircons) or whole rocks (e.g. Banded Iron Formations). 
    Data from the \textsc{pint} database \citep{bono2022pint} and \citet{nichols2024possible}. Geological eons are also shown (H: Hadean, A: Archean).}
    \label{fig:paleomag}
\end{figure}

Paleomagnetism indicates that the geomagnetic field is at least $3.4-3.5$~Gy old \citep{tarduno2010geodynamo,biggin2011palaeomagnetism}, with a paleo-amplitude that is roughly comparable with the present field until $-3.25$~Gy.
Such observations agree well with more indirect estimates from geochemistry.
Indeed, the ${}^{15}N/{}^{14}N$ isotopic composition of the Archean atmosphere $3.5$~Gy ago was found to be rather similar to the one of the present-day atmosphere \citep{marty2013nitrogen}, which would require a paleomagnetic amplitude of at least $50$~\% of the current intensity to avoid $N_2$ loss in the upper atmosphere \citep{lichtenegger2010aeronomical}. 
However, planetary scientists currently disagree on whether the Earth's magnetic field could have appeared earlier or not.
It is indeed very challenging to go further back in time, as ancient rocks have experienced multiple geological events throughout their history.
As shown in Figure \ref{fig:paleomag}, some paleomagnetic studies suggest that the geomagnetic field probably existed during the Eoarchean era with a surface amplitude $>15$~$\mu$T \citep{nichols2024possible}, and possibly $4.2$~Gy ago as inferred from Hadean silicate minerals \citep{tarduno2015hadean,tarduno2020paleomagnetism,tarduno2023hadaean}.
However, the quality of such ancient paleomagnetic data is strongly disputed.
The magnetic carriers may have a secondary origin, such that the magnetisation could post‐date the formation of the minerals by millions to billions of years \citep[e.g.][]{weiss2015pervasive,weiss2018secondary,borlina2020reevaluating,taylor2023direct}.

Finding a convincing scenario to explain the ancient Earth's magnetic field is a long-term goal in geophysical modelling \citep{landeau2022sustaining}. 
To this end, dynamo action is believed to predominantly operate in the Earth's liquid core \citep[e.g.][]{braginsky1995equations}, as the latter is surrounded by a silicate mantle having a weaker electrical conductivity (compared to that of liquid iron at core conditions) in both its upper and lower regions \citep[e.g.][for the Earth]{yoshino2010laboratory,jault2015illuminating}.
As such, the mantle is often considered as an electrical insulator on long time scales for dynamo modelling.
Note that dynamo action may be possible in a (basal) magma ocean if the electrical conductivity of molten silicate rocks is high enough \citep[e.g.][]{ziegler2013implications,scheinberg2018basal,stixrude2020silicate,dragulet2025electrical}, but this dynamical scenario may be energetically expansive \citep{schaeffer2025BMO}. 
However, there is currently no consensus within the community regarding the physical mechanisms that may have driven core flows in the ancient core.

%-----------------------------------------------------------------------
\subsection{Non-consensual dynamo scenarios}
%-----------------------------------------------------------------------
To assess the viability of a candidate dynamo scenario, we should strive to reproduce the main characteristics of the recorded field over geological time scales.
In particular, we can consider the typical amplitude of the large-scale field in the dynamo region, which can be extrapolated from the data.
Indeed, for an electrically insulating mantle, the amplitude $B_\mathrm{cmb}$ of the largest-scale dipolar field at the core-mantle boundary (\textsc{cmb}) is related to that at the planet's surface $B_s$ by \citep[e.g.][]{moffatt2019self}
\begin{equation}
    B_\mathrm{cmb} \simeq B_s \left ( \frac{R_s}{R_\mathrm{cmb}} \right )^3,
    \label{eq:Bcmbfromdata}
\end{equation}
where $R_\mathrm{cmb}$ is a mean core radius, and $R_s$ is the mean surface radius.
With the typical value of $R_\mathrm{cmb} \approx 3480$~km for the Earth's core, Figure \ref{fig:paleomag} leads to the estimate  $B_\mathrm{cmb} \sim 10^{-2}-10^{-1}$~mT for the early Earth's field before $-3.25$~Gy.
Deep in the core, the strength of the dynamo magnetic field could even be larger than at the \textsc{cmb}, as for instance reported for the current Earth \citep[with a hidden toroidal field of a few mT, e.g.][]{gillet2010fast}. 
Note that it is unclear whether these ancient dynamos have operated in a strong-field regime (as currently in the core) or not, which is a regime with a magnetic energy dominating over the kinetic energy \citep[e.g.][]{moffatt2019self}. 
If so, a strong-field behaviour would put a strong constraint to assess the viability of candidate dynamo models.

Now, do we have some ancient dynamo scenarios meeting the above constraints?
Currently, the main driver of the Earth's core flows is inner-core crystallisation \citep[e.g.][]{buffett1996thermal}.
Indeed, this scenario has proven successful in reproducing the main characteristics of the current geomagnetic field using numerical simulations \citep[e.g.][]{schaeffer2017turbulent,aubert2023state}.
However, inner-core crystallisation was missing in the distant past. 
The exact chronology remains disputed, but a nucleation starting $1 \pm 0.5$~Gy ago seems reasonable from recent thermal-evolution models \citep[e.g.][]{labrosse2015thermal} or paleomagnetism \citep{biggin2015palaeomagnetic,bono2019young,zhou2022early,li2023late}.
Prior to inner-core growth, it remains unclear which mechanism could have sustained a large-scale magnetic field.
In addition to thermal convection alone \citep[e.g.][]{aubert2009modelling,burmann2025rapidly,lin2025invariance}, various scenarios have been invoked to trigger turbulence in the core \citep[e.g.][]{landeau2022sustaining}, such as flows driven by double-diffusive convection in the core \citep[e.g.][]{monville2019rotating} or by tidal forcing \citep[e.g.][]{le2015flows}.
Indeed, the recent estimates of the thermal conductivity of liquid iron at core conditions, which do not show a consensus yet between experimental and computational values \citep[e.g.][]{williams2018thermal,hsieh2025moderate,andrault2025long}, might suggest that secular cooling was energetically less efficient than initially thought to sustain dynamo action in the Earth's core.

%-----------------------------------------------------------------------
\subsection{Outline of the manuscript}
%-----------------------------------------------------------------------
Motivated by the paleomagnetic data shown in Figure \ref{fig:paleomag}, we want to thoroughly assess whether tidal forcing could have sustained the early Earth's magnetic field. 
\citet{landeau2022sustaining} provided preliminary estimates but, as shown below, applying this scenario to the Earth is still intricate.
The extrapolation is underpinned by some arguments that must be quantitatively revisited in the light of recent geophysical models and fluid-dynamics studies. 
Thus, this manuscript also intends to explain how the fluid-dynamic community models tidally driven flows and turbulence in planetary cores.

For instance, to the best of our knowledge, there is no self-consistent numerical code to efficiently explore the dynamo capability of orbitally driven flows in realistic core geometries (i.e. with a weakly non-spherical \textsc{cmb}). 
This is a noticeable difference with convection-driven dynamos, for which efficient numerical strategies have been developed for decades \citep[e.g.][]{christensen2001numerical,schaeffer2013efficient}.
Therefore, we must currently rely on scaling arguments to extrapolate the few available results to the early Earth. 
However, planetary extrapolation is not an easy task because it requires a strong understanding of the turbulence properties, which are still debated in the fluid-dynamics community \citep[e.g. as recently reviewed in][]{vidal2024geophysical}. 
Similarly, a universal scaling law that could be applied to any tidally driven flow is unlikely to exist, because the scaling theory should be tailored to each physical mechanism. 

With the aforementioned goals in mind, the manuscript is divided as follows.
We introduce the basic fluid-dynamics ingredients for core flows driven by orbital forcings (e.g. tides) in \S\ref{sec:model}, as they may not be familiar to non-expert readers.
We then focus on the tidal forcing in \S\ref{sec:tidal}, combining constraints from geophysical models, hydrodynamic studies and new dynamo scaling laws, to extrapolate our findings to the early Earth.
We discuss the results in \ref{sec:discussion}, and we end the manuscript in \S\ref{sec:conclusion}. 

%-----------------------------------------------------------------------
\section{Modelling of orbitally driven flows}
\label{sec:model}
%-----------------------------------------------------------------------
In this section, we will present the minimal fluid-dynamics ingredients needed to model the dynamics of planetary liquid cores subject to orbital mechanical forcings (such as tides in the early Earth). 
First, we introduce in \S\ref{subsec:modelEQ} an idealised common description of orbitally driven core flows. 
Then, we briefly outline in \S\ref{subsec:modelROAD} the different steps of the (expected) flow response of a planetary liquid core to an orbital mechanical forcing, which will be revisited in \S\ref{sec:tidal} to carry out the planetary extrapolation to the early Earth.

%-----------------------------------------------------------------------
\subsection{Equations of fluid dynamics}
\label{subsec:modelEQ}
%-----------------------------------------------------------------------
We want to model the dynamics of a planetary core prior to the nucleation of a solid inner core. 
Thus, we consider a liquid core of volume $V$ with no inner core, which is surrounded by a rigid and electrically insulating mantle.
The \textsc{cmb} is denoted by $\partial V$ below. 
The core geometry is generally assumed to be spherical in most geodynamo models, as it is sufficient for convection-driven studies and allows developing very efficient numerical methods  \citep[e.g.][]{christensen2001numerical,schaeffer2013efficient}. 
However, it is important to take the small departures from a spherical \textsc{cmb} into account for tidal and precession forcings \citep[e.g.][]{le2015flows}. 
For simplicity, we assume below that $V$ is an ellipsoid, which agrees with the mathematical theory of equilibrium figures for a rotating fluid mass with an orbital partner \citep[e.g.][]{chandrasekhar1987ellipsoidal}. 
Then, it is customary to work in the frame rotating with the ellipsoidal distortion, which is rotating at the angular velocity $\boldsymbol{\Omega}_\epsilon$. 
Note that the latter generally differs from the rotation of the fluid, which will be denoted by $\boldsymbol{\Omega}_s$ below.
Working in the frame rotating at $\boldsymbol{\Omega}_\epsilon$ will ease the computations, since the \textsc{cmb} will be steady in that frame and can be written as  $({x}/{a})^2 + ({y}/{b})^2 + ({z}/{c})^2 = 1$, where $[a,b,c]$ are the ellipsoidal semi-axes and $(x,y,z)$ are the Cartesian coordinates.
Note that dynamical pressure (due to core flows) is expected to be much weaker than hydrostatic pressure on long time scales.
Consequently, in practice, the flow dynamics is usually explored in rigid ellipsoids by considering prescribed values of $[a,b,c]$.
Planetary values can be estimated from tidal theory \citep[e.g.][]{farhat2022resonant} and the theory of planetary figures \citep[e.g.][]{chambat2010flattening}.
Note that the $z-$axis is usually chosen along the rotation of the planet, such that it is customary to introduce the two parameters given by
\begin{subequations}
\begin{equation}
    \beta = \frac{|a^2-b^2|}{a^2+b^2}, \quad f= \frac{a-c}{a},
    \tag{\theequation a,b}
\end{equation}
\end{subequations}
where $ 0 < \beta \ll 1$ is the (equatorial) ellipticity and $0 < f \ll 1$ is the (polar) flattening. 

The core is filled with an electrically conducting liquid, which is assumed to have a constant kinematic viscosity $\nu$ and magnetic diffusivity $\eta$. 
The ratio of these two quantities yields the magnetic Prandtl number $Pm = \nu/\eta$, whose typical value is $Pm \sim 10^{-6}$ in a planetary liquid core when adopting the typical core values $\nu \simeq 10^{-6}$~m${}^2$.s${}^{-1}$ \citep[e.g.][]{de1998viscosity} and $\eta \simeq 1$~m${}^2$.s${}^{-1}$ \citep[e.g.][]{nataf2024dynamic}. 
As a consequence, we expect the magnetic dissipation to be much larger than the viscous one in a dynamo regime.
Note that we neglect density effects below, to focus on incompressible fluids with constant density. 
Indeed, most fluid-dynamics studies consider orbitally driven flows without buoyancy effects in the incompressible regime, assuming that the fluid has a constant density $\rho_f$. 
In the frame rotating at $\boldsymbol{\Omega}_\epsilon$, the fluid velocity $\boldsymbol{v}$ is then governed by the incompressible momentum equations given by
\begin{subequations}
\begin{align}
    \mathrm{d}_t \boldsymbol{v} + 2 \boldsymbol{\Omega}_\epsilon \times \boldsymbol{v} &= - \nabla \Pi + \nu \nabla^2 \boldsymbol{v} + \boldsymbol{f}_L + \boldsymbol{f}_P, \\
    \nabla \boldsymbol{\cdot} \boldsymbol{v} &= 0,
\end{align}
\end{subequations}
where $\mathrm{d}_t = \partial_t + \boldsymbol{v} \boldsymbol{\cdot} \nabla$ is the material derivative, $\Pi$ is a reduced pressure term (for incompressible flows), and $[\boldsymbol{f}_L,\boldsymbol{f}_P]$ are the Lorentz and Poincar\'e forces given by
\begin{subequations}
    \begin{equation}
        \boldsymbol{f}_L = \frac{1}{\rho_f \mu} (\nabla \times \boldsymbol{B}) \times \boldsymbol{B}, \quad \boldsymbol{f}_P = \boldsymbol{r} \times \dot{\boldsymbol{\Omega}}_\epsilon,
        \tag{\theequation a--b}
    \end{equation}
\end{subequations}
where $\mu$ is the magnetic permeability of the fluid\footnote{ We have $\mu \approx \mu_0$ for core conditions in practice, where $\mu_0$ is the magnetic permeability of vacuum.} and $\boldsymbol{r}$ is the position vector. 
The momentum equation is coupled, through the Lorentz force, to the magnetic field equations given by \citep[e.g.][]{moffatt2019self}
\begin{subequations}
\label{eq:induction}
\begin{align}
    \partial_t \boldsymbol{B} &= \nabla \times ( \boldsymbol{v} \times \boldsymbol{B} ) + \eta \nabla^2 \boldsymbol{B}, \\
    \nabla \boldsymbol{\cdot} \boldsymbol{B} &= 0.
\end{align}
\end{subequations}
Finally, the above equations must be supplemented with appropriate boundary conditions (BCs) at the \textsc{cmb}. 
The mantle being supposed to be an electrical insulator, the magnetic field in the core must match at the \textsc{cmb} the field in the mantle, which is given by
\begin{subequations}
\label{eq:BCmag}
\begin{equation}
    \boldsymbol{B} = \nabla \Phi, \quad \Phi \to 0 \ \text{when} \ |\boldsymbol{r}| \to +\infty.
    \tag{\theequation a,b}
\end{equation}
\end{subequations}
Next, because the mantle is supposed to be rigid, the velocity must satisfy the no-penetration BC given by $\left . \boldsymbol{v} \boldsymbol{\cdot} \boldsymbol{1}_n \right |_{\partial V} = 0$ for any forcing, where $\boldsymbol{1}_{n}$ is the (outward) unit vector normal to the boundary.
Some BCs must also be enforced on the tangential velocity components, but they are forcing-dependent.

Finally, it is customary to write down the mathematical problem using dimensionless variables. 
In particular, we introduce the Ekman number $E$ and the Rossby number $Ro$ given by
\begin{subequations}
\label{eq:EkRonumbers}
\begin{equation}
    E = \frac{\nu}{\Omega_s R_\mathrm{cmb}^2}, \quad Ro = \frac{\mathcal{U}}{\Omega_s R_\mathrm{cmb}},
    \tag{\theequation a,b}
\end{equation}
\end{subequations}
where $\Omega_s = |\boldsymbol{\Omega}_s|$ is the mean angular velocity of the fluid (e.g. $\Omega_s \simeq 7.3 \times 10^{-5}$~rad.s${}^{-1}$ currently in the Earth's core), $R_\mathrm{cmb}$ is the mean core radius, and $\mathcal{U}$ is a typical amplitude of the core flows. 
We will show in \S\ref{sec:tidal} that both $E$ and $Ro$ will be key for the planetary extrapolation of the results.
Indeed, despite $E$ is a very small quantity in planetary cores (e.g. $E\sim 10^{-15}$ in the Earth and $E\sim 10^{-12}$ in the Moon), it will mainly control the growth of turbulent flows.
Similarly, the strength of the rotating turbulence (and hence that of the dynamo magnetic field) will be influenced by the value of $Ro$.

%-----------------------------------------------------------------------
\subsection{Towards dynamo magnetic fields}
\label{subsec:modelROAD}
%-----------------------------------------------------------------------
Even with the strong physical assumptions we have employed, the model presented in \S\ref{subsec:modelEQ} is extremely difficult to solve.
One of the reasons is that, currently, there is no numerical code that can efficiently account for magnetic BC (\ref{eq:BCmag}) in an ellipsoidal geometry. 
Nonetheless, previous numerical and experimental studies have allowed the identification of a rather general pattern for the flow response to orbital forcings \citep[e.g.][for a recent review]{vidal2024geophysical}.
We briefly describe it below, as it will guide us for the extrapolation of tidal forcing in \S\ref{sec:tidal}.

To start with, we consider a non-dynamo regime characterised by negligible magnetic effects, which occurs prior to the establishment of a planetary magnetic field. 
At the leading order, an orbital forcing drives a large-scale flow $\boldsymbol{U}_0$ in an ellipsoid, which is governed by
\begin{subequations}
    \label{eq:forcedflowU0}
    \begin{align}
        \mathrm{d}_t \boldsymbol{U}_0 + 2 \boldsymbol{\Omega}_\epsilon \times \boldsymbol{U}_0 &= - \nabla \Pi_0 + \nu \nabla^2 \boldsymbol{U}_0 + \boldsymbol{f}_P, \\
        \nabla \boldsymbol{\cdot} \boldsymbol{U}_0 &= 0.
    \end{align}
\end{subequations}
This is a laminar flow, sometimes referred to as a Poincar\'e flow for precession \citep[e.g.][]{roberts2011flows}, which has a nearly uniform vorticity and can be written in the bulk as \citep{noir2013precession}
\begin{subequations}
\label{eq:formGP1}
\begin{equation}
    \boldsymbol{U}_0 \simeq \boldsymbol{\omega}_\epsilon \times \boldsymbol{r} + \nabla \Psi_\epsilon, \quad \left . \boldsymbol{U}_0 \boldsymbol{\cdot} \boldsymbol{1}_n \right |_{\partial V} = 0,
    \tag{\theequation a,b}
\end{equation}
\end{subequations}
where $\boldsymbol{\omega}_\epsilon$ is the fluid rotation vector in the $\boldsymbol{\Omega}_\epsilon-$frame, and $\nabla \Psi_\epsilon$ is a shear flow that is non-vanishing only when $a \neq b$, $a \neq c$ or $b \neq c$ (i.e. when the \textsc{cmb} is not spherical).
Note that orbital forcings also sustain other laminar flows, such as mean flows due to weak nonlinear interactions in the boundary layer below the \textsc{cmb} \citep[e.g.][]{busse1968steady,cebron2021mean}.

It turns out that the forced laminar flows are unlikely to sustain dynamo action, because their spatial structures are generally too simple \citep[e.g.][for the precession-driven forced flow]{tilgner1998models}. 
However, the flow components departing from $\boldsymbol{U}_0$ could be viable candidates for dynamo action.
Indeed, there are often unstable flow perturbations $\boldsymbol{u}$ that can grow upon $\boldsymbol{U}_0$ in the bulk of the core with an amplitude $|\boldsymbol{u}| \propto \exp(\sigma t)$ in the initial stage (i.e. when $\boldsymbol{u} \boldsymbol{\cdot} \nabla \boldsymbol{u}$ remains negligible in the momentum equation for $\boldsymbol{u}$), where $\sigma>0$ is the growth rate of the unstable flows. 
Physically, such hydrodynamic instabilities can result from couplings between the shear component $\nabla \Psi_\epsilon$ in Equation (\ref{eq:formGP1}a) and free waves existing in the bulk of the core (e.g. inertial waves, whose restoring force is the Coriolis force associated to the global rotation of the liquid core). 
The growth rate $\sigma \geq 0$ of these unstable bulk flows can be written as
\begin{equation}
    \sigma \simeq \sigma^i - \sigma^d,
    \label{eq:growthrateTH}
\end{equation}
where $\sigma^i>0$ is the diffusionless growth rate (i.e. when $\nu=\eta=0$), and $\sigma^d>0$ is the diffusive damping term.
Prior to the existence of a strong planetary magnetic field, the main dissipation mechanism is due to viscous dissipation in the Ekman boundary layer below the \textsc{cmb}, such that $\sigma^d/\Omega_s \sim \mathcal {O}(E^{1/2})$ with a numerical pre-factor that depends on the spatial complexity of $\boldsymbol{u}$ \citep[e.g.][]{greenspan1969theory}.
A first condition to reach a dynamo regime is thus given by
\begin{equation}
    \frac{\sigma^i}{\Omega_s} > \mathcal{O}(E^{1/2}).
    \label{eq:sigmaivsdamping}
\end{equation}
Condition (\ref{eq:sigmaivsdamping}), which gives a condition for bulk instability in the core, will allow us to estimate when unstable bulk flows could have been triggered by tidal forcing in the early Earth's core.
Note that other hydrodynamic instabilities may be triggered by tidal forcing, either in the bulk \citep[e.g.][]{sauret2014tide} or below the \textsc{cmb} \citep[e.g. see a discussion in][]{cebron2021mean}. 
However, they are not expected to play a key role for planetary dynamos (as their existence mostly rely on viscous effects).

After their exponential growth, these flows will saturate in amplitude due to a non-vanishing nonlinear term $\boldsymbol{u} \boldsymbol{\cdot} \nabla \boldsymbol{u}$, and become turbulent. 
As outlined in \citet{vidal2024geophysical}, various regimes of turbulent flows can be expected, leading to different scaling laws for the velocity amplitude $\mathcal{U} \sim |\boldsymbol{u}|$ as a function of the control parameters (e.g. $E$ and $Ro$).
Therefore, we need to determine the correct scaling law for the amplitude of turbulence.

Once we can estimate the amplitude of turbulent flows in planetary conditions, we can start assessing their ability to generate a self-sustained magnetic field.
In induction equation (\ref{eq:induction}a), the magnetic field can growth over time if the production term $\nabla \times ( \boldsymbol{u} \times \boldsymbol{B} )$ is much larger than the dissipation term $\eta \nabla^2 \boldsymbol{B}$.
Using orders of magnitude, this yields the condition that the magnetic Reynolds number $Rm$, given by
\begin{equation}
    Rm = \frac{\mathcal{U} R_\text{cmb}}{\eta}
    \label{eq:Rmnumber}
\end{equation}
where $\eta \simeq 1$~m${}^2$.s${}^{-1}$ is the typical value in the Earth's core, must be larger than some threshold value $Rm_c$ to have a growth of the magnetic energy.
In practice, a typical condition 
\begin{equation}
    Rm_c \sim 40-100
    \label{eq:Rmcdynamo}
\end{equation}
is often assumed from convection-driven dynamos in spherical geometries \citep{christensen2006scaling}, in good agreement with theoretical studies \citep[e.g.][]{chen2018optimal,holdenried2019trio,vidal2021kinematic}.
Yet, larger values $Rm_c \sim 200$ might be required for orbitally driven dynamos according to prior simulations \citep{reddy2018turbulent,cebron2019precessing}.

Finally, the nonlinear regime of orbitally driven dynamos remains unknown. 
Indeed, direct numerical simulations (\textsc{dns}) of dynamos in ellipsoids have only be performed using either unrealistic (ad-hoc) numerical approximations that strongly hamper the planetary extrapolation \citep{cebron2014tidally,vidal2018magnetic}, or were limited to exponential growth regimes \citep{ivers2017kinematic,reddy2018turbulent}.
Therefore, we must rely on scaling-law arguments to estimate the possible strength of dynamo fields in planetary interiors.
However, there is no consensual dynamo scaling law in the literature for tidal forcing \citep{le2011impact,barker2014non,vidal2018magnetic} or precession \citep{cebron2019precessing}. 

Finally, it is worth noting that a dynamo scenario can fail at any stage. 
For instance, condition (\ref{eq:sigmaivsdamping}) may not be fulfilled on long enough time scales to yield turbulent flows $\boldsymbol{u}$, or the resulting turbulence may have not remained vigorous enough to maintain a $Rm$ value above the dynamo onset over time. 
Similarly, the associated dynamo magnetic field may be too weak to match the paleomagnetic estimates given in Figure \ref{fig:paleomag}. 
Therefore, to assess the dynamo capability of tidally driven flows in the early Earth, we have first to estimate if conditions (\ref{eq:sigmaivsdamping}) and (\ref{eq:Rmcdynamo}) are satisfied in the distant past, and then to estimate a typical magnetic field amplitude using appropriate scaling laws.

%-----------------------------------------------------------------------
\section{Results for tidal forcing}
\label{sec:tidal}
%-----------------------------------------------------------------------
As customary in long-term evolution scenarios for the Earth-Moon system \citep[e.g.][]{farhat2022resonant}, we consider a simplified model neglecting the effects of Earth's obliquity, of the Moon's orbital eccentricity and inclination, and of the small phase lag between the tidal bulge of the Earth and the Moon.
Hence, the Earth's core is supposed to be instantaneously deformed by the tidal potential into an ellipsoid whose equatorial semi-axes can be written as
\begin{subequations}
\begin{equation}
    \frac{a}{R_\text{cmb}} =  \sqrt{1 + \beta}, \quad \frac{b}{R_\text{cmb}} = \sqrt{1 - \beta},
    \tag{\theequation a,b}
\end{equation}
\end{subequations}
where is $R_\text{cmb}$ the mean radius of the liquid core.
We work in the frame rotating with the Moon at the angular velocity $\boldsymbol{\Omega}_\epsilon = \Omega_\text{orb} \boldsymbol{1}_z$, and assume that the liquid core is co-rotating with the planet at the angular velocity $\boldsymbol{\Omega}_s = \Omega_s \boldsymbol{1}_z$ with respect to the inertial frame. 
In fluid-dynamics studies, it is tacitly assumed that the orbital parameters evolve on a time scale (denoted by $\tau$ below) that is much longer than that of the turbulent flows.
Consequently, $[\Omega_\text{orb}$, $\Omega_s]$ and the ellipsoidal geometry are always supposed to be constant for the flow dynamics. 
This allows performing parametric studies as a function of the different parameters over geological time scales, and obtaining scaling laws for extrapolation to the Earth.

\begin{figure}
    \centering
    \includegraphics[width=0.4\textwidth]{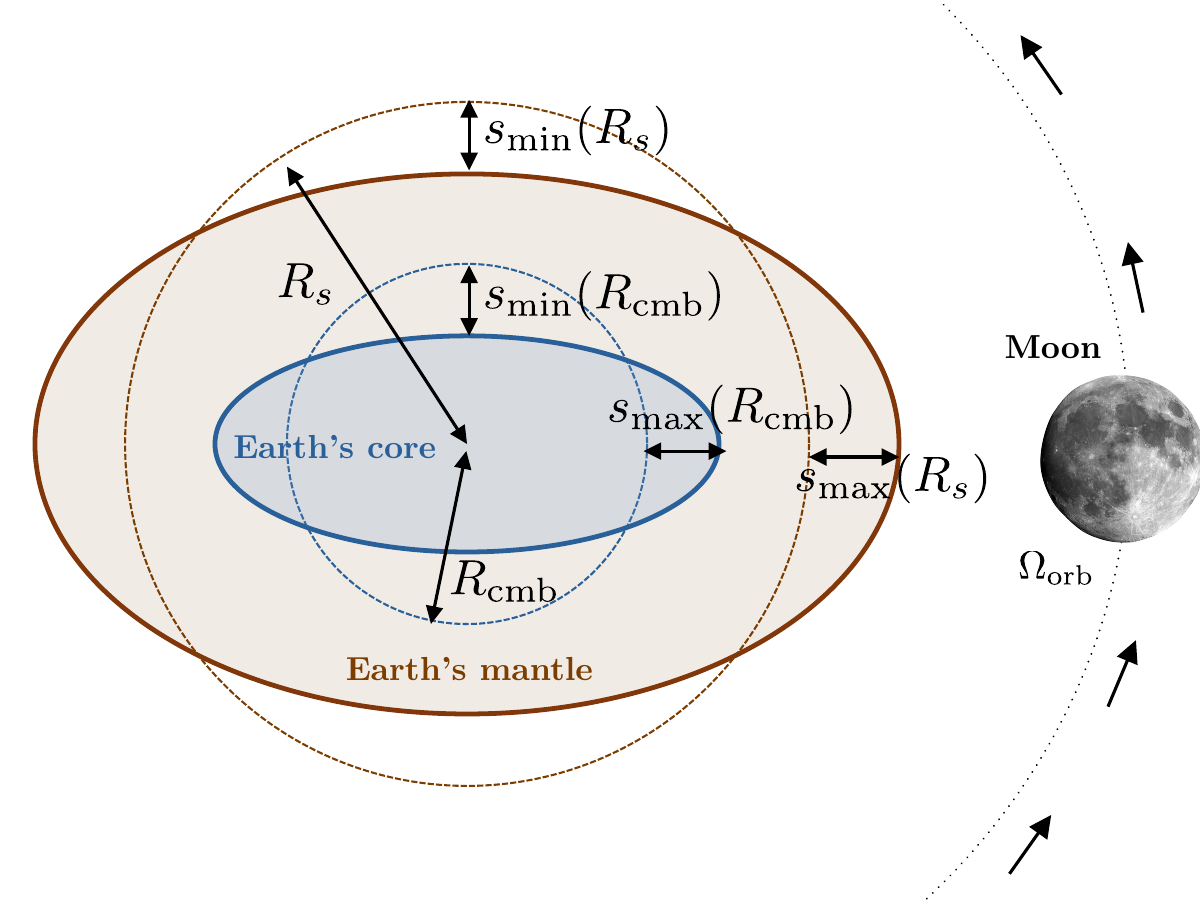}
    \caption{Sketch (not to scale) of the elliptical geometry of the tidally deformed Earth's core, as seen the orbital plane of the Moon. $R_s$ is the mean surface radius, and $R_\text{cmb}$ is the mean core radius. The radial displacement along the major axis in the equatorial plane is $s_\mathrm{max}$, and that along the minor axis is given by $s_\mathrm{min} = -s_\mathrm{max}/2$ for a tidal potential of degree $2$.}
    \label{fig:tidesmodel1}
\end{figure}

Given the above assumptions, tidal forcing first drives a laminar flow in an ellipsoidal liquid core of the form (\ref{eq:formGP1}) with \citep[e.g.][]{le2015flows}
\begin{subequations}
\label{eq:GP1tides}
\begin{equation}
    \boldsymbol{\omega}_\epsilon = \Delta \Omega \, \boldsymbol{1}_z, \quad \Psi_\epsilon = -\beta \, \Delta \Omega xy,
    \tag{\theequation a,b}
\end{equation}
\end{subequations}
where $\Delta \Omega = \Omega_s - \Omega_\mathrm{orb}$ is the differential rotation between the fluid and the orbit.
To assess the validity of the tidal scenario, we first estimate the orbital parameters in \S\ref{subsec:resultsorbit} by using geophysical models for the early Earth-Moon system.
Then, we investigate whether $\boldsymbol{U}_0$ can sustain flow instabilities and turbulence over geological time scales by using constraints from hydrodynamics studies in \S\ref{subsec:resultshydro}.
Finally, we discuss possible dynamo effects in \S\ref{subsec:resultsdynamo} and \S\ref{subsec:extrapolation} using scaling-law arguments.

%-----------------------------------------------------------------------
\subsection{Estimates from geophysical models}
\label{subsec:resultsorbit}
%-----------------------------------------------------------------------
We need to estimate how the geometry of the liquid core has changed over geological time scales, as well as the spin $\Omega_s$ and the orbital angular velocity $\Omega_\text{orb}$.
It is known that the polar flattening is mainly due to centrifugal effects such that $f \propto \Omega_s^2$ according to equilibrium hydrostatic models \citep[e.g.][]{chambat2010flattening}. 
Moreover, tidal theory for an incompressible fluid predicts that the equatorial ellipticity of the core should evolve as $\beta \propto (1/a_M)^3$ \citep[e.g.][]{cebron2012elliptical}, where $a_M$ is the Earth-Moon distance.
Therefore, assuming that the mantle's properties have not significantly changed over time, the polar flattening $f(\tau)$ and equatorial ellipticity $\beta(\tau)$ of the core at age $\tau$ before present can be estimated from the present-day values $[f(0),\beta(0)]$ as
\begin{subequations}
\label{eq:betaftime}
\begin{equation}
    \frac{f(\tau)}{f(0)} = \left ( \frac{\Omega_s (t)}{\Omega_s (0)} \right )^2, \quad \frac{\beta(\tau)}{\beta(0)} = \left ( \frac{a_M(0)}{a_M(\tau)} \right )^3,
    \tag{\theequation a,b}
\end{equation}
\end{subequations}
with the present-day values $a_M(0)/R_s \simeq 60.14$ and $\Omega_s(0) \simeq 7.27 \times 10^{-3}$~rad.s${}^{-1}$ \citep{farhat2022resonant}. 

\begin{figure}
    \centering
    \includegraphics[width=0.49\textwidth]{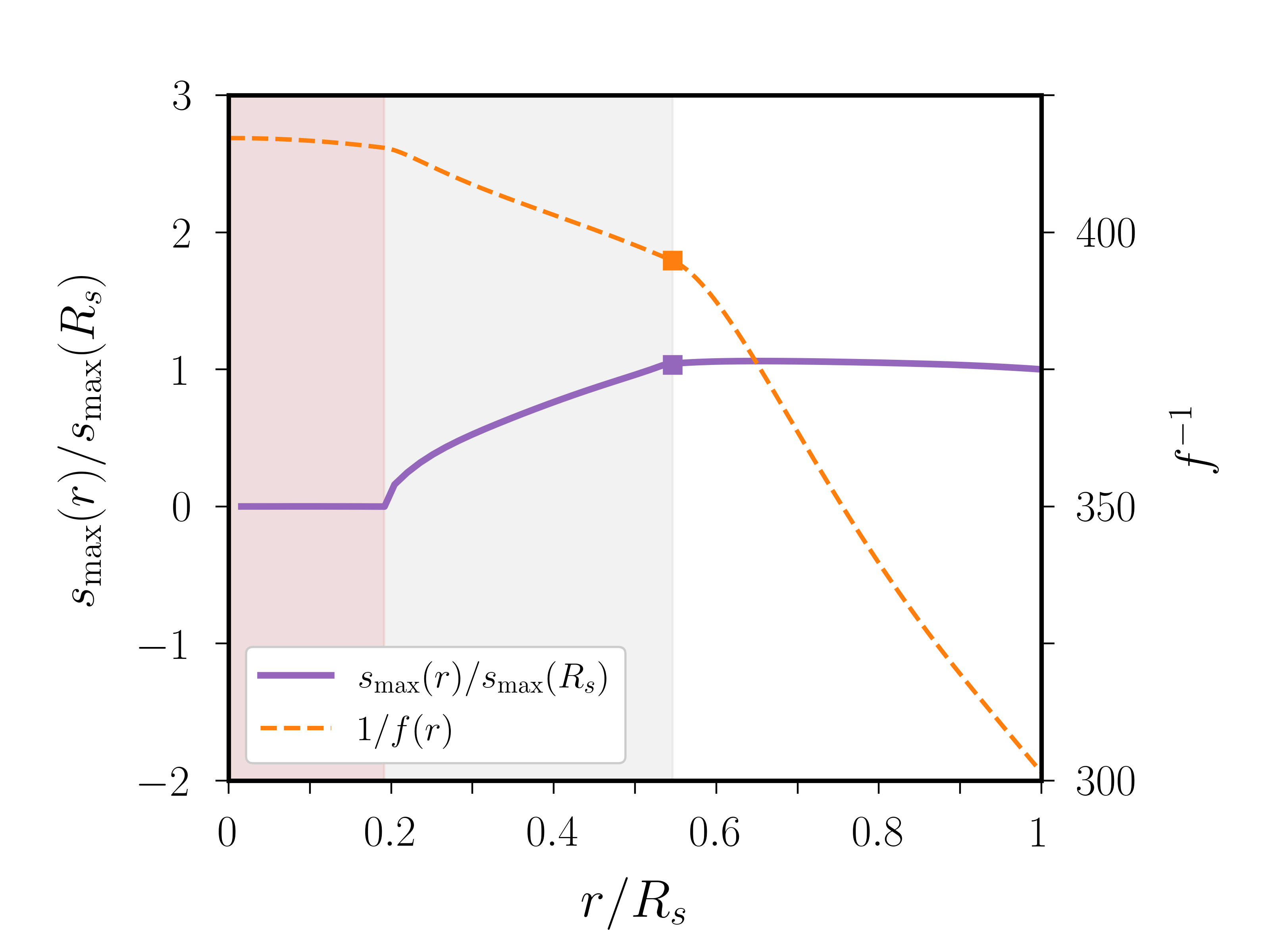}
    \caption{Maximum radial displacement $s_\mathrm{max}$ and inverse polar flattening $f^{-1}$ at present day, as a function of normalised mean radius $r/R_s$, as computed for tidal theory and hydrostatic equilibrium theory. In both cases, the same Earth's reference model is chosen \citep{dziewonski1981preliminary}. Red region shows the solid inner core, and grey one the liquid outer core.}
    \label{fig:tidesmodel2}
\end{figure}

\begin{figure}
    \centering
    \begin{tabular}{c}
    \includegraphics[width=0.49\textwidth]{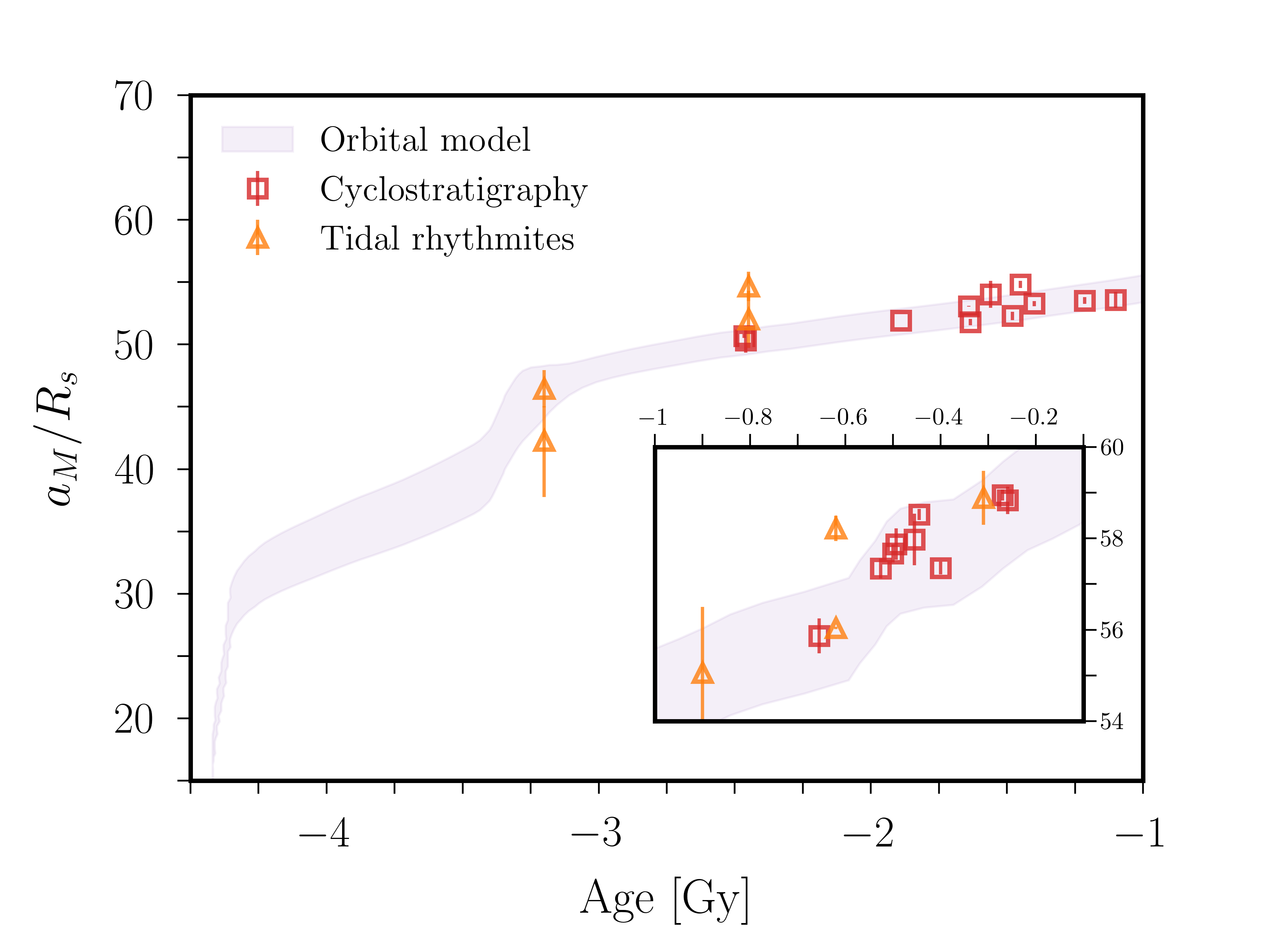} \\ (a) \\
    \includegraphics[width=0.49\textwidth]{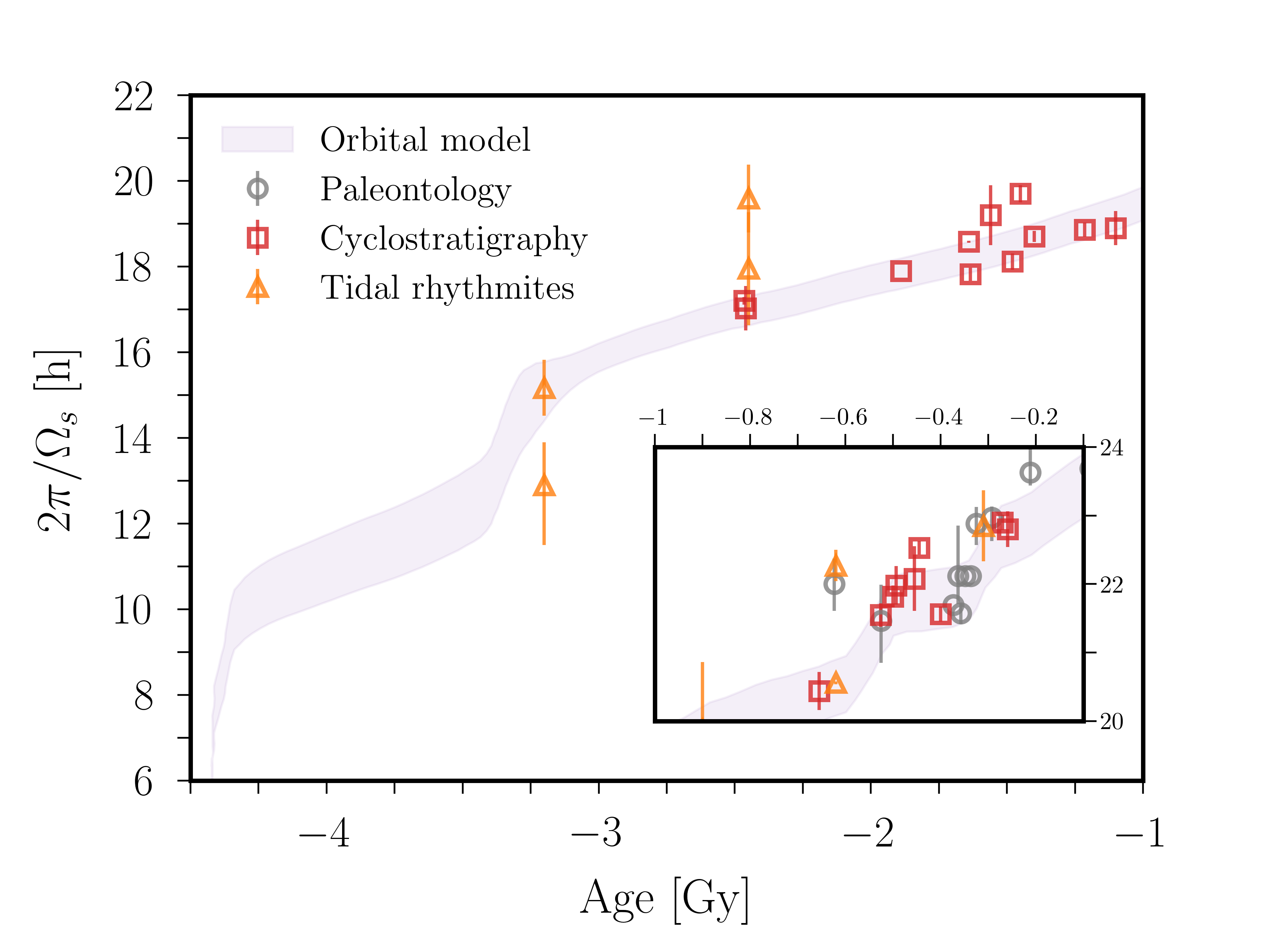} \\ (b) \\
    \end{tabular}
    \caption{Comparison between geologic data and models for the evolution of the normalised Earth-Moon distance $a_M/R_s$ in \textbf{(a)}, and of the length of day in \textbf{(b)}. Insets show the age $\tau$ between $-1$ and $-0.1$~Gy. $R_s \simeq 6378$~km is the mean value of the Earth's radius. Orbital model:  \citet{farhat2022resonant}. 
    Paleontological data: \citet{williams2000geological} and references therein. Cyclostratigraphic data: \citet{zhou2024earth} and references therein. Tidal rhythmites data: \citet{farhat2022resonant,eulenfeld2023constraints} and references therein.}
    \label{fig:forcing1}
\end{figure}

We have to estimate the present-day values $[f(0),\beta(0)]$, before extrapolating $[f(\tau),\beta(\tau)]$ back in time.
Following \citet{chambat2010flattening}, the current flattening $f(0)$ can be obtained using hydrostatic equilibrium theory, with 
\begin{equation}
    f(0) \approx 2.54 \times 10^{-3}
\end{equation}
at the \textsc{cmb}. 
Note that the Earth's nutations provide an estimate of $f(0)$ with an error margin of only a few percents \citep[e.g.][]{dehant2017understanding}, such that its value is already well constrained. 
For $\beta(0)$, we need to relate the Earth's surface maximum displacement due to Moon's tides, which is roughly $s_\text{max}(R_s) \approx22$~cm for the solid tides \citep{AGNEW2015151}, to that at the \textsc{cmb} (see Figure \ref{fig:tidesmodel1}). 
To do so, we have used the open-source code \textsc{TidalPy} \citep{renaud2023tidalpy} to solve the standard elastic-gravitational equations for a layered planet subject to a diurnal tidal potential of degree $2$ \citep[e.g.][]{alterman1959oscillations}, using \textsc{prem} as the reference Earth's internal structure \citep{dziewonski1981preliminary}.
The results are shown in Figure \ref{fig:tidesmodel2}.
We observe that the maximum tidal displacement at the \textsc{cmb} is nearly equal to that at the Earth's surface.
By writing the equatorial semi-axes as $a = R_\mathrm{cmb} + s_\mathrm{max}(R_\mathrm{cmb})$ and $b = R_\mathrm{cmb} - s_\mathrm{min}(R_\mathrm{cmb})$, with $s_\mathrm{min} = -s_\mathrm{max}/2$ for a tidal potential of degree $2$ \citep[e.g.][]{AGNEW2015151}, the \textsc{cmb} ellipticity at present time $\beta(0)$ is thus given by (at leading order in the small displacement)
\begin{equation}
    \beta (0) \simeq \frac{3}{2} \frac{s_\mathrm{max} (R_\text{cmb})}{R_\text{cmb}} \approx 9.8 \times 10^{-8}.
    \label{eq:betaradius}
\end{equation}
In the following, we will use the above present-day values to constrain the extrapolation back in time. 
Note that such values are subject to uncertainties, whose influence will be touched upon below in \S\ref{sec:discussion}.  
To do so, we use the orbital model provided in \citet{farhat2022resonant}.
This is a semi-analytical model that fits the most accurate constraints in the Earth-Moon evolution (i.e. the present tidal dissipation rate and the age of the Moon). 
As discussed below in \S\ref{sec:discussion}, this is not the case for other orbital models.
As illustrated in Figure \ref{fig:forcing1}, the physical model by \citet{farhat2022resonant} also agrees very well with most available geological data (e.g. tidal rhythmites and cyclostratigraphic records).

\begin{figure}
    \centering
    \includegraphics[width=0.49\textwidth]{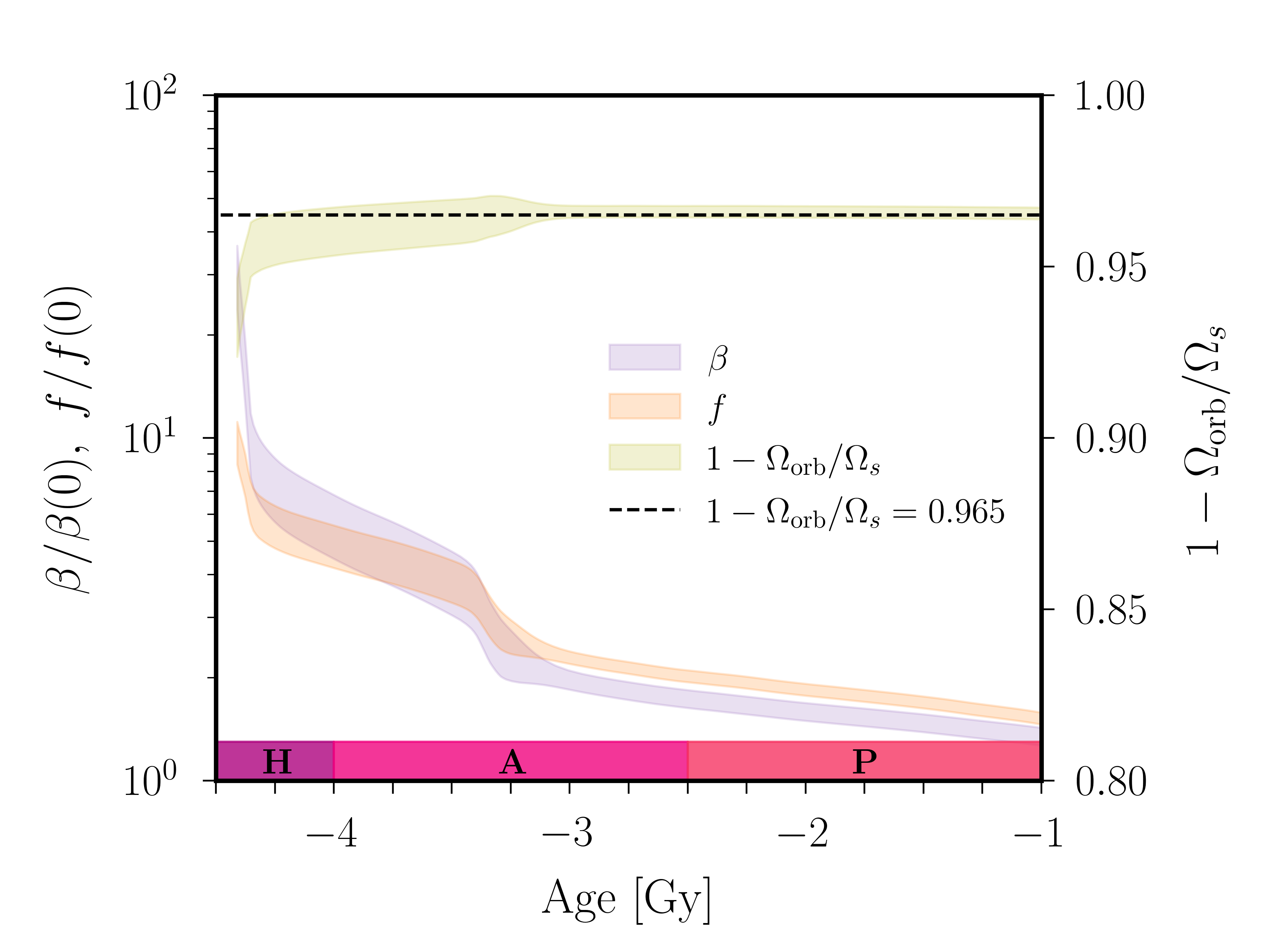}
    \caption{Evolution of the Earth's core ellipticity $\beta$, polar flattening $f$, and of $\Omega_\text{orb}/\Omega_s$, as a function of age. Dashed line shows the frequency value associated with the least-damped mode in resonance condition (\ref{eq:resonancecondition}). Geological eons are also shown (H: Hadean, A: Archean, P: Proterozoic).}
    \label{fig:betaftime}
\end{figure}

The corresponding values of $\beta(\tau)$ and $f(\tau)$, computed as a function of $\tau$ from equations (\ref{eq:betaftime}a,b), are shown in Figure \ref{fig:betaftime}.
We see that $\beta$ only varies from one order of magnitude over the Earth's history, that is from $\beta \approx 10^{-7}$ nowadays to $\beta \approx 10^{-6}$ at $-4.25$~Gy.
This narrow range of values will put severe constraints for the viability of the tidal scenario (as explained below in \S\ref{subsec:resultshydro}).
Finally, the Moon's orbital frequency is reconstructed using Kepler's third law as
\begin{subequations}
\label{eq:kepler3law}
\begin{equation}
    \frac{\Omega_\mathrm{orb}(\tau)}{\Omega_\mathrm{orb}(0)} = \left ( \frac{a_M(0)}{a_M(\tau)} \right )^{3/2}, \quad \Omega_\mathrm{orb}(\tau) = \frac{2 \pi}{T_\mathrm{orb}(\tau)},
    \tag{\theequation a,b}
\end{equation}
\end{subequations}
where $T_\mathrm{orb}(\tau)$ is the Moon's orbital period (whose current value is $T_\mathrm{orb}(0) \simeq 27.3217$~days).
As shown in Figure \ref{fig:betaftime}, we find that $\Omega_\text{orb}/\Omega_s$ only varied weakly in the distant past, a typical value being $\Omega_\text{orb}/\Omega_s \sim 0.04 \pm 0.01$ during the Hadean and Archean eons.

%-----------------------------------------------------------------------
\subsection{Hydrodynamic constraints}
\label{subsec:resultshydro}
%-----------------------------------------------------------------------
\begin{figure*}
    \centering
    \begin{tabular}{cc}
    \includegraphics[width=0.47\textwidth]{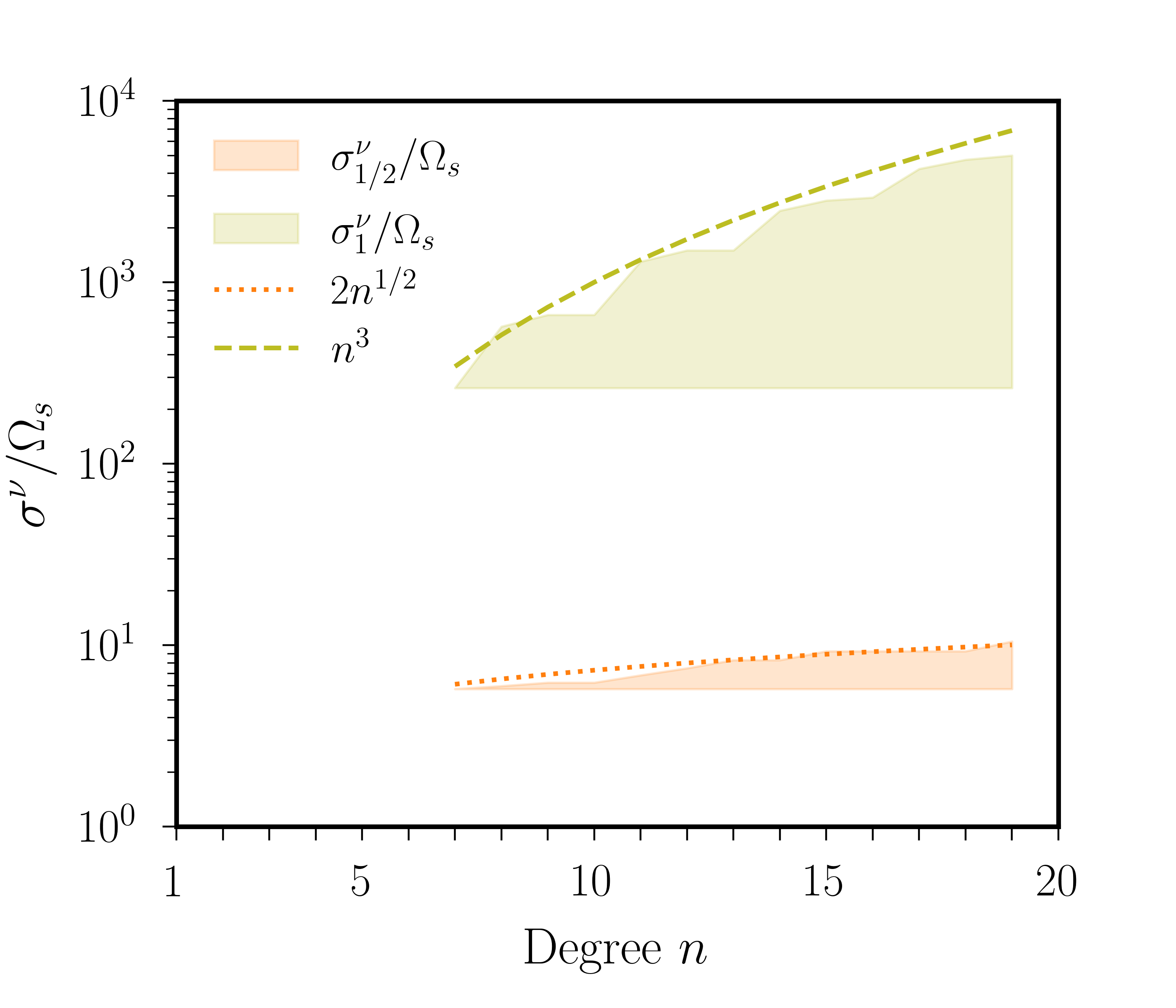} &
    \includegraphics[width=0.47\textwidth]{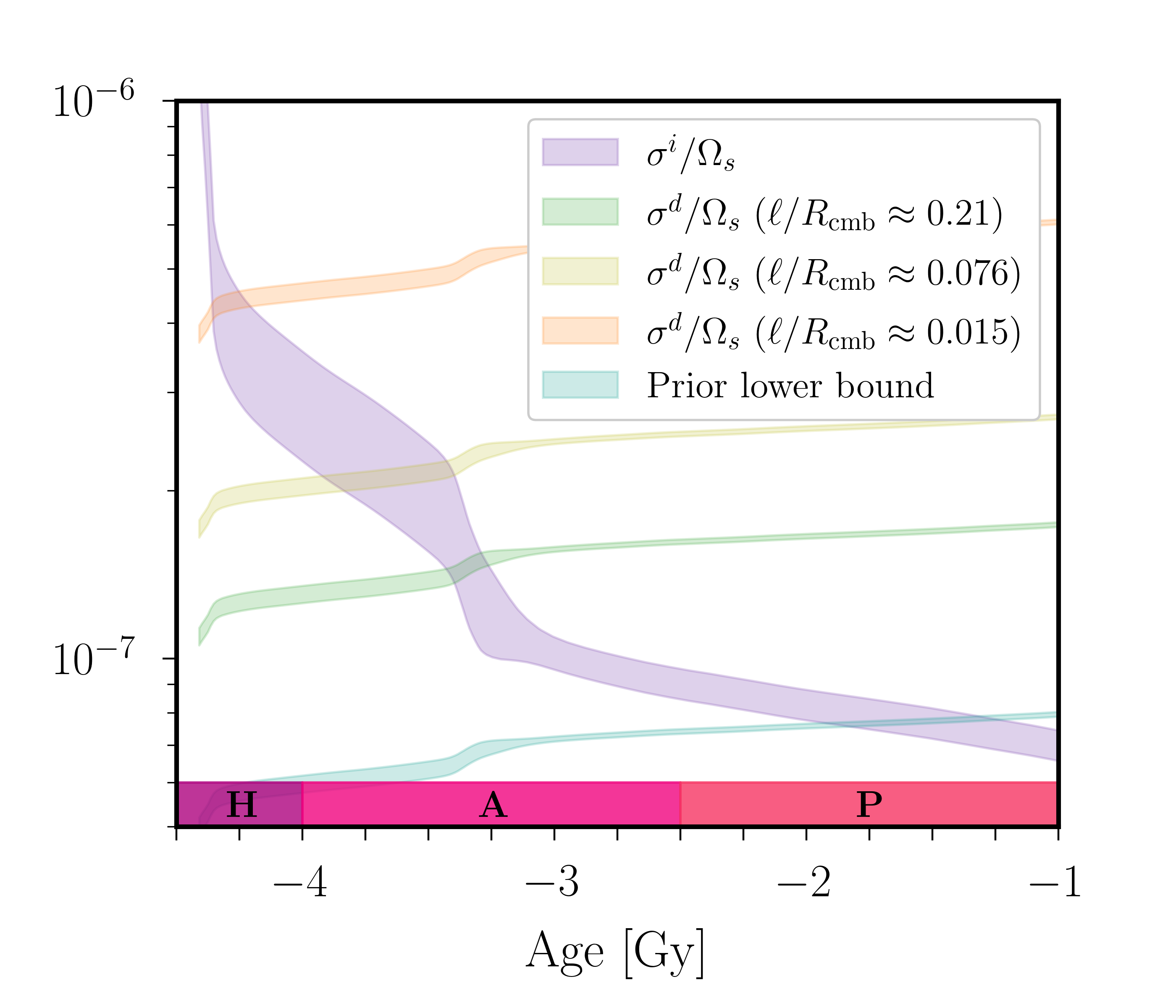} \\
    (a) & (b) \\
    \end{tabular}
    \caption{\textbf{(a)}~Surface and bulk contributions to the viscous damping $\sigma^\nu$, as a function of the polynomial degree $n$ of inertial modes with $\beta = 10^{-6}$ and $f = 10^{-2}$. 
    Only the modes with $0.94 \leq |\omega_i/\Omega_s| \leq 0.98$ that could satisfy resonance condition (\ref{eq:resonancecondition}) are shown. \textbf{(b)}~Diffusionless growth rate  $\sigma^i/\Omega_s$ of unstable tidal flows as a function of time, given by formula (\ref{eq:sigmatdei}), and leading-order damping term $\sigma^d \gtrsim \sigma^\nu$ as a function of the typical length scale of expected unstable flows. We have used the standard value $\nu = 10^{-6}$~m${}^2$.s${}^{-1}$ of the core viscosity \citep[e.g.][]{de1998viscosity}. We have included the prior lower bound obtained with $\sigma^\nu_{1/2}/\Omega_s =2.62$ (blue region). Geological eons are also shown (H:  Hadean, A: Archean, P: Proterozoic).}
    \label{fig:tdei}
\end{figure*}

The geophysical models discussed above have allowed us to estimate the parameters that need to be prescribed in the fluid-dynamics models.
Hence, we can move on the hydrodynamic constraints we have about tidal flows. 
Before we can estimate the strength of dynamo action in \S\ref{subsec:resultsdynamo}, we need to estimate whether turbulent tidal flows could have been triggered in the early Earth's core in \S\ref{subsec:resultssigma} and estimate their amplitude in \S\ref{subsec:resultsturbulence}.

%-----------------------------------------------------------------------
\subsubsection{Onset of turbulent flows}
\label{subsec:resultssigma}
%-----------------------------------------------------------------------
As outlined in \S\ref{subsec:modelROAD}, unstable tidal flows can possibly grow upon the flow $\boldsymbol{U}_0$ with an exponentially increasing amplitude $\propto \exp(\sigma t)$ in the initial stage. 
The underlying mechanism is that of a sub-harmonic (parametric) instability, known as the elliptical instability \citep{kerswell2002elliptical,le2015flows}.
This instability results from couplings between some normal modes $\boldsymbol{u}_i$ sustained by the global rotation of the core, called inertial modes and oscillating with an angular frequency $\omega_i$, and the shear component of the forced flow $\boldsymbol{U}_0$ through the linearised term $(\boldsymbol{u}_i \boldsymbol{\cdot} \nabla) \boldsymbol{U}_0 + (\boldsymbol{U}_0 \boldsymbol{\cdot} \nabla) \boldsymbol{u}_i$ in the momentum equation.

For this instability to exist, the modes and the tidal forcing must satisfy a resonance condition in time given by \citep[e.g.][]{vidal2017inviscid}
\begin{equation}
    \omega_i \simeq \pm |\Omega_s-\Omega_\text{orb}|.
    \label{eq:resonancecondition}
\end{equation}
Imperfect resonances can only occur in condition (\ref{eq:resonancecondition}) for finite values of the ellipticity $\beta$ or the Ekman number $E$. 
Hence, the latter should be negligible in the early Earth's core with $\beta \ll 1$ and $E \to 0$.
Moreover, we emphasise that Equation (\ref{eq:resonancecondition}) is only a necessary condition, because some resonance conditions must also be fulfilled in space to have a non-zero growth rate $\sigma$ (not shown here).
Then, theory predicts that $\sigma$ is of the form
(\ref{eq:growthrateTH}), in which the diffusionless part $\sigma^i$ is given by \citep[e.g.][]{vidal2019fossil}
\begin{equation}
    \frac{\sigma^i}{\Omega_s} \lesssim \frac{(2 \tilde{\Omega} + 3)^2}{16(1+\tilde{\Omega})^2} |1 - \Omega_0| \beta
    \label{eq:sigmatdei}
\end{equation}
with $\Omega_0 = \Omega_\mathrm{orb}/\Omega_s$ and $\tilde{\Omega} = \Omega_0/(1-\Omega_0)$.
Note that the upper bound in Equation (\ref{eq:sigmatdei}) is reached when the unstable flows have a large enough spatial complexity \citep{vidal2017inviscid}.
Moreover, in the absence of strong magnetic fields, we can estimate the damping term $\sigma^d$ from the viscous damping $\sigma^\nu$ of the inertial modes involved in resonance condition (\ref{eq:resonancecondition}). 
Boundary-layer theory shows that the viscous damping of inertial modes in an ellipsoid is given by 
\begin{equation}
     \sigma^\nu\simeq \Omega_s \left (\sigma^\nu_{1/2} \, E^{1/2} + \sigma^\nu_{1} \, E \right ),
    \label{eq:dampingBLT}
\end{equation}
where $\sigma^\nu_{1/2}>0$ results from the viscous friction in the Ekman layer below the \textsc{cmb} \citep[e.g.][]{greenspan1969theory}, and $\sigma^\nu_{1}\geq 0$ is a bulk contribution \citep[e.g.][]{liao2001viscous,lemasquerier2017libration}. 

Equation (\ref{eq:dampingBLT}) is illustrated in Figure \ref{fig:tdei}~(a).
Note that the inertial modes can be expressed in terms of polynomial functions of degree $\leq n$ in rotating ellipsoids \citep[e.g.][]{backus2017completeness,CdV2025spectrum}, such that they can be computed using dedicated numerical methods \citep[e.g.][]{vidal2024inertia}.
We have only shown the modes for which we could expect resonances from condition (\ref{eq:resonancecondition}), that is with 
$|\omega_i| \sim 0.96 \pm 0.02$ for the early Earth according to Figure \ref{fig:betaftime}.
Interestingly, we deduce that no modes of degree $n<7$ could be triggered in the early Earth's core. 
Indeed, the first modes that possibly satisfy resonance condition (\ref{eq:resonancecondition}) in the early Earth are the two $n=7$ modes whose angular frequencies are
\begin{equation}
    \frac{\omega_i}{\Omega_s} \approx \pm 0.965.
\end{equation}
Among all modes satisfying the resonance condition, the latter $n=7$ modes also have the lowest surface and bulk damping contributions given by 
\begin{equation}
    \frac{\sigma^{\nu}_{1/2}}{\Omega_s} \approx 5.73, \quad \frac{\sigma^{\nu}_{1}}{\Omega_s} \approx 261.
\end{equation}
Hence, we can conclude that the damping term had a lower bound in the early Earth given by
\begin{equation}
    {\sigma^d}/{\Omega_s} \geq 5.73 \, E^{1/2} + 261 \, E.
    \label{eq:dampingTDEIEarlyEarth}
\end{equation}
This is an improvement with respect to the prior lower bound ${\sigma^d}/{\Omega_s} \geq 2.62 \, E^{1/2}$ \citep[e.g.][]{landeau2022sustaining}, which corresponds to the damping term of the $n=1$ spin-over modes with $\omega_i/\Omega_s \approx \pm 1$ \citep{greenspan1969theory}. 
However, theory shows that such modes cannot yield an elliptical instability for the \textsc{cmb} geometry with $a \geq b \geq c$ \citep{cebron2010systematic}. 
The damping term in Equation (\ref{eq:growthrateTH}) is thus at least two times larger than previously thought  for the early Earth. 
Moreover, given the observed scalings of $\sigma^\nu_{1/2}$ and $\sigma^\nu_{1}$ with $n$ in Figure \ref{fig:tdei}~(a), perturbations with $n \gg 1000$ would be required to have a bulk contribution larger than the surface one in Equation (\ref{eq:dampingTDEIEarlyEarth}).

Next, we compare $\sigma^i$ given by Equation (\ref{eq:sigmatdei}) and $\sigma^\nu$ in Figure \ref{fig:tdei}~(b).
We have estimated $\sigma^\nu$ from Figure \ref{fig:tdei}~(a) for unstable flows with the typical length scale $\ell(n)$ estimated as \citep{nataf2024dynamic}
\begin{equation}
    \frac{\ell}{R_\text{cmb}} \simeq  \frac{1}{2} \frac{\pi}{n + 1/2},
    \label{eq:lton}
\end{equation}
where $n$ is the degree of the corresponding modes in the resonance condition.
First, we see that small-scale flows (e.g. with $n > 100$) were certainly entirely damped by viscosity over the entire Earth's history. 
Therefore, only some large-scale flows may have become unstable in the early Earth's core.
Moreover, we see that the largest-scale modes with $n=7$ can only be triggered before $-3.25$~Gy.  
These results provides a strong constraint on the age of a possible tidally driven dynamo in the core.
Indeed, we can assert that paleomagnetic measurements younger than $-3.25$~Gy, which had a strong magnetic field amplitude (Figure \ref{fig:paleomag}a), did record an ancient large-scale magnetic field driven by another mechanism than tidal forcing.
Yet, it remains possible so far that older points could evidence an ancient tidally driven geodynamo because tides may have injected energy into some large-scale unstable flows with $7 \leq n \ll 100$ between $-4.25$ and $-3.25$~Gy.

%-----------------------------------------------------------------------
\subsubsection{Turbulent flows}
\label{subsec:resultsturbulence}
%-----------------------------------------------------------------------
\begin{figure}
    \centering
    \includegraphics[width=0.49\textwidth]{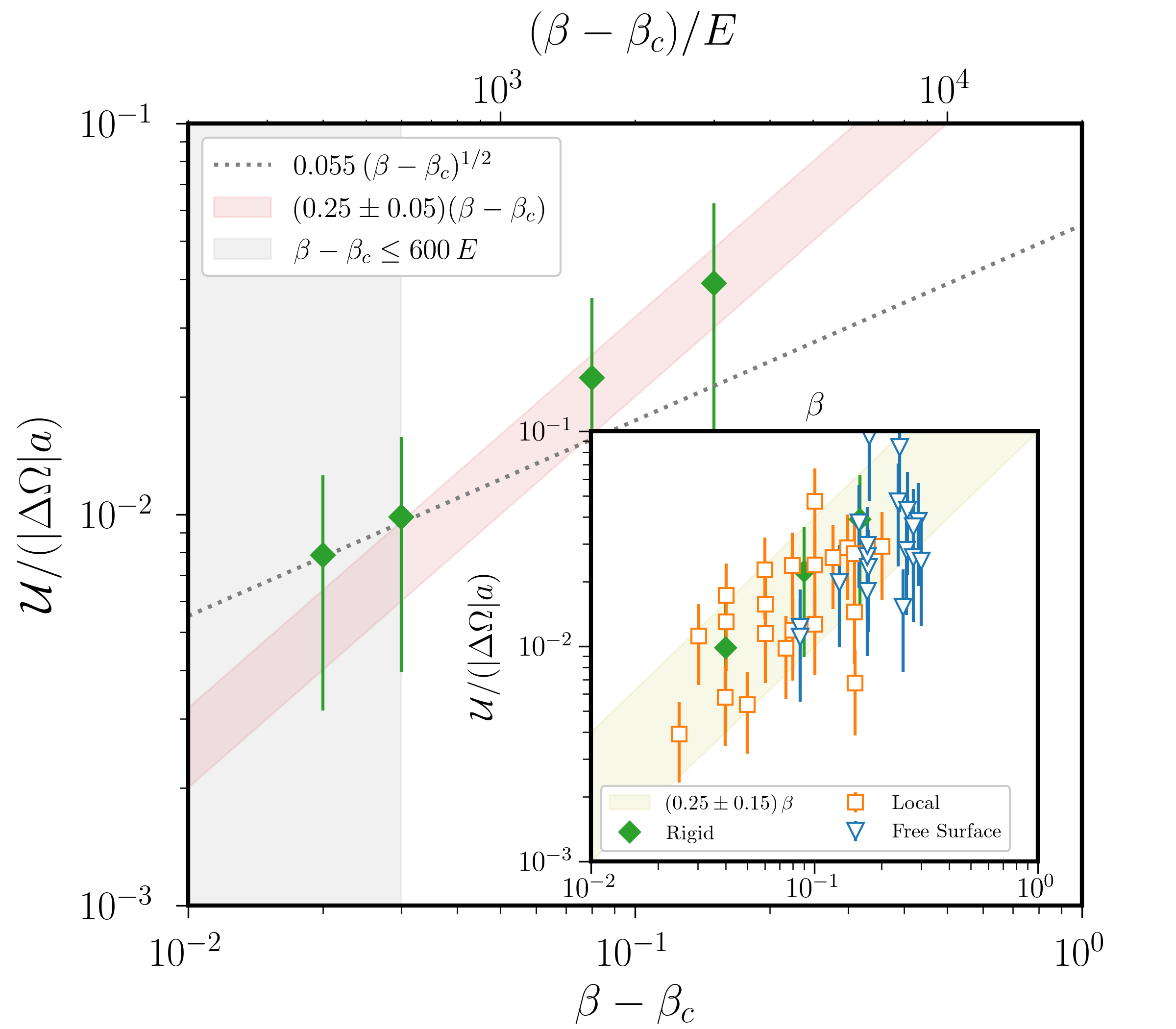}
    \caption{Normalised velocity $\mathcal{U}/(|\Delta \Omega| a)$ with $\Delta \Omega = \Omega_s - \Omega_\mathrm{orb}$, as a function of $\beta - \beta_c$ (with $\beta_c\approx10^{-2}$) in \textsc{dns} of tidally driven flows in rigid ellipsoids \citep{grannan2017tidally}. 
    Inset also shows the velocity but as a function of $\beta$ when $\beta \gg \beta_c$ for additional simulations in ellipsoids with a free-surface condition \citep{barker2016non}, and in Cartesian boxes \citep{barker2014non}.
    In the latter case, we have defined $a = \sqrt{1+\beta}$ for the normalisation. 
    Coloured region shows scaling law $\mathcal{U}/(|\Delta \Omega| a) = (0.25 \pm 0.15) \, \beta$, which broadly agrees with simulations.}
    \label{fig:turbulence}
\end{figure}

Linear analysis yields predictions for the time window where turbulent flows may be triggered by elliptical instabilities (sustained by tidal forcing).
Now, we have to estimate the typical velocity amplitude $\mathcal{U}$, as this quantity will play an important role in the planetary extrapolation.
To do so, we can have a look at simulations of tidally driven flows without magnetic fields. 
Actually, several turbulence regimes can be expected.
Weakly nonlinear analysis shows that the normal form of the elliptical instability is that of a supercritical Hopf bifurcation \citep{knobloch1994normal,kerswell2002elliptical}.
Hence, the saturation amplitude should scale as $\mathcal{U}/(|\Delta \Omega| a) \propto \sqrt{\beta - \beta_c}$ near the onset, where $\beta_c$ is the critical ellipticity (i.e. such that $\sigma = 0$).
On the contrary, we expect $\mathcal{U}/(|\Delta \Omega| a) \propto \beta - \beta_c \sim \beta$ far enough from the onset according to phenomenological arguments \citep{barker2014non}.

We estimate $\mathcal{U}$ below from the volume average of the axial velocity component $u_z$, which has been reported in prior numerical and experimental studies.
Since we have $\boldsymbol{U}_0 \boldsymbol{\cdot} \boldsymbol{1}_z = 0$, any departure from zero will be associated to tidally driven turbulence and mixing. 
Moreover, it is often assumed to be a good proxy for the mixing in the core driven by tides, which is key dynamo action driven by tides \citep[e.g.][]{vidal2018magnetic}.
We show in Figure \ref{fig:turbulence} the normalised velocity amplitude $\mathcal{U}$, obtained from simulations in ellipsoids with a rigid boundary at $E=1.5 \times 10^{-5}$ and $|1-\Omega_\text{orb}/\Omega_s| \approx 1.98$ \citep{grannan2017tidally}.
We do recover the two expected regimes in the simulations. 
The transition is believed to occur when $\beta - \beta_c \sim \mathcal{O}(E)$ from theory \citep{kerswell2002elliptical}.
This is again in broad agreement with the simulations, but we report here a quite large numerical pre-factor since it occurs at $\beta - \beta_c \sim 600 \, E$.
The second regime with $\beta \gg \beta_c$ can be more efficiently probed by relaxing the no-penetration BC in the model.
This can be achieved by using a free-surface condition in an ellipsoid \citep{barker2016non}, or by performing simulations of turbulent flows growing upon $\boldsymbol{U}_0$ in Cartesian periodic boxes \citep{barker2014non}.
Such simulations, performed for different values of $E$ and forcing frequencies $|1-\Omega_\text{orb}/\Omega_s|$, are gathered in the inset of Figure \ref{fig:turbulence}. 
Although they have been performed for different parameters, almost all simulations are well reproduced by the linear scaling law $u_z/(|\Delta \Omega| a) \propto (0.25 \pm 0.15) \ \beta$ in the second regime (see the inset in Figure \ref{fig:turbulence}).

To apply these results to the early Earth, we need to estimate how supercritical the early Earth's core was before $-3.25$~Gy.
Going back to Figure \ref{fig:tdei}~(b), we see that super-criticality is never very large (i.e. $\sigma^i/\sigma^\nu \ll 10$).
We can estimate a lower bound for the critical value of the ellipticity at the onset from equations (\ref{eq:sigmatdei}) and (\ref{eq:dampingTDEIEarlyEarth}). 
This yields
\begin{equation}
    \beta_c \gtrsim \frac{5.73  \, E^{1/2}}{|1 - \Omega_\text{orb}/\Omega_s|} \frac{16(1+\tilde{\Omega})^2}{(2 \tilde{\Omega} + 3)^2},
\end{equation}
from which we obtain the typical value $\beta_c \geq 2.5 \times 10^{-7}$ before $-3.25$~Gy.
Hence, we estimate that we had at most $\beta/\beta_c \lesssim 2$ after $-4$~Gy from Figure \ref{fig:betaftime}.

Assuming that the transition between the two regimes still occurs at $\beta - \beta_c \sim 600 \, E$ when $E\ll1$, the early Earth's core would have been in the second regime for most of the Hadean and Archean eons (since $\beta - \beta_c \gg 600 \, E$).
Therefore, we can consider that the tidally driven velocity amplitude (in planetary cores) is given by
\begin{subequations}
\label{eq:scalinglawUtides}
\begin{equation}
    \mathcal{U} \simeq \alpha_1 \beta |\Delta \Omega| R_\text{cmb}, \quad \alpha_1 = 0.25 \pm 0.15,
    \tag{\theequation a,b}
\end{equation}
\end{subequations}
as deduced from the simulations above. 
It is worth noting here that \citet{landeau2022sustaining} assumed that $\alpha_1=1$ in their velocity scaling law, which is at odds with the numerical results gathered in Figure \ref{fig:turbulence}.

Finally, we point out that scaling law (\ref{eq:scalinglawUtides}) says almost nothing about the characteristics of the underlying turbulence.
Indeed, the properties of tidally driven turbulence remains largely disputed for planetary conditions \citep[e.g.][]{vidal2024geophysical}.  
Scaling law (\ref{eq:scalinglawUtides}) may apply for tidally driven flows characterised by weakly nonlinear interactions of three-dimensional waves
when $Ro \ll 1$ \citep{le2017inertial,le2019experimental,le2021evidence}, as well as with strong geostrophic flows when $Ro \gtrsim 10^{-2}$ \citep[e.g.][]{barker2014non,barker2016non,vidal2018magnetic}.

%-----------------------------------------------------------------------
\subsection{A dynamo scaling law}
\label{subsec:resultsdynamo}
%-----------------------------------------------------------------------
\begin{figure}
    \centering
    \includegraphics[width=0.49\textwidth]{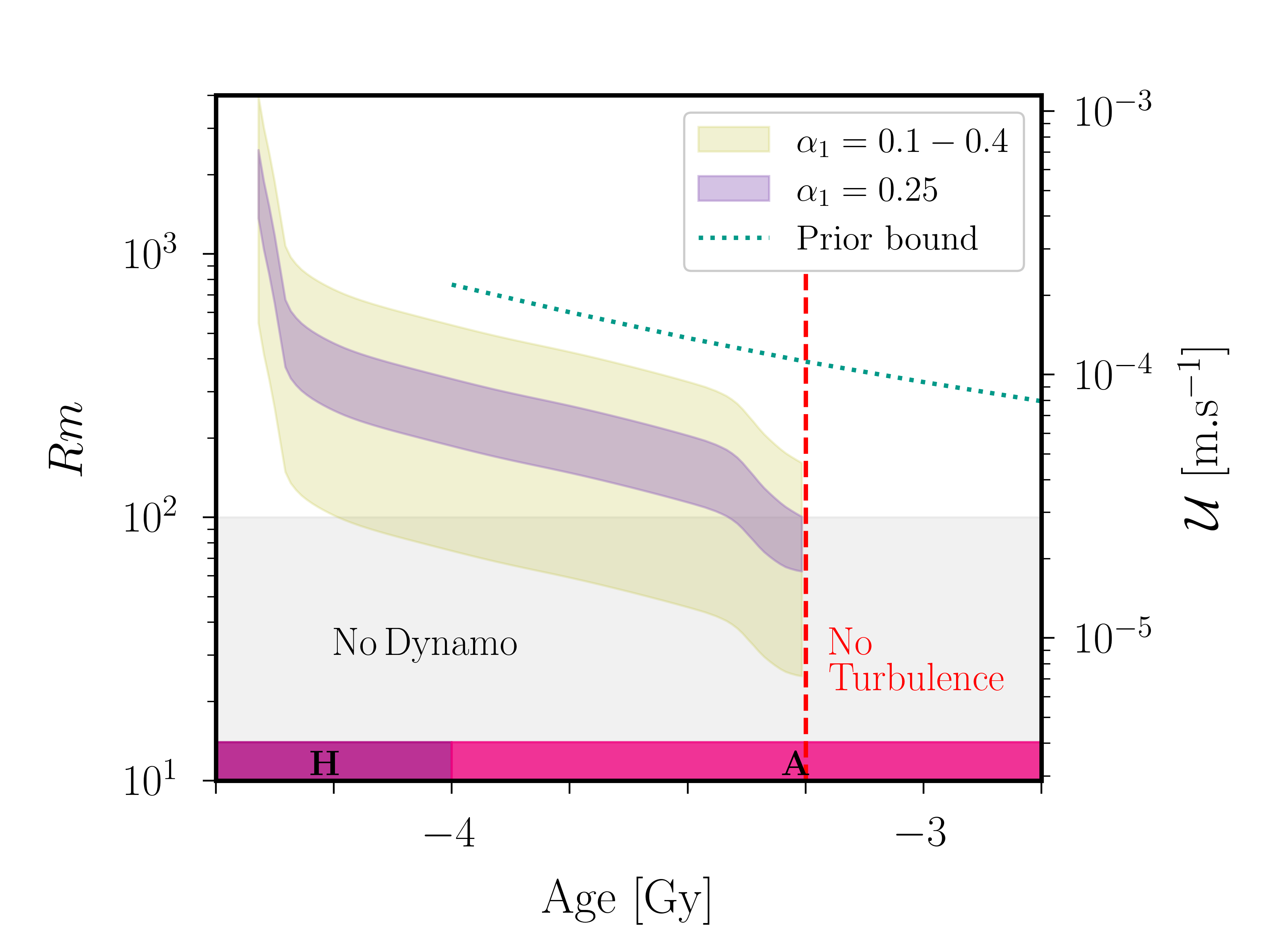}
    \caption{Magnetic Reynolds number $Rm$ as a function of age $\tau$. Second $y-$axis shows the amplitude $\mathcal{U}$ of the expected turbulence according to formula (\ref{eq:scalinglawUtides}) when $\tau \leq -3.25$~Gy. Gray zone shows the non-dynamo region $Rm \leq Rm_c$ with $Rm_c = 100$. Predictions younger than $-3.25$~Gy are hidden, since the elliptical instability was not triggered (see Figure \ref{fig:tdei}b). Prior bound from \citet{landeau2022sustaining} has been included for comparison. Geological eons are also shown (H: Hadean, A: Archean).}
    \label{fig:Rmevolution}
\end{figure}

Given the scaling law for the velocity amplitude, we evaluate in Figure \ref{fig:Rmevolution} the value of $Rm$, defined in Equation (\ref{eq:Rmnumber}), as a function of time. 
We have chosen a critical value $Rm_c = 100$ for the onset of dynamo action, which is standard in dynamo studies.
Our results show a more pessimistic view than the one presented in \citet{landeau2022sustaining}, in which the $Rm$ values seem overestimated due to the chosen upper bound value $\alpha_1 = 1$ in the velocity scaling law. 
Indeed, taking all uncertainties into account (e.g. on $\alpha_1$ in the scaling law), we see that the $Rm$ value is rather loosely constrained during most of the Hadean and Archean eons. 
The uncertainties in the Earth-Moon model yield $Rm$ values that can vary by a factor $2$, and those in scaling law (\ref{eq:scalinglawUtides}) even yield more larger variations.
As such, tidally driven dynamo action might have well never existed or ceased $-4.25$~Gy ago, or even have operated until the flow turbulence ceased near $-3.25$~Gy. 
Therefore, a putative tidally driven dynamo was likely less super-critical than previously thought. 
Smaller $Rm$ values not only narrow the time window for a tidally driven dynamo, but also weaken the magnetic field possibly sustained by such a mechanism. 
In particular, since the amplitude of the tidal forcing decreases over time (see Figure \ref{fig:tdei}b), $Rm-Rm_c$ decreases during the Hadean and Archean eons.
Hence, we expect the magnetic field driven by tidal forcing to weaken over time.
This may be at odds with the paleomagnetic measurements shown in Figure \ref{fig:paleomag}~(a), which may suggest that the (maximum) amplitude of the geomagnetic field did not vary much between $-3.5$ and $-2.5$~Gy. 
In the following, we focus on dynamo action during the late Hadean and Archean eras (where more powerful tidally driven dynamos may be expected).

Since state-of-the-art \textsc{dns} cannot properly investigate dynamo action for realistic \textsc{cmb} geometries and in geophysical conditions (even for convective flows in spherical geometries), appropriate scaling laws must be developed to establish a connection between dynamo modelling and geophysical parameters.
We assume that $Rm - Rm_c$ was large enough at that time, to render the proposed scaling laws for dynamo action in the vicinity of the onset invalid \citep[e.g.][]{fauve2007scaling}.
A fruitful approach in dynamo theory is to consider power-based scaling laws \citep{christensen2006scaling,christensen2010dynamo,davidson2013scaling}. 
For convection-driven dynamos, such laws are often tested against numerical results regardless of the parameters  \citep[e.g.][]{oruba2014predictive,schwaiger2019force}, and have provided useful insight into planetary extrapolation.  
In such scaling theories, the saturated magnetic energy density per unit of mass, which is given by
\begin{equation}
    \mathcal{E}(\boldsymbol{B}) = \frac{1}{\rho_f V} \int_V \frac{\boldsymbol{B}^2}{2 \mu} \, \mathrm{d} V \sim \frac{1}{2} \frac{B^2}{\rho_f \mu}
    \label{eq:MagNRJ}
\end{equation}
where $\rho_f$ is the mean fluid density and $B$ is a typical magnetic field strength in the dynamo region, should be somehow related to the available power per unit of mass for dynamo action $\mathcal{P}_M = \mathcal{P}/(\rho_f V)$, where $\mathcal{P}$ is the mean energy production rate (in W), and to the Joule dissipation per unit of mass given by
\begin{equation}
    \epsilon_\eta = \frac{1}{\rho_f V} \int_V \frac{\eta}{\mu} |\nabla \times \boldsymbol{B}|^2 \, \mathrm{d} V \sim \frac{\eta}{\rho_f \mu} \frac{B^2}{\ell_B^2},
    \label{eq:epsilon_eta}
\end{equation}
where $\ell_B \sim \sqrt{2 \eta \, \mathcal{E}(\boldsymbol{B})/\epsilon_\eta}$ is a magnetic dissipation length scale.
Note that $\epsilon_\eta$ is expected to dominate over the viscous dissipation $\epsilon_\nu$ for a turbulent dynamo when $Pm \ll 1$. 
Here, we follow \citet{davidson2013scaling} to assume that, in the low-$Ro$ regime characterising the Earth's core, the magnetic field solely scales with $\mathcal{P}_M$ and $R_\mathrm{cmb}$. 
If so, we obtain from dimensional analysis that
\begin{equation}
    B \propto \sqrt{\rho_f \mu} \, \left ( R_\text{cmb} \mathcal{P}_M \right )^{1/3}.
    \label{eq:lawBdavidson}
\end{equation}
Plugging typical values for the current Earth's liquid core (i.e. $\rho_f \sim 10^4$~kg.m${}^{-3}$, $\mathcal{P} \sim 1-10$~TW) into Equation (\ref{eq:lawBdavidson}) yields $B \sim 1-3$~mT, which is a satisfactory upper bound for the current magnetic field amplitude atop the core \citep[e.g.][]{gillet2010fast}.
This law generalises previous laws for buoyancy-driven dynamos  \citep[e.g.][]{christensen2009energy}, which assume that $B$ varies with the advected energy flux to the power $1/3$.
As such, scaling law (\ref{eq:lawBdavidson}) is often a cornerstone for extrapolation to natural dynamos.
However, in practice, other laws may also be relevant (e.g. see in Appendix \ref{sec:appendix} for weakly turbulent dynamos). 
Finally, for the comparison with paleomagnetic data, note that only upper bounds are generally obtained from scaling law (\ref{eq:lawBdavidson}).
Indeed, the surface field atop the dynamo region is usually only a fraction of typical field amplitude $B$ \citep[e.g.][]{aubert2017spherical}. Moreover, the surface field is not always purely dipolar.
This is measured by introducing the dipolar fraction $0 \leq f_\mathrm{dip} \leq 1$ as a pre-factor in the dynamo scaling law \citep{christensen2006scaling}. 
Numerical simulations show that self-sustained dynamos can have very different values of $f_\mathrm{dip}$ over the parameter space, as reported for convection-driven \citep[e.g.][]{oruba2014predictive,schwaiger2019force} or precession-driven dynamos \citep{cebron2019precessing}. 
Therefore, only a fraction of the produced dynamo field has a dipolar morphology at the \textsc{cmb}. 
Without further knowledge, we will discard such prefactors for tidally driven dynamos below, to focus on upper-bound estimates.

\begin{figure}
    \centering
    \includegraphics[width=0.49\textwidth]{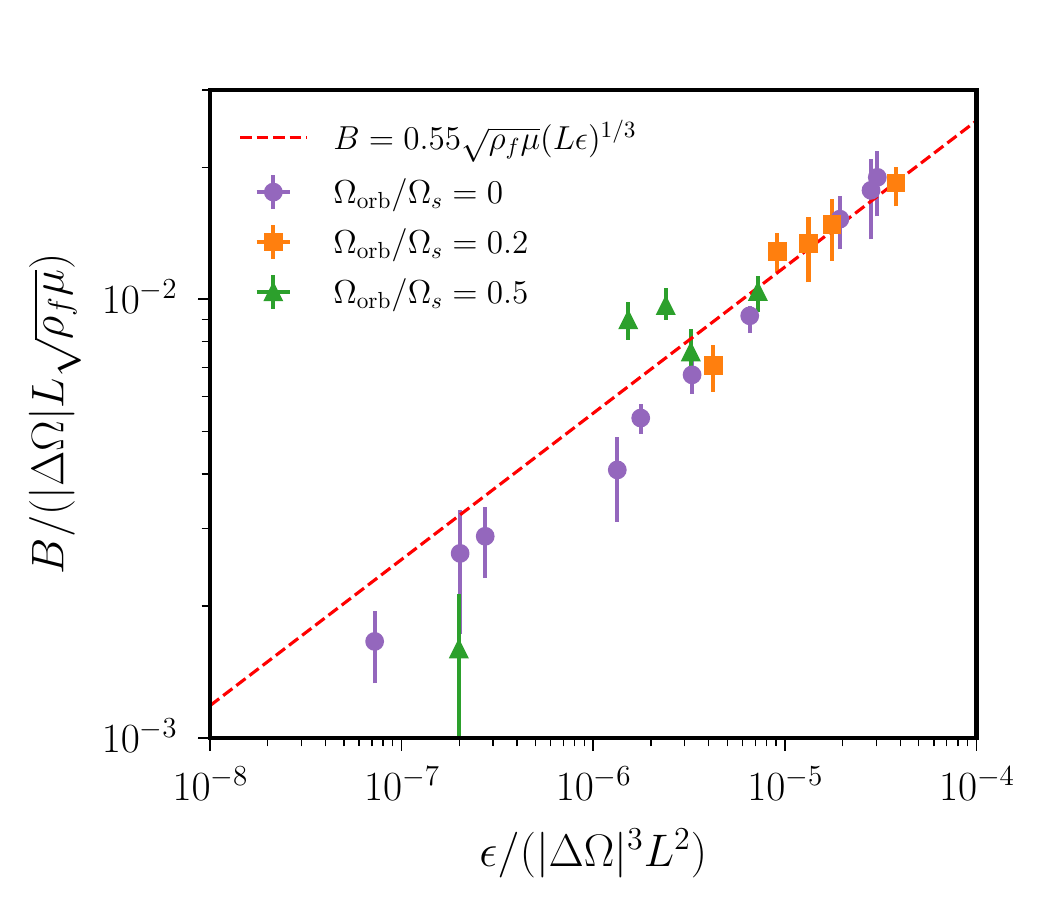}
    \caption{Simulations of tidally driven flows at moderate values of $Ro$, performed in Cartesian periodic boxes of unit length $L$ at $Pm=1$. Data from \citet{barker2014non}. Typical magnetic field amplitude $B$ defined from Equation (\ref{eq:MagNRJ}), as a function of total dissipation $\epsilon = \epsilon_\nu + \epsilon_\eta$.}
    \label{fig:scalingB2}
\end{figure}

Next, could we faithfully use scaling law (\ref{eq:lawBdavidson}) for tidally driven dynamos?
Since there are no dynamo simulations of tidal flows in ellipsoids against which we can compare theoretical predictions, we cannot be assertive.
However, we expect the above power-based arguments to remain largely valid for orbitally driven flows. 
We have re-analysed in Figure \ref{fig:scalingB2} magnetohydrodynamics simulations of tidally driven flows performed in Cartesian periodic boxes at $Ro \gtrsim 10^{-2}$ \citep{barker2014non}.
Note that it was not possible to separate $\epsilon_\nu$ and $\epsilon_\eta$ in the re-analysis of the published magnetohydrodynamic simulations. 
Yet, a good agreement is found with scaling law (\ref{eq:lawBdavidson}), assuming that $\mathcal{P}_M \sim \epsilon_\nu + \epsilon_\eta$ in a statistically steady state. 
Such observations are very promising, but quantitative applications to planets remain somehow speculative at present for tidal forcing. Indeed, it is difficult to safely estimate $\mathcal{P}_M$ for geophysical conditions.

%-----------------------------------------------------------------------
\subsection{Towards planetary conditions}
\label{subsec:extrapolation}
%-----------------------------------------------------------------------
We have found that dynamo scaling law (\ref{eq:lawBdavidson}) is likely valid for tidally driven flows.
However, it remains difficult to apply this law in practice, because tidally driven turbulence is still poorly understood at planetary core conditions. 
To make progress in this direction, we first present in \S\ref{subsubsec:rationale} the arguments that underpin our planetary extrapolation in \S\ref{subsubsec:cartoon}.

%-----------------------------------------------------------------------
\subsubsection{Heuristic rationale}
\label{subsubsec:rationale}
\begin{figure}
    \centering
    \includegraphics[width=0.49\textwidth]{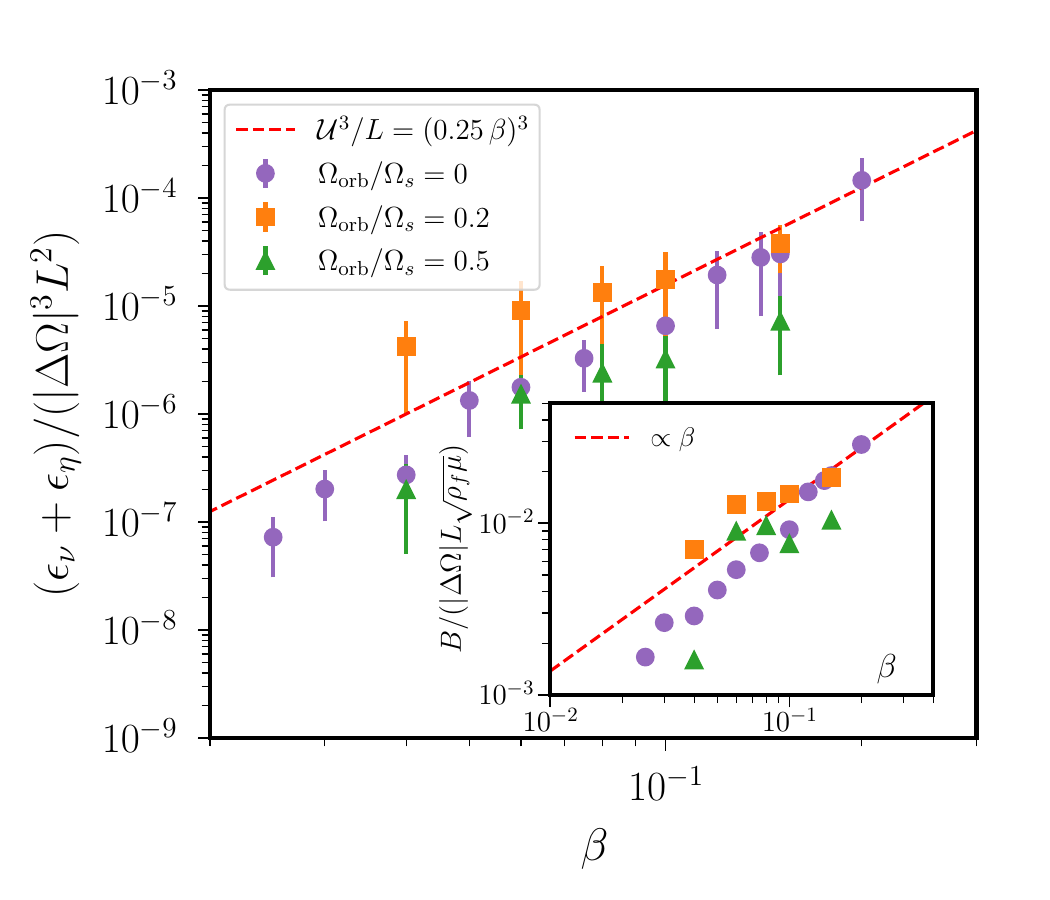}
    \caption{Simulations of tidally driven flows at moderate values of $Ro$, performed in Cartesian periodic boxes of unit length $L$ at $Pm=1$. Data from \citet{barker2014non}. Total dissipation $\epsilon_\nu + \epsilon_\eta$ as a function of equatorial ellipticity $\beta$. Inset shows the magnetic field amplitude as a function of $\beta$.}
    \label{fig:heuristicBarker}
\end{figure}

%-----------------------------------------------------------------------
We can further analyse the dynamo simulations presented in Figure \ref{fig:scalingB2}, as they can guide us for the planetary extrapolation below.
Indeed, these simulations were performed for moderate values of the Rossby number $Ro \gtrsim 10^{-2}$, for which scaling arguments have been proposed in rotating turbulence. 
In such a regime, rotating turbulence usually exhibits strong nearly two-dimensional (geostrophic) flows \citep[e.g.][]{le2019experimental,le2020near}. 
Two different scaling theories have been proposed when $Ro \lesssim 1$, such that that the mean viscous dissipation could either scale as
\citep{nazarenko2011critical,baqui2015phenomenological}
\begin{subequations}
\begin{equation}
    \epsilon_{\nu} \sim \frac{u^3_{\ell_\parallel,\ell_\perp}}{\ell_\perp} \quad \text{or} \quad \epsilon_{\nu} \sim  \frac{u^3_{\ell_\parallel,\ell_\perp}}{\ell_\parallel},
    \tag{\theequation a,b}
\end{equation}
\end{subequations}
where $u_{\ell_\parallel,\ell_\perp}$ is the velocity amplitude at the length scales $\ell_\parallel$ and $\ell_\perp$ (where $\ell_\parallel$ is the length scale parallel to the rotation axis, and $\ell_\perp$ is the one perpendicular to it). 
We see that we recover from the two formulas the usual Kolmogorov prediction $\epsilon_\nu \sim u^3_\ell/\ell$ for homogeneous isotropic turbulence when $l_\perp = l_\parallel = \ell$. 
Assuming that $u_{\ell_\parallel, \ell_\perp} \lesssim \mathcal{U}$ and $\ell_\perp \sim \ell_\parallel \sim \ell$ at large scales, the two laws should reduce at large scales to
\begin{equation}
    \epsilon_{\nu} \sim \frac{\mathcal{U}^3}{\ell} \propto \beta^3
    \label{eq:epsU3/ltides}
\end{equation}
where $\ell$ is some length scale and $\mathcal{U}$ is the flow amplitude given by equations (\ref{eq:scalinglawUtides}a,b). 
As illustrated in Figure \ref{fig:heuristicBarker}, it turns out that the total dissipation $\epsilon = \epsilon_\nu + \epsilon_\eta$ in the dynamo simulations is in very good quantitative agreement with the above scaling law (i.e. when $\epsilon_\nu$ is replaced by $\epsilon$). 
Moreover, dynamo action operates in a weak-field regime (not shown) with $\mathcal{E}(\boldsymbol{B})/\mathcal{E}(\boldsymbol{u}) < 1$, where $\mathcal{E}(\boldsymbol{u})$ is the kinetic energy.

These results show that, for planetary extrapolation, the injected power $\mathcal{P}_M$ for dynamo action can be estimated from the bulk dissipation sustained by the turbulent flows in a statistically steady state. 
Moreover, the theoretical prediction laws could hold in a dynamo regime by considering the total dissipation (instead of the viscous one). 
Such heuristic findings are consistent with the fact that, in the nonlinear regime, the elliptical-instability mechanism involves inertial-wave motions. 
Indeed, inertial waves are barely affected by magnetic effects in realistic core conditions, and are such that $\mathcal{E}(\boldsymbol{B})/\mathcal{E}(\boldsymbol{u}) < 1$. 
This results from the dispersion relation of magnetohydrodynamic waves in unbounded fluids \citep[e.g.][]{moffatt2019self}, and have also been obtained numerically in an ellipsoid \citep{vidal2019fossil,gerick2020pressure}. 
It is also worth noting that mean-field dynamo theory shows that turbulent interactions of inertial waves could sustain weak-field dynamo action \citep{moffatt1970dynamo,moffatt1970turbulent}. 
Therefore, we assume below that inertial wave motions play a key dynamical role in sustaining a weak-field dynamo regime driven by tidal forcing, which will allow extrapolating our theoretical laws to the Earth.

%-----------------------------------------------------------------------
\subsubsection{A plausible extrapolation}
\label{subsubsec:cartoon}
%-----------------------------------------------------------------------
The regime $Ro \lesssim 1$ described above could apply to short-period Hot Jupiters \citep[e.g.][]{barker2014non,barker2016non} or binary systems \citep[e.g.][]{vidal2019fossil}, in which tidal forcing can be much stronger such that large values $\beta \to 10^{-2}$ could be obtained.
On the contrary, the Earth's core is characterised by much smaller values $Ro \sim 10^{-7}-10^{-6}$. 
For such small values $Ro \ll 1$, tidal forcing is believed to sustain a regime of inertial-wave turbulence \citep{le2017inertial,le2019experimental,le2021evidence}.
This is a regime of weak turbulence, characterised by weakly nonlinear interactions of three-dimensional inertial waves. 
Like in Kolmogorov turbulence, inertial-wave turbulence involves energy being injected at some large scale, denoted by $\ell$ below. 
This energy is then transmitted to smaller scales via a direct cascade in an inertial range, in which the input power is balanced by dissipation at every scale, until energy is finally dissipated at sufficiently small scales at a rate $\epsilon$.
However, contrary to isotropic homogeneous turbulence, inertial-wave turbulence is described by an anisotropic energy spectrum \citep{galtier2003weak,galtier2023multiple}, depending on the two length scales $\ell_\parallel \geq \ell_\perp$ introduced above.

By analogy with the regime $Ro \lesssim 1$ described in \S\ref{subsubsec:rationale}, we assume that the injected power $\mathcal{P}_M$ available for dynamo action can be estimated in a statistically steady state from the dissipation $\epsilon$ of turbulent flows given by wave-turbulence theory when $Ro \ll 1$. 
This rests on the fact that inertial waves are barely modified by magnetic effects at core conditions, such nonlinear interactions of almost pure inertial waves will still be triggered in a weak-field dynamo regime.  
The main difference with the pure hydrodynamic regime would be that, for small values $Pm \ll 1$, the dissipation would occur on a diffusive magnetic length scale larger than the viscous one, such that the width of the inertial range (in the wavenumber space) would be shortened compared to the hydrodynamic case. 
Hence, we estimate the effective dissipation as \citep{galtier2003weak,galtier2023multiple}
\begin{equation}
    \epsilon_{Ro \ll 1} \sim \frac{\ell_\parallel}{\ell_\perp} \frac{u_{\ell_\parallel,\ell_\perp}^4}{\Omega_s \ell_\perp^2}
    \label{eq:IWT}
\end{equation}
with an expected $\mathcal{O}(1)$ pre-factor from theory \citep{zeman1994note,zhou1995phenomenological}, where $u_{\ell_\parallel,\ell_\perp}$ is the velocity amplitude at the length scales $\ell_\parallel \geq \ell_\perp$
Since the dissipation is a constant in the theory, it can be estimated from the knowledge of $u_{\ell_\parallel,\ell_\perp}$ at some length scales.

Figure \ref{fig:tdei}~(b) shows that tidal forcing can only inject energy at rather large scales $\ell$ (i.e. at fraction of the core radius denoted by $\alpha_2$ below), for which we may assume $\ell_\perp \sim \ell_\parallel \sim \ell$.
Moreover, scaling law (\ref{eq:scalinglawUtides}) shows that  $u_{\ell_\parallel,\ell_\perp} \lesssim \mathcal{U}$ at large scales.
Altogether, this allows us to estimate the mean dissipation in a regime of inertial-wave turbulence as
\begin{equation}
    \epsilon_{Ro \ll 1} \lesssim \frac{\mathcal{U}^4}{\Omega_s \ell^2}
    \label{eq:IWT2}
\end{equation}
for tidal forcing, with $\ell = \alpha_2 R_\mathrm{cmb}$ and $\alpha_2 \simeq 0.01 - 1$.
Formula (\ref{eq:IWT2}) is compatible with an energy spectrum scaling as $\ell^2$ at large scales \citep{galtier2003weak}, which is consistent with prior studies in rotating turbulence \citep[e.g.][]{baroud2002anomalous,thiele2009structure,li2025energy}. 
For completeness, we remind the reader that another predictive law ought to be used for tidally driven turbulence with moderately small values $Ro \lesssim 1$. 
This should be given by dissipation law (\ref{eq:epsU3/ltides}) as explained above, together with an energy injection at the large scale $\ell = \alpha_2 R_\mathrm{cmb}$ with $\alpha_2 = 0.01 - 1$ that is consistent with Figure \ref{fig:heuristicBarker}.

Finally, we can combine equations (\ref{eq:epsU3/ltides})-(\ref{eq:IWT2}) with dynamo scaling law (\ref{eq:lawBdavidson}) to obtain
\begin{equation}
    B \propto \sqrt{\rho_f \mu} \begin{cases}
     \alpha_2^{-1/3} \, \mathcal{U} & \text{if} \quad Ro \lesssim 1 \\
    \alpha_2^{-2/3} \, Ro^{1/3} \mathcal{U} & \text{if} \quad Ro \ll 1 \\
    \end{cases}
    \label{eq:BIWTtides}
\end{equation}
for tidal forcing in the early core, which yields
\begin{equation}
    B \propto \begin{cases}
     \beta & \text{if} \quad Ro \lesssim 1 \\
    \beta^{4/3} & \text{if} \quad Ro \ll 1 \\
    \end{cases}
\end{equation}
by using scaling law (\ref{eq:scalinglawUtides}) for the amplitude of turbulence. 
Note that the magnetic field amplitude is expected to be $Ro^{1/3}$ smaller when $Ro \ll 1$ than in a nearly two-dimensional (geostrophic) regime at $Ro \lesssim 1$.
We can also calculate from law (\ref{eq:BIWTtides}) the ratio of the magnetic energy to the kinetic energy per unit of mass  $\mathcal{E}(\boldsymbol{u})$ as
\begin{equation}
    \mathcal{E}(\boldsymbol{B})/\mathcal{E}(\boldsymbol{u}) \propto \begin{cases}
     \alpha_2^{-2/3} & \text{if} \quad Ro \lesssim 1 \\
    \alpha_2^{-4/3} \, Ro^{2/3} & \text{if} \quad Ro \ll 1 \\
    \end{cases}.
\end{equation}
Hence, only weak-field dynamos with $\mathcal{E}(\boldsymbol{B})/\mathcal{E}(\boldsymbol{u}) \ll 1$ are expected for a wave-turbulence when $Ro \ll 1$. 

\begin{figure}
    \centering
    \includegraphics[width=0.49\textwidth]{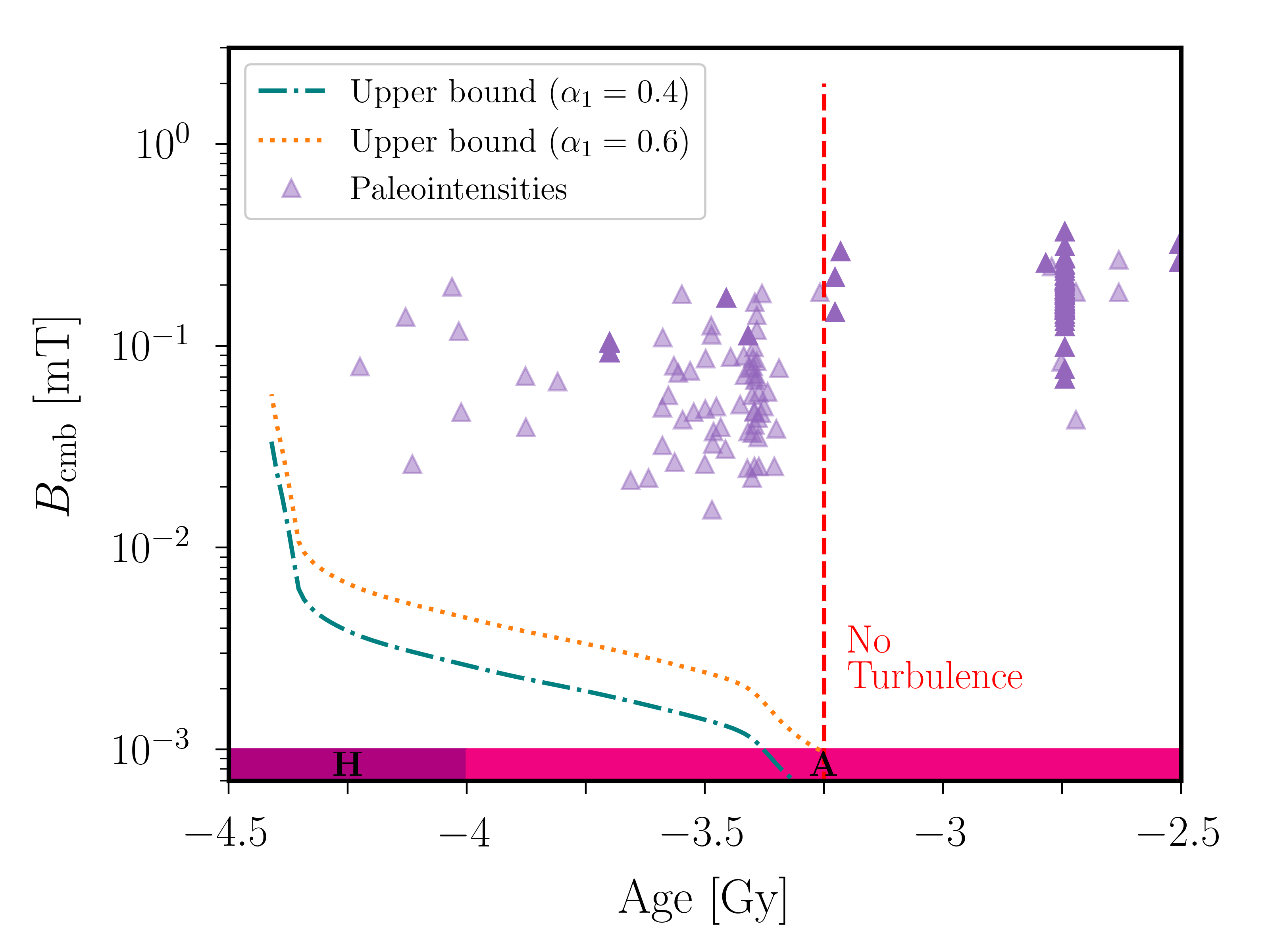}
    \caption{Paleointensity at the \textsc{cmb} $B_\mathrm{cmb}$, reconstructed Figure \ref{fig:paleomag} using Equation (\ref{eq:Bcmbfromdata}) with $R_s = 6378$~km and $R_\mathrm{cmb} = 3480$~km, and upper bounds for $B$ (dotted and dashdot lines) from scaling law (\ref{eq:BIWTtides}) when $Ro \ll 1$. Numerical estimates using $\rho_f \simeq 1.1 \times 10^4$~kg.m${}^{-3}$ for a mean density (accounting for the mass of the outer and inner cores), $\eta = 1$~m${}^{2}$.s${}^{-1}$, and $\mu = 4 \pi \times 10^{-7}$~H.m${}^{-1}$. Geological eons are also shown (H: Hadean, A: Archean).}
    % Same legend as in Figure \ref{fig:paleomag} for the paleomagnetic measurements.
    \label{fig:takehomeB}
\end{figure}

The predictions from scaling law (\ref{eq:BIWTtides}) as a function of the age are illustrated in Figure \ref{fig:takehomeB}. 
We obtain as a typical estimate $B \sim 10^{-5}-10^{-3}$~mT during the Hadean period $4.25-4$~Gy ago (i.e. when tidal forcing was maximal), and the field amplitude would then ultimately decrease until $-3.25$~Gy (since tidal forcing had a decreasing amplitude during the Archean era). 
The predicted amplitudes are thus at least ten times smaller than the \textsc{cmb} field $B_\mathrm{cmb}$, which is estimated from surface measurements with Equation (\ref{eq:Bcmbfromdata}). 
Therefore, it seems unlikely that the ancient geodynamo was solely sustained by a wave-turbulence regime at $Ro \ll 1$ driven by tidal forcing.

%-----------------------------------------------------------------------
\section{Geophysical discussion}
\label{sec:discussion}
%-----------------------------------------------------------------------
Our extrapolation suggests that tidal forcing may have been too weak to generate a dynamo magnetic field with an amplitude matching the (scarce) Hadean and Archean paleomagnetic measurements, at least for flows in a wave-turbulence regime.
Since several assumptions were made to arrive at this conclusion, we discuss below if our main results could be modified or not by adopting  other modelling choices.

%-----------------------------------------------------------------------
\subsection{Influence of the orbital scenario}
%-----------------------------------------------------------------------
Obviously, one source of uncertainty arises from the parameters given by the orbital scenario, which was less constrained in the distant past. 
We have here employed the semi-analytical model by \citet{farhat2022resonant}, as it fits most of the available geological proxies for the history of the Earth-Moon system and reproduce the age of the Moon's formation fairly well. 
However, we should assess how the results could be affected by adopting other orbital scenarios. 
Most models reasonably well agree on the recent Earth-Moon evolution (i.e. after $-1$~Gy), as they are constrained by geological data. 
However, the models can significantly differ further back in time.

\begin{figure}
    \centering
    \includegraphics[width=0.49\textwidth]{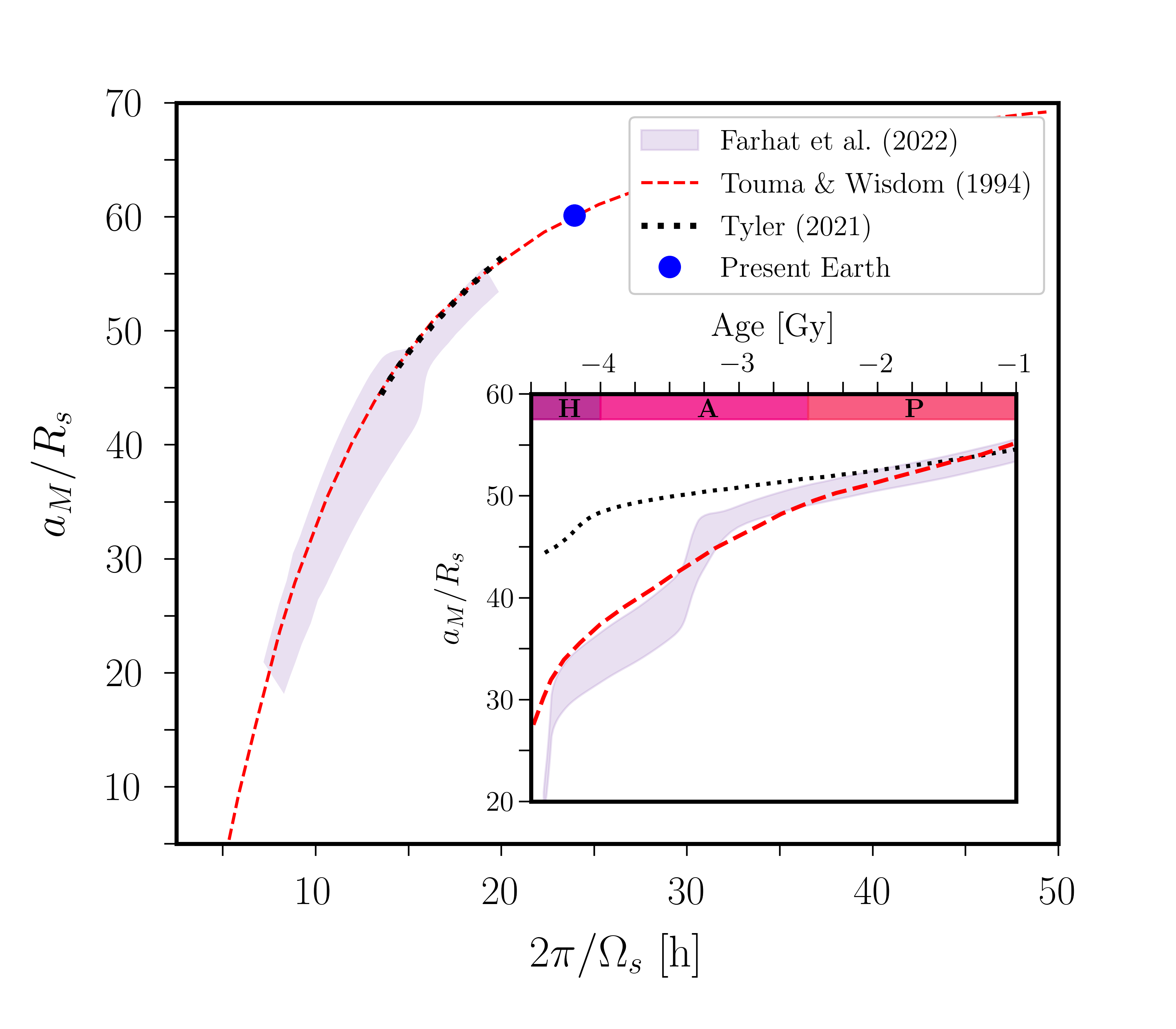}
    \caption{Uncertainties in the orbital model. Comparison between the predictions in the distant past from different orbital models \citep{touma1994evolution,tyler2021tidal,farhat2022resonant} for the Earth-Moon distance $a_M/R_s$ and the length of day. Inset is analogous to Figure \ref{fig:forcing1}~(a), and to Figure 6 in \citet{eulenfeld2023constraints}. Geological eons are also shown (H: Hadean, A: Archean, P: Proterozoic).}
    \label{fig:orbitaluncertainties}
\end{figure}

The comparison between different orbital models is shown in Figure \ref{fig:orbitaluncertainties}.
Note that we have discarded models that cannot be extrapolated during the Hadean and the Archean eras \citep[e.g.][]{green2017explicitly,zeeden2023earth,zhou2024earth}. 
We find that the presented models fall within the error bars of the model by \citet{farhat2022resonant}.
This is probably due to the conservation of angular momentum that is enforced in all the models, which gives good constraints on the Earth–Moon distance for a given value of the Earth's spin rate.
Yet, the evolution curves can differ in time between the models. 
In particular, the model by \citet{tyler2021tidal} does not reproduce the estimated age of the Moon, overestimating the Earth–Moon distance during the Hadean and Archean eras. 
A similar conclusion can be drawn for \citet{daher2021long}, as shown in Figure 6 of \citet{eulenfeld2023constraints}.
On the contrary, the model by \citet{touma1994evolution} is found to be quite close to the orbital model by \citet{farhat2022resonant}. 
Consequently, our results already account for most of the uncertainties of the community regarding the orbital scenario. 
However, the next generation of orbital models may change the overall picture. 

%-----------------------------------------------------------------------
\subsection{Rheological uncertainties}
%-----------------------------------------------------------------------
Another key quantity in the model is the present-day value of the equatorial ellipticity $\beta(0)$ at the \textsc{cmb}. 
Seismological observations of the peak-to-peak amplitude of the \textsc{cmb} topography \citep[e.g.][]{koper2003constraints,sze2003core} suggest that the current value of the equatorial ellipticity may be order-of-magnitude larger than our considered value in Equation (\ref{eq:betaradius}). 
However, as this elliptical deformation results from mantle dynamics, it is nearly in phase with the spin frequency of the Earth (i.e. $\Delta \Omega \approx 0$). 
As such, we do not expect this elliptical deformation to play a role in the elliptical-instability mechanism.
This is only the asynchronous component of the ellipticity, which is driven by tidal forcing, that is able to drive an elliptical instability inside the liquid core. 
The amplitude of the tidal potential is quite well constrained at the present time \cite{AGNEW2015151}, such that the main uncertainties on the value of $\beta(0)$ probably come from the mantle's rheology.

We have employed the \textsc{prem} model \citep{dziewonski1981preliminary} to account for the Earth's rheology in Figure \ref{fig:tidesmodel2}. 
More recent Earth models could naturally be used together with the open-source code \textsc{TidalPy} \citep{renaud2023tidalpy}, but this is beyond the scope of the present study. 
More importantly, we have assumed that the mantle remained rigid over time. 
This seems to be a reasonable assumption throughout most the Earth's history, but the Earth's mantle was probably molten after the giant impact that formed the Moon \citep[at least partially, e.g.][]{nakajima2015melting}.
Then, a crystallising basal magma ocean (\textsc{bmo}) probably survived in the aftermath for millions of years to a billion of years \citep[e.g.][]{labrosse2007crystallizing,boukare2025solidification}.
How such a \textsc{bmo} could have interacted with tidal forcing remains largely unknown \citep[e.g.][]{korenaga2025tidala,korenaga2025tidalb,korenaga2025tidalc}, as well as its possible interactions with core flows.
A crude extrapolation of old fluid-dynamics experiments on the spin-up of rotating immiscible fluids \citep{pedlosky1967spin,o1992spin} might suggest that \textsc{cmb} dissipation is not significantly reduced in the presence of a low-viscosity \textsc{bmo} (due to interfacial friction with the core, and Ekman friction with the solid mantle above).
However, further work is needed to elucidate this question.

%-----------------------------------------------------------------------
\subsection{Early core's conditions remain elusive}
%-----------------------------------------------------------------------
In addition to the orbital parameters and mantellic properties, there are also uncertainties regarding the physical state of the liquid core in the distant past. 
In particular, additional dissipation mechanisms may have operated within the core, hindering the onset of tidally driven turbulence over geological timescales.

%-----------------------------------------------------------------------
\subsubsection{Hydrodynamic dissipation in the core}
%-----------------------------------------------------------------------
The value of the core's viscosity $\nu$ appears to be another physical quantity of interest when determining the occurrence of tidally driven turbulence.
Indeed, as shown in Figure \ref{fig:tdei}, it controls the leading-order damping inhibiting the growth of unstable flows.  
We have chosen the standard value $\nu=10^{-6}$~m${}^2$.s${}^{-1}$ \citep[e.g.][]{de1998viscosity}, but the core value may be between $10^{-7}$~m${}^2$.s${}^{-1}$ and $3 \times 10^{-6}$~m${}^2$.s${}^{-1}$ \citep{mineev2004viscosity}.
Thus, in Figure \ref{fig:tdei}, the damping term may be decreased by a factor of $\sqrt{10} \approx 3$ with the lowest value, or increased by a factor $\sqrt{3} \approx 2$ with the largest one.
Similar effects could be obtained if the liquid core were rotating in the bulk faster or slower than the mantle. 
We have here assumed that the liquid core is co-rotating with the mantle at the angular velocity $\Omega_s$, but the core may be rotating a bit faster than the mantle according to some tidal models \citep[e.g.][]{wahr1981effect}.
Our model would then remain largely unchanged, except that the value of $E$ would be smaller as it must be based on the fluid rotation \citep[see in][]{greenspan1969theory}.

Note that possible interactions between tidal forcing and buoyancy effects, such as with either convective flows or density stratification, could also provide additional dissipation mechanisms in the core.
Convection would essentially sustain slowly varying flows in the core, whereas tidal forcing may mainly trigger high-frequency motions with inertial-wave turbulence.
Because of this separation of time scales, strong interactions are not expected between both flows
\citep[unless convective flows could locally cancel out the rotation of the core, e.g.][]{de2023interactions,de2023tidal}.
Convection would then provide an additional damping, but the latter might be rather weak for fast tidal forcing \citep[e.g.][]{duguid2020convective,duguid2020tidal,vidal2020efficiency,vidal2020turbulent}. 
Note that the early core may have been instead (at least partially) stably stratified in density, according to thermal evolution modelling \citep[only for large values of the thermal conductivity of liquid iron at core conditions, e.g.][]{labrosse2015thermal} or if the early core had been insufficiently mixed after giant impacts \citep{landeau2016core}.
In this case, density stratification would not modify the largest growth rate of the elliptical instability $\sigma^i$ \citep{vidal2019fossil}.
Similarly, boundary-layer theory suggests that the value of the damping rate would not change much with a stable density stratification \citep{friedlander1989asymptotic}, such that the predictions shown in Figure \ref{fig:tdei}~(b) may remain quantitatively accurate. 
However, preliminary simulations show that density stratification could significantly weaken the strength of radial flows and mixing \citep{vidal2018magnetic}. 
We may thus expect dynamo action to be less favourable with stratification, but further work is needed to carefully investigate the interplay with density stratification.

%-----------------------------------------------------------------------
\subsubsection{What about magnetic damping?}
%-----------------------------------------------------------------------
It is unclear whether the early core was subject to strong magnetic effects during the Hadean era, that is before an undoubted dynamo action was recorded in paleomagnetic data.
However, the Earth's core had a magnetic field since from at least $-3.5$~Gy, whatever its dynamical origin. 
Hence, we may wonder if the presence of an ambient magnetic field (possibly of different origin) could alter the onset of tidally driven flows within the core during the Archean era .

Actually, the elliptical-instability mechanism would be largely unchanged in the presence of a background field.
To quantify magnetic effects, we usually introduce the Lehnert number defined as
\begin{equation}
    Le = \frac{B_\mathrm{cmb}}{\Omega_s R_\mathrm{cmb} \sqrt{\rho_f \mu}}.
\end{equation}
A typical value is $Le \sim 10^{-4}-10^{-3}$ in the early core, which is similar to the current value in the Earth's core.
For such low values, the (high-frequency) inertial waves responsible for the elliptical instability are only weakly affected by the magnetic field \citep[only in the absence of an inner core, otherwise see in][]{lin2018tidal}.
Given the expected forcing frequency (in dimensionless units) $|1-\Omega_\mathrm{orb}/\Omega_s| \gg Le$ of tidal forcing inside the early core, the injection of energy can only come from resonances with nearly inertial waves in the resonance condition \citep[e.g.][]{kerswell1994tidal,vidal2019fossil}.
Then, the diffusionless growth rate $\sigma^i$ of the elliptical instability, given in Equation (\ref{eq:sigmatdei}), will remained unchanged at the leading order in $Le$.
However, Ohmic diffusion below the \textsc{cmb} will provide an additional magnetic damping term $\sigma^\eta$ in the growth rate equation.
\citet{kerswell1994tidal} showed that the magnetic damping $\sigma^\eta$ and the viscous one $\sigma^\nu$ are probably of comparable order of magnitude for plausible core fields.
Therefore, the effective damping term in Figure \ref{fig:tdei} may be increased by a factor of (up to) two.
To summarise, if we consider an ambient magnetic field in the early core, tidal forcing may not have been strong enough to sustain turbulence during the Hadean and Archean eras.

%-----------------------------------------------------------------------
\subsection{Scaling-law uncertainties}
%-----------------------------------------------------------------------
The presented tidal scenario heavily relies on different scaling laws, which must be used to extrapolate the results in the turbulent regime to core conditions.
We have done our best to constrain the various scaling laws as much as possible, by uniquely combining an Earth-Moon evolution scenario with theoretical results and re-analyses of the most up-to-date numerical simulations of tidally driven turbulent flows.
Yet, other modelling uncertainties also call the dynamo extrapolation into question.
In particular, the outcome of turbulent tidally driven flows and dynamo action remains uncertain as discussed below.

%-----------------------------------------------------------------------
\subsubsection{Comparison with earlier works}
%-----------------------------------------------------------------------
We have shown in Figure \ref{fig:Rmevolution} that the prior predictions presented in \citet{landeau2022sustaining} are likely overestimated, since we have used the same modelling assumptions in our study. 
This mainly results from the chosen numerical prefactors in the different extrapolation laws, which were set to unity for a first proof-of-concept study.

Notably, we have revisited the value of $\alpha_1$ in velocity scaling law (\ref{eq:scalinglawUtides}), showing that $\alpha_1 = 1$ disagrees with the available numerical results gathered in Figure \ref{fig:turbulence}.
Instead, the numerical simulations suggest that $\alpha_1 \simeq 0.25 \pm 0.15$ for planetary extrapolation.
Note that we have estimated the value of $\mathcal{U}$ directly from $u_z$.
Actually, we found that $|u_z| \sim 0.7 - 0.8 \, |\boldsymbol{u}|$ for most of the dynamo simulations presented in Figures \ref{fig:scalingB2} and \ref{fig:heuristicBarker}, where $|\boldsymbol{u}|$ is estimated from the kinetic energy $\mathcal{E}(\boldsymbol{u})$.
This is in good agreement with a preliminary estimate of the radial mixing induced by tidally driven flows \citep[see Figure 4 in][]{vidal2018magnetic}.
However, if the horizontal mixing were also important for dynamo action, the velocity prefactor $\alpha_1$ may be increased up to $0.6$. 
Yet, this would only lead to a small increase of the magnetic field amplitude (see the dashdot line in Figure \ref{fig:takehomeB}).
Future research work may thus shed new light on this estimate.
For instance, it is unknown whether the value of $\alpha_1$ could be increased or not when $E$ is lowered (and similarly $\beta$) but, heuristically, we always expect $\alpha_1 < 1$ to sustain tidally driven turbulence in the core over long time scales.
Otherwise, tidally driven turbulent flows would become of the same amplitude as the shear component of the forced flow $\boldsymbol{U}_0$ when $\alpha_1 \to 1$, which would temporarily stop the injection of energy and hinder the development of a wave-turbulence regime. 
This may echo some peculiar regimes in rotating turbulence, which are sometimes observed in experiments involving growth-and-collapse phases \citep[e.g.][]{mcewan1970inertial,malkus1989experimental}.
We might also obtain during the energy growth regimes of self-killing dynamos \citep[e.g.][]{reuter2009wave,fuchs1999self}, which have already been reported in simulations of precession-driven flows in a sphere \citep{cebron2019precessing}. 
If such exotic regimes were obtained, our theoretical predictions for the flow turbulence and its dynamo capability would be invalid.

Finally, there are also strong uncertainties associated with dynamo scaling law (\ref{eq:lawBdavidson}). 
Obviously, the lack of numerical methods for simulating self-sustained dynamos in ellipsoids currently hampers our ability to assess its validity for planetary relevant regimes. 
For instance, the inset of Figure \ref{fig:heuristicBarker} shows that the simulations are compatible with
\begin{equation}
    B \approx 0.55 \sqrt{\rho_f \mu} \, \left ( R_\text{cmb} \epsilon \right )^{1/3},
\end{equation}
where $\epsilon$ is given by scaling law (\ref{eq:epsU3/ltides}) in the $Ro \lesssim 1$ regime. 
The observed numerical prefactor in the above law shows that the dynamo scaling laws generally come with non-unit numerical prefactors. 
For instance, the dynamo predictions could be overestimated by a factor of nearly $2$ if the observed prefactor were the same in the $Ro \ll 1$ regime.

%-----------------------------------------------------------------------
\subsubsection{Disputed wave-turbulence regime}
%-----------------------------------------------------------------------
We have assumed that tidal forcing establishes a regime of inertial-wave turbulence,  as often postulated after \citet{le2017inertial,le2019experimental,le2021evidence}.
However, inertial-wave turbulence is a research topic that is far from being well understood, especially because it is very challenging to obtain in simulations or in laboratory experiments.
As such, current investigations still strive observing predictions of wave-turbulence theory in set-ups mimicking as much as possible the theoretical model \citep[e.g.][]{yarom2014experimental,monsalve2020quantitative}.
Therefore, it is still unclear whether the quantitative predictions of wave turbulence theory could be directly applied to rotating turbulent flows in bounded geometries at $Ro \ll 1$, such as in the early Earth's core with $Ro \sim 10^{-7}$.
On the contrary, Figure \ref{fig:heuristicBarker} shows that simulations at moderate values $Ro \sim 10^{-2}$ already agree fairly well with the dissipation law (\ref{eq:epsU3/ltides}).

Moreover, how magnetic effects modify inertial-wave turbulence remains speculative so far. 
We conjecture that inertial-wave turbulence can persist in a weak-field dynamo regime, since high-frequency inertial waves are barely affected by magnetic fields in an ellipsoid. 
The main difference would be that the dissipation would occur on a diffusive magnetic length scale for small values $Pm \ll 1$. 
This may agree with the qualitatively view drawn from magnetohydrodynamics simulations of tides \citep{barker2014non,vidal2018magnetic} and precession \citep{barker2016turbulence,kumar2024dynamo}, showing that the obtained turbulence could be largely unchanged in a weak-field dynamo regime. 
Future numerical works, for instance using local simulations at smaller values of $Pm$, may shed new light on this point.

%-----------------------------------------------------------------------
\subsubsection{A magnetostrophic dynamo regime?}
%-----------------------------------------------------------------------
Apart from wave turbulence, another possibility could be that tidally driven turbulence at $Ro \ll 1$ could be in a regime of geostrophic turbulence  characterised by the presence nearly two-dimensional (geostrophic) flows. 
This would challenge the applicability of wave-turbulence theory when $Ro \ll 1$ \citep[e.g.][]{gallet2015exact}, since geostrophic flows are filtered out in the wave-turbulence theory. 
However, a regime of hydrodynamic geostrophic turbulence is likely to be modified by magnetic effects. 
This notably rests on the properties of low-frequency waves in rotating systems subject to magnetic effects. 
Indeed, slow quasi-geostrophic wave motions morph into various low-frequency waves shaped by the magnetic field, such as torsional Alfv\'en waves \citep[e.g.][]{luo2022waves1} and magneto-Coriolis waves \citep[e.g.][]{luo2022waves2,gerick2024interannual}. 
Consequently, a magnetostrophic regime is expected to supersede purely geostrophic turbulence when $Ro \ll 1$ \citep[e.g.][]{hollerbach1996theory}.
In such a regime, the Coriolis force could balance the Lorentz force such that $B$ would be given by \citep[e.g.][]{christensen2010dynamo}
\begin{equation}
    B \propto \sqrt{\rho_f \mu} \sqrt{l_B \Omega_s \mathcal{U}}.
    \label{eq:Bmagnetostrophic}
\end{equation}
Formula (\ref{eq:Bmagnetostrophic}) would then give the field estimate $B \lesssim 5$~mT in the early Earth's core at $-4$~Gy, using the same physical parameters as in Figure \ref{fig:takehomeB} and with $\ell_B/R_\text{cmb} = 0.02$ as in \citep[as in][]{starchenko2002typical}.
A magnetostrophic regime may be a better candidate than inertial-wave turbulence to reach a strong-field dynamo regime.
Unfortunately, there is no evidence so far that magnetostrophy could be achieved with tidal forcing.

%-----------------------------------------------------------------------
\section{Concluding remarks}
\label{sec:conclusion}
\subsection{Summary}
%-----------------------------------------------------------------------
We have thoroughly explored the capability of tidal forcing to explain the ancient geodynamo.
We have combined geophysical constraints from Earth-Moon evolution models, theoretical predictions, and re-analyses of recent results on tidally driven turbulent flows. 
This has allowed us to show that tidal forcing was likely strong enough to sustain turbulence prior to $-3.25$~Gy, and possibly dynamo action.
However, the self-sustained magnetic field was certainly too weak to explain paleomagnetic measurements if tides were only sustaining an inertial-wave turbulence regime in the ancient core. 
We hope that our results will guide future studies of tidally driven flows, to possibly strive beyond the limits we have identified.
For example, we outlined that magnetostrophic turbulence driven by tidal forcing might produce a strong-field dynamo consistent with the observations.
However, evidence of such a tidally driven regime remains to be found.

Alternatively, other mechanisms in the core could be invoked to explain the ancient geodynamo \citep[see in][]{landeau2022sustaining}.
For instance, the \textsc{cmb} heat flow extracted by mantle convection could power thermal convection in the ancient core \citep[e.g.][]{al2024coupled}, but this requires low-to-moderate values of the thermal conductivity of liquid iron at core conditions \citep[e.g.][]{hsieh2025moderate,andrault2025long}.
Alternatively, exsolution (or precipitation) of light elements below the \textsc{cmb} could sustain double-diffusive turbulent convection in the core \citep[e.g.][]{monville2019rotating}.
Yet, a difficulty with such mechanisms may be to obtain vigorous enough turbulence, or to sustain magnetic fields with a dipolar morphology. 
Actually, a weak tidally driven dynamo might have been important in providing an optimal magnetic seed of finite amplitude to kick-start an efficient convection-driven geodynamo \citep{cattaneo2022earth}.

%-----------------------------------------------------------------------
\subsection{Towards precession in the Moon?}
%-----------------------------------------------------------------------
\begin{figure}
    \centering
    \includegraphics[width=0.49\textwidth]{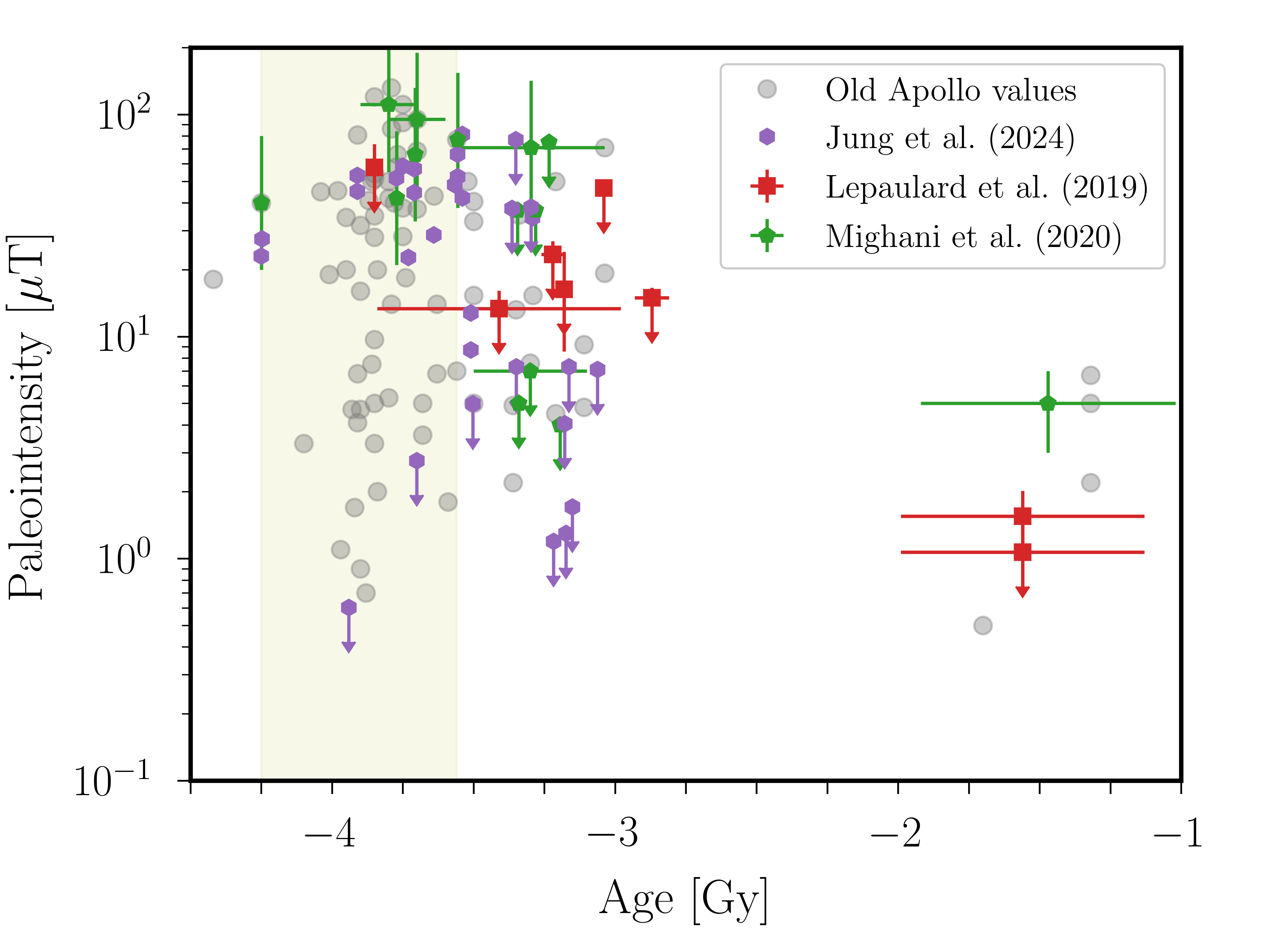} \\
    \caption{Paleointensity at the Moon's surface, inferred from Apollo rocks. Olive region: high-field dynamo epoch. Old Apollo values extracted from \cite{lepaulard2019survey}.}
    \label{fig:Moonpaleomag}
\end{figure}

Beyond Earth, the Moon is another planetary body for which we have geological samples that could help to understand planetary dynamos over long time scales.
The analysis of Apollo samples (Figure \ref{fig:Moonpaleomag}) has revealed that the Moon had a planetary dynamo field in the distant past \citep[e.g.][]{wieczorek2023lunar}, with a (possibly intermittent) magnetic activity from about $-4.2$~Gy until at least $-1.9$~Gy.
Recent analyses confirmed that the Moon was first characterised by a high-field epoch \citep{lepaulard2019survey,jung2024assessing}, which persisted from $\sim 3.9$ until $\sim 3.5$~Gy ago, with measured surface field intensities of $40-110$~$\mu$T.
This high-field period was then succeeded by an epoch with a declining field \citep{tikoo2014decline,strauss2021constraining}, whose surface amplitude fell to below $10$~$\mu$T after $- 3.5$~Gy. 
Note that it remains unclear whether a lunar core dynamo was long-lived \citep[e.g.][]{cai2024reinforced,cai2025persistent}, episodic \citep[e.g.][]{evans2022episodic}, or instead limited to the first hundred millions of years of the Moon's life \citep{zhou2024lunar}.
In any case, the magnetic activity certainly ceased between $-1.9$ and $-0.8$~Gy \citep{tikoo2017two,mighani2020end}.

Actually, such paleomagnetic data put very tough constraints for dynamo action inside the ancient Moon.  
As inferred from Equation (\ref{eq:Bcmbfromdata}), any viable dynamo scenario should be capable of generating a magnetic field that is $10$ larger in the Moon's core than in the Earth's one, despite its core radius being about $10$ times smaller (e.g. $B_\mathrm{cmb} \sim 10 - 70$~mT during the high-field epoch, with $R_\mathrm{cmb} \approx 200-380$~km).
However, standard dynamo scenarios currently fail to explain the observed field values \citep{wieczorek2023lunar}. 
It has been proposed that the ancient Moon's dynamo could result from precession-driven flows \citep[e.g.][]{dwyer2011long}, which have many dynamical similarities with tidal flows \citep[e.g.][]{vidal2024geophysical}. 
The \textsc{g\'eodynamo} team and its collaborators have worked on precession for a long time, making pioneering contributions to the hydrodynamics \citep[e.g.][]{noir2001numerical,lin2015shear} and magnetohydrodynamics \citep{lin2016precession,cebron2019precessing} of such flows.
The present work, at the crossroad of the \textsc{g\'eodynamo}'s research activities, may thus also guide future studies of precession-driven flows.

%-----------------------------------------------------------------------

%% Final declarations after the Conclusion
%-----------------------------------------------------------------------
\section*{CRediT authorship contribution statement}
%-----------------------------------------------------------------------
\textbf{J\'er\'emie Vidal:} conceptualisation, formal analysis, methodology, software, visualisation, funding, writing -- original draft, writing -- review and editing. \textbf{David C\'ebron:} funding, writing -- review and editing.
Both authors gave final approval for submission, and agreed to be held accountable for the work performed therein.

%-----------------------------------------------------------------------
\section*{Acknowledgements}
%-----------------------------------------------------------------------
The authors warmly acknowledge an anonymous referee for the thorough revision of the manuscript, which helped to greatly improve its quality. 
JV thanks Les Houches School of Physics for the hospitality and stimulating discussions during the workshop “Physics of Wave Turbulence and beyond”, which occurred in September 2024 and where the main idea of the study first emerged. 
JV also acknowledges the organisers of the Advanced Summer School “Mathematical Fluid Dynamics”, which was held in Corsica in April 2025 and where part of the work was finalised.

%-----------------------------------------------------------------------
\section*{Declaration of interests}
%-----------------------------------------------------------------------
The authors do not work for, advise, own shares in, or receive funds from any organisation that could benefit from this article, and have declared no affiliations other than their research organisations.

%-----------------------------------------------------------------------
\section*{Funding}
%-----------------------------------------------------------------------
JV received funding from \textsc{ens} de Lyon under the programme "Terre \& Plan\`etes".
DC received funding from the European Research Council (\textsc{erc}) under the European Union's Horizon $2020$ research and innovation programme (grant agreement No $847433$, \textsc{theia} project). 
As part of the \textsc{g\'eodynamo} team, DC greatly acknowledges the support from the French Academy of Sciences \& Electricit\'e de France. 
% JV acknowledges support from the CBPsmn (\textsc{psmn}, P\^ole Scientifique de Mod\'elisation Num\'erique) of the \textsc{ens} de Lyon for the computing resources. 
% The platform operates the \textsc{sidus} solution \citep{quemener2013sidus}.

%-----------------------------------------------------------------------
\section*{Supplementary materials}
%-----------------------------------------------------------------------
The \textsc{matlab} code used to compute the Earth's flattening in Figure \ref{fig:tidesmodel2} is available at \url{http://frederic.chambat.free.fr/hydrostatic/HYDROSTATIC_dec2011.zip}.
The code \textsc{TidalPy} used to compute the tidal deformations in Figure \ref{fig:tidesmodel2} is available at \url{https://doi.org/10.5281/zenodo.14867405}.

%-----------------------------------------------------------------------
\appendix
%-----------------------------------------------------------------------
\section{A weakly turbulent dynamo law?}
\label{sec:appendix}
%-----------------------------------------------------------------------
Scaling law (\ref{eq:lawBdavidson}) has proven to fairly well reproduce the dynamo simulations shown in Figure \ref{fig:scalingB2}.
However, other scaling laws may also be appropriate for orbitally driven dynamos.
For instance, we can assume that the magnetic energy saturates when the injected power is statistically balanced by Joule heating (i.e. $\mathcal{P}_M \sim \epsilon_\eta$).
This simple reasoning yields \citep{christensen2004power}
\begin{equation}
    B \propto \sqrt{\rho_f \mu} \,  \left ( \frac{\ell_B^2}{\eta} \epsilon_\eta \right)^{1/2},
    \label{eq:lawBchristensen}
\end{equation}
where $\ell_B$ is defined as in Equation (\ref{eq:epsilon_eta}).
This scaling law was found to agree quite well with numerical results \citep[e.g.][]{christensen2010dynamo,oruba2014predictive}.
However, the above scaling law cannot generally be employed as a predictive dynamo law for the extrapolation to planets. 
Indeed, $\ell_B$ is largely unknown for core conditions, since it could be non-constant and regime-dependent.
Yet, it is sometimes assumed to be some fraction of the core radius \citep[as found in some simulations, e.g.][]{starchenko2002typical}.
Similarly, relating $\epsilon_\eta$ to the input parameters of the problem is often elusive.

\begin{figure}
    \centering
    \includegraphics[width=0.49\textwidth]{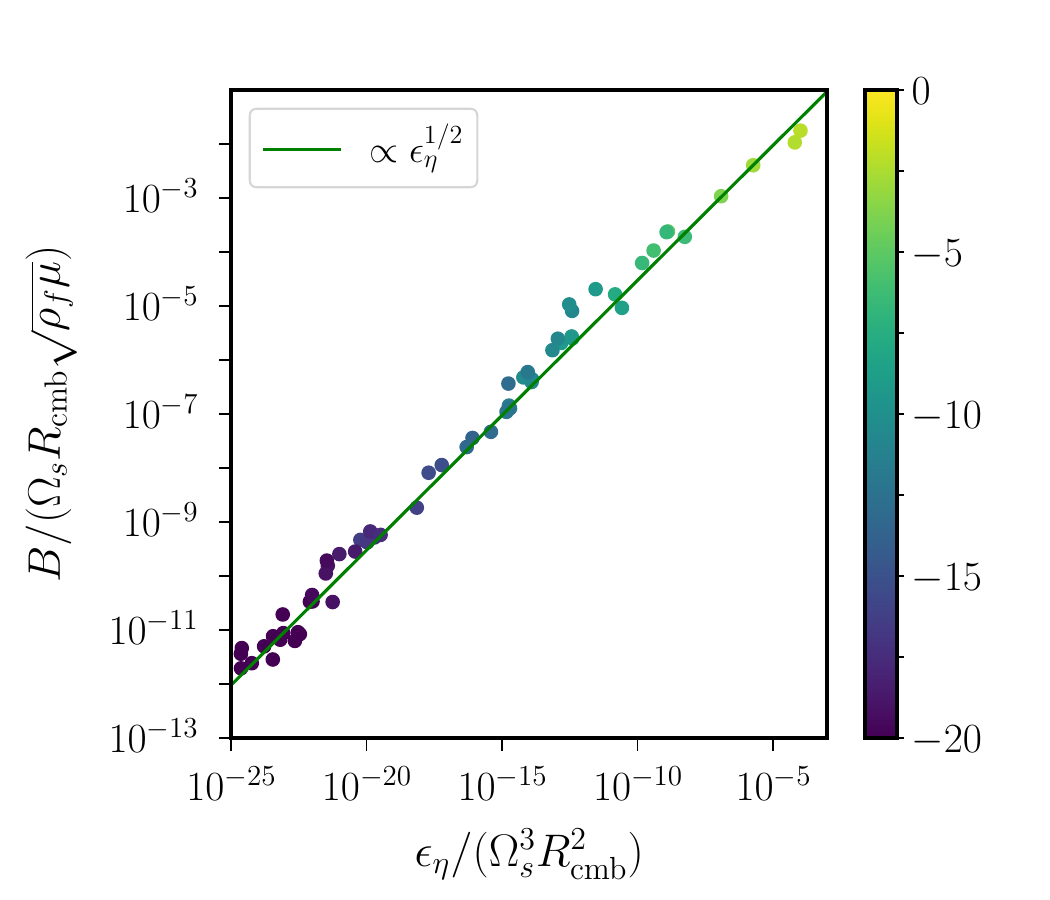}
    \caption{Typical magnetic field amplitude $B$ defined from Equation (\ref{eq:MagNRJ}), as a function of magnetic dissipation $\epsilon_\eta$ in numerical simulations of precession-driven flows inside a sphere \citep[data from][]{cebron2019precessing}. 
    The calculated magnetic dissipation length scale is $\ell_B/R_\mathrm{cmb} = 0.077 \pm 0.046$. 
    Colour bar shows the value of $\log_{10}(\epsilon_\eta/(\epsilon_\eta + \epsilon_\nu))$ for each simulation.}
    \label{fig:scalingB1}
\end{figure}

We have re-analysed in Figure \ref{fig:scalingB1} the dynamo simulations of precession-driven flows in a sphere reported in \citet{cebron2019precessing}.
The simulations show that $B \propto \epsilon_\eta^{1/2}$ when $B$ is estimated from Equation (\ref{eq:MagNRJ}), which agrees with scaling law (\ref{eq:lawBchristensen}).
Moreover, the magnetic length scale is found to only weakly vary across the parameter space, which suggests that the dynamos are not very turbulent.

%-----------------------------------------------------------------------

%% For the database
\CDRGrant[ERC]{847433}

%-----------------------------------------------------------------------
% The next command determines the bibliography style. Please do not
% change this.
\bibliographystyle{crgeos}
\bibliography{main.bib}

@article{fauve2007scaling,
  title={Scaling laws of turbulent dynamos},
  author={Fauve, S. and P{\'e}tr{\'e}lis, F.},
  journal={C. R. Phys.},
  volume={8},
  number={1},
  pages={87--92},
  year={2007}
}

@article{davidson2013scaling,
  title={Scaling laws for planetary dynamos},
  author={Davidson, P. A.},
  journal={Geophys. J. Int.},
  volume={195},
  number={1},
  pages={67--74},
  year={2013},
  publisher={Oxford University Press},
}

@article{landeau2022sustaining,
  title={Sustaining {Earth's} magnetic dynamo},
  author={Landeau, M. and Fournier, A. and Nataf, H.-C. and C{\'e}bron, D. and Schaeffer, N.},
  journal={Nat. Rev. Earth Environ.},
  volume={3},
  number={4},
  pages={255--269},
  year={2022},
  publisher={Nature Publishing Group},
}

@article{farhat2022resonant,
  title={The resonant tidal evolution of the {Earth}--{Moon} distance},
  author={Farhat, M. and Auclair-Desrotour, P. and Bou{\'e}, G. and Laskar, J.},
  journal={Astron. Astrophys.},
  volume={665},
  pages={L1},
  year={2022},
  publisher={EDP Sciences},
}

@article{nichols2024possible,
  title={Possible {Eoarchean} records of the geomagnetic field preserved in the {Isua} {Supracrustal} {Belt}, southern west {Greenland}},
  author={Nichols, C. I. O. and Weiss, B. P. and Eyster, A. and Martin, C. R. and Maloof, A. C. and Kelly, N. M. and Zawaski, M. J. and Mojzsis, S. J. and Watson, E. B. and Cherniak, D. J.},
  journal={J. Geophys. Res. Solid Earth},
  volume={129},
  number={4},
  pages={e2023JB027706},
  year={2024},
  publisher={Wiley Online Library},
}

@article{tarduno2020paleomagnetism,
  title={Paleomagnetism indicates that primary magnetite in zircon records a strong {Hadean} geodynamo},
  author={Tarduno, J. A. and Cottrell, R. D. and Bono, R. K. and Oda, H. and Davis, W. J. and Fayek, M. and Erve, O. v. and Nimmo, F. and Huang, W. and Thern, E. R. and Fearn, S. and Mitra, G. and Smirnov, A V. and Blackman, E. G.},
  journal={Proc. Natl. Acad. Sci.},
  volume={117},
  number={5},
  pages={2309--2318},
  year={2020},
  publisher={National Acadademy of Sciences},
}

@article{lepaulard2019survey,
  title={A survey of the natural remanent magnetization and magnetic susceptibility of {Apollo} whole rocks},
  author={Lepaulard, C. and Gattacceca, J. and Uehara, M. and Rochette, P. and Quesnel, Y. and Macke, R. J. and Kiefer, S. J. W.},
  journal={Phys. Earth Planet. Int.},
  volume={290},
  pages={36--43},
  year={2019},
  publisher={Elsevier},
}

@article{barker2014non,
  title={Non-linear evolution of the elliptical instability in the presence of weak magnetic fields},
  author={Barker, A. J. and Lithwick, Y.},
  journal={Mon. Not. R. Astron. Soc.},
  volume={437},
  number={1},
  pages={305--315},
  year={2014},
  publisher={The Royal Astronomical Society},
}

@article{barker2016non,
  title={Non-linear tides in a homogeneous rotating planet or star: global simulations of the elliptical instability},
  author={Barker, A. J},
  journal={Mon. Not. R. Astron. Soc.},
  volume={459},
  number={1},
  pages={939--956},
  year={2016},
  publisher={Oxford University Press},
}

@article{grannan2017tidally,
  title={Tidally forced turbulence in planetary interiors},
  author={Grannan, A. M. and Favier, B. and Le Bars, M. and Aurnou, J. M.},
  journal={Geophys. J. Int.},
  volume={208},
  number={3},
  pages={1690--1703},
  year={2017},
  publisher={Oxford University Press},
}

@article{zhou2024earth,
  title={{Earth}-{Moon} dynamics from cyclostratigraphy reveals possible ocean tide resonance in the {Mesoproterozoic} era},
  author={Zhou, M. and Wu, H. and Hinnov, L. A and Fang, Q. and Zhang, S. and Yang, T. and Shi, M.},
  journal={Sci. Adv.},
  volume={10},
  number={31},
  pages={eadn7674},
  year={2024},
  publisher={American Association for the Advancement of Science},
}

@article{williams2000geological,
  title={Geological constraints on the {Precambrian} history of {Earth's} rotation and the {Moon's} orbit},
  author={Williams, G. E.},
  journal={Rev. Geophys.},
  volume={38},
  number={1},
  pages={37--59},
  year={2000},
  publisher={Wiley Online Library},
}

@article{eulenfeld2023constraints,
  title={Constraints on {Moon's} orbit 3.2 billion years ago from tidal bundle data},
  author={Eulenfeld, T. and Heubeck, C.},
  journal={J. Geophys. Res. Planets},
  volume={128},
  number={1},
  pages={e2022JE007466},
  year={2023},
  publisher={Wiley Online Library},
}

@article{bono2022pint,
  title={The {PINT} database: {A} definitive compilation of absolute palaeomagnetic intensity determinations since 4 billion years ago},
  author = {Bono, R. K. and Paterson, G. A. and van der Boon, A. and Engbers, Y.  A. and Michael Grappone, J. and Handford, B. and Hawkins, L. M. A. and Lloyd, S. J. and Sprain, C.  J  and Thallner, D. and Biggin, A. J.},
  journal={Geophys. J. Int.},
  volume={229},
  number={1},
  pages={522--545},
  year={2022},
  publisher={Oxford University Press}
}

@article{lemasquerier2017libration,
  title={Libration-driven flows in ellipsoidal shells},
  author={Lemasquerier, D. and Grannan, A. M. and Vidal, J. and C{\'e}bron, D. and Favier, B. and Le Bars, M. and Aurnou, J. M.},
  journal={J. Geophys. Res. Planets},
  volume={122},
  number={9},
  pages={1926--1950},
  year={2017},
  publisher={Wiley Online Library}
}

@article{chambat2010flattening,
  title={Flattening of the {Earth}: further from hydrostaticity than previously estimated},
  author={Chambat, F. and Ricard, Y. and Valette, B.},
  journal={Geophys. J. Int.},
  volume={183},
  number={2},
  pages={727--732},
  year={2010},
  publisher={Oxford University Press},
}

@article{alterman1959oscillations,
  title={Oscillations of the {Earth}},
  author={Alterman, Z. and Jarosch, H. and Pekeris, C.L.},
  journal={Proc. R. Soc. Lond. A.},
  volume={252},
  number={1268},
  pages={80--95},
  year={1959},
  publisher={The Royal Society London},
}

@article{vidal2019fossil,
  title={Fossil field decay due to nonlinear tides in massive binaries},
  author={Vidal, J. and C\'ebron, D. and ud-Doula, A. and Alecian, E.},
  journal={Astron. Astrophys.},
  volume={629},
  pages={A142},
  year={2019},
  publisher={EDP Sciences},
}

@article{macouin2004long,
  title={Long-term evolution of the geomagnetic dipole moment},
  author={Macouin, M. and Valet, J.-P. and Besse, J.},
  journal={Phys. Earth Planet. Int.},
  volume={147},
  number={2-3},
  pages={239--246},
  year={2004},
  publisher={Elsevier}
}

@article{larmor1919could,
  title={How could a rotating body such as the Sun become a magnet},
  author={Larmor, J.},
  journal={Rep. Brit. Adv. Sci.},
  pages={159--160},
  volume={87},
  year={1919}
}

@article{roberts2013genesis,
  title={On the genesis of the {Earth's} magnetism},
  author={Roberts, P. H. and King, E. M.},
  journal={Rep. Prog. Phys.},
  volume={76},
  number={9},
  pages={096801},
  year={2013},
  publisher={IOP Publishing}
}

@article{halliday2023accretion,
  title={The accretion of planet {Earth}},
  author={Halliday, A. N. and Canup, R. M.},
  journal={Nat. Rev. Earth Environ.},
  volume={4},
  number={1},
  pages={19--35},
  year={2023},
  publisher={Nature Publishing Group UK London}
}

@article{dormy2025rapidly,
  title={Rapidly rotating magnetohydrodynamics and the geodynamo},
  author={Dormy, E.},
  journal={Annu. Rev. Fluid Mech.},
  volume={57},
  year={2025},
  pages={335--362},
  publisher={Annual Reviews}
}

@article{marty2013nitrogen,
  title={Nitrogen isotopic composition and density of the {Archean} atmosphere},
  author={Marty, B. and Zimmermann, L. and Pujol, M. and Burgess, R. and Philippot, P.},
  journal={Science},
  volume={342},
  number={6154},
  pages={101--104},
  year={2013},
  publisher={American Association for the Advancement of Science},
}

@article{rogers2025effects,
  title={Effects of geodynamo priors and geomagnetic data on inverted core surface flows},
  author={Rogers, H. F. and Gillet, N. and Aubert, J. and Personnettaz, P. and Mandea, M.},
  journal={Phys. Earth Planet. Int.},
  pages={107323},
  volume = {364},
  year={2025},
  publisher={Elsevier}
}

@article{driscoll2009effects,
  title={Effects of buoyancy and rotation on the polarity reversal frequency of gravitationally driven numerical dynamos},
  author={Driscoll, P. and Olson, P.},
  journal={Geophys. J. Int.},
  volume={178},
  number={3},
  pages={1337--1350},
  year={2009},
  publisher={Oxford University Press},
}

@article{frasson2025geomagnetic,
  title={Geomagnetic dipole stability and zonal flows controlled by mantle heat flux heterogeneities},
  author={Frasson, T. and Schaeffer, N. and Nataf, H. C. and Labrosse, S.},
  journal={Geophys. J. Int.},
  volume={240},
  number={3},
  pages={1481--1504},
  year={2025},
  publisher={Oxford University Press},
}

@article{labrosse2015thermal,
  title={Thermal evolution of the core with a high thermal conductivity},
  author={Labrosse, S.},
  journal={Phys. Earth Planet. Int.},
  volume={247},
  pages={36--55},
  year={2015},
  publisher={Elsevier}
}

@article{burmann2025rapidly,
  title={Rapidly rotating early-{Earth} dynamos in a full-sphere geometry},
  author={Burmann, F. and Luo, J. and Marti, P. and Jackson, A.},
  journal={Geophys. J. Int.},
  volume={241},
  number={1},
  pages={715--727},
  year={2025},
  publisher={Oxford University Press}
}

@article{tarduno2010geodynamo,
  title={Geodynamo, solar wind, and magnetopause 3.4 to 3.45 billion years ago},
  author={Tarduno, J. A. and Cottrell, R. D. and Watkeys, M. K. and Hofmann, A. and Doubrovine, P. V. and Mamajek, E. E. and Liu, D. and Sibeck, D. G. and Neukirch, L. P. and Usui, Y.},
  journal={Science},
  volume={327},
  number={5970},
  pages={1238--1240},
  year={2010},
  publisher={American Association for the Advancement of Science},
}

@article{lichtenegger2010aeronomical,
  title={Aeronomical evidence for higher CO${}_2$ levels during {Earth's} {Hadean} epoch},
  author={Lichtenegger, H. I. M. and Lammer, H. and Grie{\ss}meier, J.-M. and Kulikov, Y. N. and von Paris, P. and Hausleitner, W. and Krauss, S. and Rauer, H.},
  journal={Icarus},
  volume={210},
  number={1},
  pages={1--7},
  year={2010},
  publisher={Elsevier},
}

@article{cattaneo2022earth,
  title={How was the {Earth}--{Moon} system formed? {New} insights from the geodynamo},
  author={Cattaneo, F. and Hughes, D. W.},
  journal={Proc. Natl. Acad. Sci. U.S.A.
},
  volume={119},
  number={44},
  pages={e2120682119},
  year={2022},
  publisher={National Academy of Sciences},
}

@article{tarduno2015hadean,
  title={A {Hadean} to {Paleoarchean} geodynamo recorded by single zircon crystals},
  author={Tarduno, J. A. and Cottrell, R. D. and Davis, W. J. and Nimmo, F. and Bono, R. K.},
  journal={Science},
  volume={349},
  number={6247},
  pages={521--524},
  year={2015},
  publisher={American Association for the Advancement of Science},
}

@article{tarduno2023hadaean,
  title={{Hadaean} to {Palaeoarchaean} stagnant-lid tectonics revealed by zircon magnetism},
  author={Tarduno, J. A. and Cottrell, R. D. and Bono, R. K. and Rayner, N. and Davis, W. J. and Zhou, T. and Nimmo, F. and Hofmann, A. and Jodder, J. and Ibañez-Mejia, M. and Watkeys, M. K. and Oda, H. and Mitra, G.},
  journal={Nature},
  volume={618},
  number={7965},
  pages={531--536},
  year={2023},
  publisher={Nature Publishing Group UK London},
}

@article{biggin2011palaeomagnetism,
  title={Palaeomagnetism of {Archaean} rocks of the {Onverwacht} {Group}, {Barberton Greenstone Belt} (southern {Africa}): Evidence for a stable and potentially reversing geomagnetic field at ca. $3.5$~Ga},
  author={Biggin, A. J. and de Wit, M. J. and Langereis, C. G. and Zegers, T. E. and Vo{\^u}te, S. and Dekkers, M. J. and Drost, K.},
  journal={Earth Planet. Sci. Lett.},
  volume={302},
  number={3-4},
  pages={314--328},
  year={2011},
  publisher={Elsevier}
}

@article{borlina2020reevaluating,
  title={Reevaluating the evidence for a Hadean-Eoarchean dynamo},
  author = {Borlina, C. S.  and Weiss, B. P. and Lima, E. A. and Tang, F. and Taylor, R. J. M.  and Einsle, J. F. and Harrison, R. J. and Fu, R. R. and Bell, E. A. and Alexander, E. W. and Kirkpatrick, H. M. and Wielicki, M. M. and Harrison, T. M. and Ramezani, J. and Maloof, A. C.},
  journal={Sci. Adv.},
  volume={6},
  number={15},
  pages={eaav9634},
  year={2020},
  publisher={American Association for the Advancement of Science},
}

@article{taylor2023direct,
  title={Direct age constraints on the magnetism of {Jack Hills} zircon},
  author={Taylor, R. J. M. and Reddy, S. M. and Saxey, D. W. and Rickard, W. D. A. and Tang, F. and Borlina, C. S. and Fu, R. R. and Weiss, B. P. and Bagot, P. and Williams, H. M. and Harrison, R. J.},
  journal={Sci. Adv.},
  volume={9},
  number={1},
  pages={eadd1511},
  year={2023},
  publisher={American Association for the Advancement of Science},
}

@article{weiss2015pervasive,
  title={Pervasive remagnetization of detrital zircon host rocks in the {Jack Hills}, Western {Australia} and implications for records of the early geodynamo},
  author={Weiss, B. P. and Maloof, A. C. and Tailby, N. and Ramezani, J. and Fu, R. R. and Hanus, V. and Trail, D. and Bruce Watson, E. and Harrison, T. M. and Bowring, S. A. and Kirschvink, J. L. and Swanson-Hysell, N. L. and Coe, R. S.},
  journal={Earth Planet. Sci. Lett.},
  volume={430},
  pages={115--128},
  year={2015},
  publisher={Elsevier}
}

@article{weiss2018secondary,
  title={Secondary magnetic inclusions in detrital zircons from the {Jack Hills}, {Western Australia}, and implications for the origin of the geodynamo},
  author={Weiss, B. P. and Fu, R. R. and Einsle, J. F. and Glenn, D. R. and Kehayias, P. and Bell, E. A. and Gelb, J. and Araujo, J. F. D. F. and Lima, E. A. and Borlina, C. S. and Boehnke, P. and Johnstone, D. N. and Harrison, T. M. and Harrison, R. J. and Walsworth, R. L.},
  journal={Geology},
  volume={46},
  number={5},
  pages={427--430},
  year={2018},
  publisher={Geological Society of America},
}

@article{wieczorek2023lunar,
  title={Lunar magnetism},
  author={Wieczorek, M. A. and Weiss, B. P. and Breuer, D. and Cébron, D. and Fuller, M. and Garrick-Bethell, I. and Gattacceca, J. and Halekas, J. S. and Hemingway, D. J. and Hood, L. L. and Laneuville, M. and Nimmo, F. and Oran, R. and Purucker, M. E. and Rückriemen, T. and Soderlund, K. M. and Tikoo, S. M.},
  journal={Rev. Mineral. Geochem.},
  volume={89},
  number={1},
  pages={207--241},
  year={2023},
  publisher={Mineralogical Society of America},
}

@article{jung2024assessing,
  title={Assessing lunar paleointensity variability during the $3.9$-$3.5$~{Ga} high field epoch},
  author={Jung, J.-I. and Tikoo, S. M. and Burns, D. and V{\'a}ci, Z. and Krawczynski, M. J.},
  journal={Earth Planet. Sci. Lett.},
  volume={638},
  pages={118757},
  year={2024},
  publisher={Elsevier},
}

@article{mighani2020end,
  title={The end of the lunar dynamo},
  author={Mighani, S. and Wang, H. and Shuster, D. L. and Borlina, C. S. and Nichols, C. I. O. and Weiss, B. P.},
  journal={Sci. Sdv.},
  volume={6},
  number={1},
  pages={eaax0883},
  year={2020},
  publisher={American Association for the Advancement of Science},
}

@article{tikoo2017two,
  title={A two-billion-year history for the lunar dynamo},
  author={Tikoo, S. M. and Weiss, B. P. and Shuster, D. L. and Suavet, C. and Wang, Huapei and Grove, T. L.},
  journal={Sci. Adv.},
  volume={3},
  number={8},
  pages={e1700207},
  year={2017},
  publisher={American Association for the Advancement of Science},
}

@article{strauss2021constraining,
  title={Constraining the decline of the lunar dynamo field at $\approx 3.1$~{Ga} through paleomagnetic analyses of {Apollo} $12$ mare basalts},
  author={Strauss, B. E. and Tikoo, S. M. and Gross, J. and Setera, J. B. and Turrin, B.},
  journal={J. Geophys. Res. Planets},
  volume={126},
  number={3},
  pages={e2020JE006715},
  year={2021},
  publisher={Wiley Online Library},
}

@article{tikoo2014decline,
  title={Decline of the lunar core dynamo},
  author={Tikoo, S. M. and Weiss, B. P. and Cassata, W. S. and Shuster, D. L. and Gattacceca, J. and Lima, E. A. and Suavet, C. and Nimmo, F. and Fuller, M. D.},
  journal={Earth Planet. Sci. Lett.},
  volume={404},
  pages={89--97},
  year={2014},
  publisher={Elsevier},
}

@article{jault2015illuminating,
  title={Illuminating the electrical conductivity of the lowermost mantle from below},
  author={Jault, D.},
  journal={Geophys. J. Int.},
  volume={202},
  number={1},
  pages={482--496},
  year={2015},
  publisher={Oxford University Press}
}

@article{yoshino2010laboratory,
  title={Laboratory electrical conductivity measurement of mantle minerals},
  author={Yoshino, T.},
  journal={Surv. Geophys.},
  volume={31},
  pages={163--206},
  year={2010},
  publisher={Springer},
}

@article{evans2022episodic,
  title={An episodic high-intensity lunar core dynamo},
  journal={Nat. Astron.},
  author={Evans, A. J. and Tikoo, S. M.},
  volume={6},
  number={3},
  pages={325--330},
  year={2022},
  publisher={Nature Publishing Group},
}

@article{zhou2024lunar,
  title={A lunar core dynamo limited to the {Moon's} first $140$ million years},
  author={Zhou, T. and Tarduno, J. A. and Cottrell, R. D. and Neal, C. R. and Nimmo, F. and Blackman, E. G. and Iba{\~n}ez-Mejia, M.},
  journal={Commun. Earth Environ.},
  volume={5},
  number={1},
  pages={456},
  year={2024},
  publisher={Nature Publishing Group},
}

@article{cai2024reinforced,
  title={A reinforced lunar dynamo recorded by {Chang'e}-$6$ farside basalt},
  author={Cai, S. and Qi, K. and Yang, S. and Fang, J. and Shi, P. and Shen, Z. and Zhang, M. and Qin, H. and Zhang, C. and Li, X. and Chen, F. and Chen, Y. and Li, J. and He, H. and Deng, C. and Li, C. and Pan, Y. and Zhu, R.},
  journal={Nature},
  pages={1--3},
  volume={643},
  year={2024},
  publisher={Nature Publishing Group},
}

@article{cai2025persistent,
  title={Persistent but weak magnetic field at the {Moon's} midstage revealed by {Chang'e}-5 basalt},
  author={Cai, S. and Qin, H. and Wang, H. and Deng, C. and Yang, S. and Xu, Y. and Zhang, C. and Tang, X. and Gu, L. and Li, X. and Shen, Z. and Zhang, M. and He, K. and Qi, K. and Fan, Y. and Dong, L. and Hou, Y. and Shi, P. and Liu, S. and Su, F. and Chen, Y. and Li, Q. and Li, J. and Mitchell, R. N. and He, H. and Li, C. and Pan, Y. and Zhu, R.},
  journal={Sci. Adv.},
  volume={11},
  number={1},
  pages={eadp3333},
  year={2025},
  publisher={American Association for the Advancement of Science},
}

@article{stixrude2020silicate,
  title={A silicate dynamo in the early {Earth}},
  author={Stixrude, L. and Scipioni, R. and Desjarlais, M. P.},
  journal={Nat. Comm.},
  volume={11},
  number={1},
  pages={935},
  year={2020},
  publisher={Nature Publishing Group},
}

@article{ziegler2013implications,
  title={Implications of a long-lived basal magma ocean in generating {Earth's} ancient magnetic field},
  author={Ziegler, L. B. and Stegman, D. R.},
  journal={Geochem. Geophys. Geosyst.},
  volume={14},
  number={11},
  pages={4735--4742},
  year={2013},
  publisher={Wiley Online Library},
}

@article{scheinberg2018basal,
  title={A basal magma ocean dynamo to explain the early lunar magnetic field},
  author={Scheinberg, A. L. and Soderlund, K. M. and Elkins-Tanton, L. T.},
  journal={Earth Planet. Sci. Lett.},
  volume={492},
  pages={144--151},
  year={2018},
  publisher={Elsevier},
}

@article{gillet2010fast,
  title={Fast torsional waves and strong magnetic field within the {Earth's} core},
  author={Gillet, N. and Jault, D. and Canet, E. and Fournier, A.},
  journal={Nature},
  volume={465},
  number={7294},
  pages={74--77},
  year={2010},
  publisher={Nature Publishing Group UK London}
}

@book{moffatt2019self,
  title={Self-exciting fluid dynamos},
  author={Moffatt, K. H. and Dormy, E.},
  year={2019},
  publisher={Cambridge University Press (Cambridge, UK)}
}

@article{aubert2023state,
  title={State and evolution of the geodynamo from numerical models reaching the physical conditions of {Earth's} core},
  author={Aubert, J.},
  journal={Geophys. J. Int.},
  volume={235},
  number={1},
  pages={468--487},
  year={2023},
  publisher={Oxford University Press}
}

@article{schaeffer2017turbulent,
  title={Turbulent geodynamo simulations: a leap towards Earth’s core},
  author={Schaeffer, N. and Jault, D. and Nataf, H.-C. and Fournier, A.},
  journal={Geophys. J. Int.},
  volume={211},
  number={1},
  pages={1--29},
  year={2017},
  publisher={Oxford University Press},
}

@article{li2023late,
  title={Late {Cambrian} geomagnetic instability after the onset of inner core nucleation},
  author={Li, Y.-X. and Tarduno, J. A. and Jiao, W. and Liu, X. and Peng, S. and Xu, S. and Yang, A. and Yang, Z.},
  journal={Nat. Comm.},
  volume={14},
  number={1},
  pages={4596},
  year={2023},
  publisher={Nature Publishing Group},
}

@article{zhou2022early,
  title={Early {Cambrian} renewal of the geodynamo and the origin of inner core structure},
  author={Zhou, T. and Tarduno, J. A. and Nimmo, F. and Cottrell, R. D. and Bono, R. K. and Ibanez-Mejia, M. and Huang, W. and Hamilton, M. and Kodama, K. and Smirnov, A. V. and Crummins, B. and Padgett, F.},
  journal={Nat. Comm.},
  volume={13},
  number={1},
  pages={4161},
  year={2022},
  publisher={Nature Publishing Group},
}

@article{bono2019young,
  title={Young inner core inferred from {Ediacaran} ultra-low geomagnetic field intensity},
  author={Bono, R. K. and Tarduno, J. A. and Nimmo, F. and Cottrell, R. D.},
  journal={Nat. Geosci.},
  volume={12},
  number={2},
  pages={143--147},
  year={2019},
  publisher={Nature Publishing Group UK London}
}

@article{biggin2015palaeomagnetic,
  title={Palaeomagnetic field intensity variations suggest {Mesoproterozoic} inner-core nucleation},
  author={Biggin, A. J. and Piispa, E. J. and Pesonen, L. J. and Holme, R. and Paterson, G. A. and Veikkolainen, T. and Tauxe, L.},
  journal={Nature},
  volume={526},
  number={7572},
  pages={245--248},
  year={2015},
  publisher={Nature Publishing Group},
}

@article{williams2018thermal,
  title={The thermal conductivity of {Earth's} core: A key geophysical parameter's constraints and uncertainties},
  author={Williams, Q.},
  journal={Annu. Rev. Earth Planet. Sci.},
  volume={46},
  number={1},
  pages={47--66},
  year={2018},
  publisher={Annual Reviews}
}

@article{aubert2009modelling,
  title={Modelling the palaeo-evolution of the geodynamo},
  author={Aubert, J. and Labrosse, S. and Poitou, C.},
  journal={Geophys. J. Int.},
  volume={179},
  number={3},
  pages={1414--1428},
  year={2009},
  publisher={Oxford University Press},
}

@article{monville2019rotating,
  title={Rotating double-diffusive convection in stably stratified planetary cores},
  author={Monville, R. and Vidal, J. and C{\'e}bron, D. and Schaeffer, N.},
  journal={Geophys. J. Int.},
  volume={219},
  number={Supplement\_1},
  pages={S195--S218},
  year={2019},
  publisher={Oxford University Press}
}

@article{vidal2024geophysical,
  title={Geophysical flows over topography, a playground for laboratory experiments},
  author={Vidal, J. and Noir, J. and C{\'e}bron, D. and Burmann, F. and Monville, R. and Giraud, V. and Charles, Y.},
  journal={C. R. Phys.},
  volume={25},
  number={S3},
  pages={1--52},
  year={2024}
}

@article{le2015flows,
  title={Flows driven by libration, precession, and tides},
  author={Le Bars, M. and C{\'e}bron, D. and Le Gal, P.},
  journal={Annu. Rev. Fluid Mech.},
  volume={47},
  number={1},
  pages={163--193},
  year={2015},
  publisher={Annual Reviews}
}

@article{dwyer2011long,
  title={A long-lived lunar dynamo driven by continuous mechanical stirring},
  author={Dwyer, C. A. and Stevenson, D. J. and Nimmo, F.},
  journal={Nature},
  volume={479},
  number={7372},
  pages={212--214},
  year={2011},
  publisher={Nature Publishing Group},
}

@article{le2011impact,
  title={An impact-driven dynamo for the early {Moon}},
  author={Le Bars, M. and Wieczorek, M. A. and Karatekin, {\"O}. and C{\'e}bron, D. and Laneuville, M.},
  journal={Nature},
  volume={479},
  number={7372},
  pages={215--218},
  year={2011},
  publisher={Nature Publishing Group},
}

@article{cebron2019precessing,
  title={Precessing spherical shells: flows, dissipation, dynamo and the lunar core},
  author={C{\'e}bron, D. and Laguerre, R. and Noir, J. and Schaeffer, N.},
  journal={Geophys. J. Int.},
  volume={219},
  number={Supplement 1},
  pages={S34--S57},
  year={2019},
  publisher={Oxford University Press},
}

@article{reddy2018turbulent,
  title={Turbulent kinematic dynamos in ellipsoids driven by mechanical forcing},
  author={Reddy, K. S. and Favier, Be. and Le Bars, M.},
  journal={Geophys. Res. Lett.},
  volume={45},
  number={4},
  pages={1741--1750},
  year={2018},
  publisher={Wiley Online Library},
}

@article{cebron2014tidally,
  title={Tidally driven dynamos in a rotating sphere},
  author={C{\'e}bron, D. and Hollerbach, R.},
  journal={Astrophys. J. Lett.},
  volume={789},
  number={1},
  pages={L25},
  year={2014},
  publisher={IOP Publishing},
}

@article{vidal2018magnetic,
  title={Magnetic fields driven by tidal mixing in radiative stars},
  author={Vidal, J. and C{\'e}bron, D. and Schaeffer, N. and Hollerbach, R.},
  journal={Mon. Not. R. Astron. Soc.},
  volume={475},
  number={4},
  pages={4579--4594},
  year={2018},
  publisher={Oxford University Press},
}

@article{kumar2024dynamo,
  title={Dynamo action driven by precessional turbulence},
  author={Kumar, V. and Pizzi, F. and Mamatsashvili, G. and Giesecke, A. and Stefani, F. and Barker, A. J.},
  journal={Phys. Rev. E},
  volume={109},
  number={6},
  pages={065101},
  year={2024},
  publisher={APS},
}

@article{christensen2001numerical,
  title={A numerical dynamo benchmark},
  author = {Christensen, U. R. and Aubert, J. and Cardin, P. and Dormy, E. and Gibbons, S. and Glatzmaier, G. A. and Grote, E. and Honkura, Y. and Jones, C. and Kono, M. and Matsushima, M. and Sakuraba, A. and Takahashi, F. and Tilgner, A. and Wicht,  J. and Zhang, K.},
  journal={Phys. Earth Planet. Int.},
  volume={128},
  number={1-4},
  pages={25--34},
  year={2001},
  publisher={Elsevier},
}

@article{christensen2010dynamo,
  title={Dynamo scaling laws and applications to the planets},
  author={Christensen, U. R.},
  journal={Space Sci. Rev.},
  volume={152},
  pages={565--590},
  year={2010},
  publisher={Springer},
}

@article{buffett1996thermal,
  title={On the thermal evolution of the {Earth's} core},
  author={Buffett, B. A. and Huppert, H. E. and Lister, J. R. and Woods, A. W.},
  journal={J. Geophys. Res. Solid Earth},
  volume={101},
  number={B4},
  pages={7989--8006},
  year={1996},
  publisher={Wiley Online Library},
}

@article{chandrasekhar1987ellipsoidal,
  title={Ellipsoidal figures of equilibrium},
  author={Chandrasekhar, S.},
  journal={Dover Publications (New York, USA)},
  year={1987}
}

@article{schaeffer2013efficient,
  title={Efficient spherical harmonic transforms aimed at pseudospectral numerical simulations},
  author={Schaeffer, N.},
  journal={Geochem. Geophys. Geosyst.},
  volume={14},
  number={3},
  pages={751--758},
  year={2013},
  publisher={Wiley Online Library}
}

@article{noir2013precession,
  title={Precession-driven flows in non-axisymmetric ellipsoids},
  author={Noir, J. and C{\'e}bron, D.},
  journal={J. Fluid Mech.},
  volume={737},
  pages={412--439},
  year={2013},
  publisher={Cambridge University Press},
}

@article{roberts2011flows,
  title={On flows having constant vorticity},
  author={Roberts, P. H. and Wu, C.-C.},
  journal={Phys. D},
  volume={240},
  number={20},
  pages={1615--1628},
  year={2011},
  publisher={Elsevier},
}

@article{tilgner1998models,
  title={On models of precession driven core flow},
  author={Tilgner, A.},
  journal={Stud. Geophys. Geod.},
  volume={42},
  pages={232--238},
  year={1998},
  publisher={Springer}
}

@article{vidal2021kinematic,
  title={Kinematic dynamos in triaxial ellipsoids},
  author={Vidal, J. and C{\'e}bron, D.},
  journal={Proc. R. Soc. A},
  volume={477},
  number={2252},
  pages={20210252},
  year={2021},
  publisher={The Royal Society Publishing},
}

@article{chen2018optimal,
  title={The optimal kinematic dynamo driven by steady flows in a sphere},
  author={Chen, L. and Herreman, W. and Li, K. and Livermore, P. W. and Luo, J. W. and Jackson, A.},
  journal={J. Fluid Mech.},
  volume={839},
  pages={1--32},
  year={2018},
  publisher={Cambridge University Press}
}

@article{christensen2006scaling,
  title={Scaling properties of convection-driven dynamos in rotating spherical shells and application to planetary magnetic fields},
  author={Christensen, U. R. and Aubert, J.},
  journal={Geophys. J. Int.},
  volume={166},
  number={1},
  pages={97--114},
  year={2006},
  publisher={Oxford University Press},
}

@article{holdenried2019trio,
  title={A trio of simple optimized axisymmetric kinematic dynamos in a sphere},
  author={Holdenried-Chernoff, D. and Chen, L. and Jackson, A.},
  journal={Proc. R. Soc. A},
  volume={475},
  number={2229},
  pages={20190308},
  year={2019},
  publisher={The Royal Society Publishing},
}

@article{nataf2024dynamic,
  title={Dynamic regimes in planetary cores: $\tau$--$\ell$ diagrams},
  author={Nataf, H.-C. and Schaeffer, N.},
  journal={C. R. Geosci.},
  volume={356},
  number={G1},
  pages={1--30},
  year={2024}
}

@article{cebron2012elliptical,
  title={Elliptical instability in terrestrial planets and moons},
  author={C{\'e}bron, D. and Le Bars, M. and Moutou, C. and Le Gal, P.},
  journal={Astron. Astrophys.},
  volume={539},
  pages={A78},
  year={2012},
  publisher={EDP Sciences}
}

@article{wahr1981effect,
  title={Effect of the fluid core on changes in the length of day due to long period tides},
  author={Wahr, J. M. and Sasao, T. and Smith, M. L.},
  journal={Geophys. J. Int.},
  volume={64},
  number={3},
  pages={635--650},
  year={1981},
  publisher={Oxford University Press}
}

@article{dziewonski1981preliminary,
  title={Preliminary reference Earth model},
  author={Dziewonski, A. M. and Anderson, D. L.},
  journal={Phys. Earth Planet. Int.},
  volume={25},
  number={4},
  pages={297--356},
  year={1981},
  publisher={Elsevier},
}

@article{vidal2017inviscid,
  title={Inviscid instabilities in rotating ellipsoids on eccentric {Kepler} orbits},
  author={Vidal, J. and C{\'e}bron, D.},
  journal={J. Fluid Mech.},
  volume={833},
  pages={469--511},
  year={2017},
  publisher={Cambridge University Press},
}

@inbook{AGNEW2015151,
title = {3.06 - {Earth} tides},
editor = {Schubert, G.},
booktitle = {Treatise on Geophysics},
publisher = {Elsevier: Amsterdam},
edition = {Second},
pages = {151--178},
year = {2015},
author = {Agnew, D. C.},
}

@article{kerswell2002elliptical,
  title={Elliptical instability},
  author={Kerswell, R. R.},
  journal={Annu. Rev. Fluid Mech.},
  volume={34},
  number={1},
  pages={83--113},
  year={2002},
  publisher={Annual Reviews},
}

@book{greenspan1969theory,
  title={The theory of rotating fluids},
  author={Greenspan, H. P.},
  year={1968},
  publisher={Cambridge University Press (Cambridge, UK)},
}

@article{liao2001viscous,
  title={On the viscous damping of inertial oscillation in planetary fluid interiors},
  author={Liao, X. and Zhang, K. and Earnshaw, P.},
  journal={Phys. Earth Planet. Int.},
  volume={128},
  number={1-4},
  pages={125--136},
  year={2001},
  publisher={Elsevier},
}

@article{backus2017completeness,
  title={Completeness of inertial modes of an incompressible inviscid fluid in a corotating ellipsoid},
  author={Backus, G. and Rieutord, M.},
  journal={Phys. Rev. E},
  volume={95},
  number={5},
  pages={053116},
  year={2017},
  publisher={APS},
}

@article{vidal2024inertia,
  title={Inertia-gravity waves in geophysical vortices},
  author={Vidal, J. and Colin de Verdi{\`e}re, Y.},
  journal={Proc. R. Soc. A},
  volume={480},
  number={2285},
  pages={20230789},
  year={2024},
  publisher={The Royal Society},
}

@article{knobloch1994normal,
  title={Normal forms for three-dimensional parametric instabilities in ideal hydrodynamics},
  author={Knobloch, E.r and Mahalov, A. and Marsden, J. E.},
  journal={Phys. D},
  volume={73},
  number={1-2},
  pages={49--81},
  year={1994},
  publisher={Elsevier},
}

@article{cebron2010systematic,
  title={A systematic numerical study of the tidal instability in a rotating triaxial ellipsoid},
  author={C{\'e}bron, D. and Le Bars, M. and Leontini, J. and Maubert, P. and Le Gal, P.},
  journal={Phys. Earth Planet. Int.},
  volume={182},
  number={1-2},
  pages={119--128},
  year={2010},
  publisher={Elsevier},
}

@article{plunian2025three,
  title={Three-body anisotropic dynamo: the rotor, the gap and the stator},
  author={Plunian, F. and Gomez, P. and Alboussi{\`e}re, T.},
  journal={C. R. Phys.},
  volume={26},
  number={G1},
  pages={295--315},
  year={2025},
}

@article{deguen2024fluid,
  title={Fluid dynamics of planetary differentiation},
  author={Deguen, R. and Huguet, L. and Landeau, M. and Lherm, V. and Maller, A. and Wacheul, J.-B.},
  journal={C. R. Phys.},
  volume={25},
  number={S3},
  pages={1--45},
  year={2024},
}

@article{le2017inertial,
  title={Inertial wave turbulence driven by elliptical instability},
  author={Le Reun, T. and Favier, B. and Barker, A. J. and Le Bars, M.},
  journal={Physical Review Letters},
  volume={119},
  number={3},
  pages={034502},
  year={2017},
  publisher={APS},
}

@article{le2021evidence,
  title={Evidence of the {Zakharov}-{Kolmogorov} spectrum in numerical simulations of inertial wave turbulence},
  author={Le Reun, T. and Favier, B. and Le Bars, M.},
  journal={Europhys. Lett.},
  volume={132},
  number={6},
  pages={64002},
  year={2021},
  publisher={IOP Publishing}
}

@article{le2019experimental,
  title={Experimental study of the nonlinear saturation of the elliptical instability: inertial wave turbulence versus geostrophic turbulence},
  author={Le Reun, T. and Favier, B. and Le Bars, M.},
  journal={J. Fluid Mech.},
  volume={879},
  pages={296--326},
  year={2019},
  publisher={Cambridge University Press},
}

@article{christensen2004power,
  title={Power requirement of the geodynamo from ohmic losses in numerical and laboratory dynamos},
  author={Christensen, U. R. and Tilgner, A.},
  journal={Nature},
  volume={429},
  number={6988},
  pages={169--171},
  year={2004},
  publisher={Nature Publishing Group}
}

@article{oruba2014predictive,
  title={Predictive scaling laws for spherical rotating dynamos},
  author={Oruba, L. and Dormy, E.},
  journal={Geophys. J. Int.},
  volume={198},
  number={2},
  pages={828--847},
  year={2014},
  publisher={Oxford University Press},
}

@article{starchenko2002typical,
  title={Typical velocities and magnetic field strengths in planetary interiors},
  author={Starchenko, S. V. and Jones, C. A.},
  journal={Icarus},
  volume={157},
  number={2},
  pages={426--435},
  year={2002},
  publisher={Elsevier}
}

@article{zhou1995phenomenological,
  title={A phenomenological treatment of rotating turbulence},
  author={Zhou, Y.},
  journal={Phys. Fluids},
  volume={7},
  number={8},
  pages={2092--2094},
  year={1995},
  publisher={American Institute of Physics},
}

@article{galtier2003weak,
  title={Weak inertial-wave turbulence theory},
  author={Galtier, S.},
  journal={Phys. Rev. E},
  volume={68},
  number={1},
  pages={015301},
  year={2003},
  publisher={APS},
}

@article{galtier2023multiple,
  title={A multiple time scale approach for anisotropic inertial wave turbulence},
  author={Galtier, S.},
  journal={J. Fluid Mech.},
  volume={974},
  pages={A24},
  year={2023},
  publisher={Cambridge University Press},
}

@article{baroud2002anomalous,
  title={Anomalous self-similarity in a turbulent rapidly rotating fluid},
  author={Baroud, C. N. and Plapp, B. B. and She, Z.-S. and Swinney, H. L.},
  journal={Phys. Rev. Lett.},
  volume={88},
  number={11},
  pages={114501},
  year={2002},
  publisher={APS}
}

@article{thiele2009structure,
  title={Structure and decay of rotating homogeneous turbulence},
  author={Thiele, M. and M{\"u}ller, W.-C.},
  journal={J. Fluid Mech.},
  volume={637},
  pages={425--442},
  year={2009},
  publisher={Cambridge University Press},
}

@article{nazarenko2011critical,
  title={Critical balance in magnetohydrodynamic, rotating and stratified turbulence: towards a universal scaling conjecture},
  author={Nazarenko, S. V. and Schekochihin, A. A.},
  journal={J. Fluid Mech.},
  volume={677},
  pages={134--153},
  year={2011},
  publisher={Cambridge University Press},
}

@article{baqui2015phenomenological,
  title={A phenomenological theory of rotating turbulence},
  author={Baqui, Y. B. and Davidson, P. A.},
  journal={Phys. Fluids},
  volume={27},
  number={2},
  year={2015},
  pages={025107},
  publisher={AIP Publishing},
}

@article{le2020near,
  title={Near-resonant instability of geostrophic modes: beyond {Greenspan's} theorem},
  author={Le Reun, T. and Gallet, B. and Favier, B. and Le Bars, M.},
  journal={J. Fluid Mech.},
  volume={900},
  pages={R2},
  year={2020},
  publisher={Cambridge University Press},
}

@article{gallet2015exact,
  title={Exact two-dimensionalization of rapidly rotating large-Reynolds-number flows},
  author={Gallet, B.},
  journal={J. Fluid Mech.},
  volume={783},
  pages={412--447},
  year={2015},
  publisher={Cambridge University Press},
}

@article{kerswell1994tidal,
  title={Tidal excitation of hydromagnetic waves and their damping in the {Earth}},
  author={Kerswell, R. R.},
  journal={J. Fluid Mech.},
  volume={274},
  pages={219--241},
  year={1994},
  publisher={Cambridge University Press},
}

@article{gerick2020pressure,
  title={Pressure torque of torsional {Alfv{\'e}n} modes acting on an ellipsoidal mantle},
  author={Gerick, F. and Jault, D. and Noir, J. and Vidal, J.},
  journal={Geophys. J. Int.},
  volume={222},
  number={1},
  pages={338--351},
  year={2020},
  publisher={Oxford University Press},
}

@article{lin2018tidal,
  title={Tidal dissipation in rotating fluid bodies: the presence of a magnetic field},
  author={Lin, Y. and Ogilvie, G. I.},
  journal={Mon. Not. R. Astron. Soc.},
  volume={474},
  number={2},
  pages={1644--1656},
  year={2018},
  publisher={Oxford University Press},
}

@article{monsalve2020quantitative,
  title={Quantitative experimental observation of weak inertial-wave turbulence},
  author={Monsalve, E. and Brunet, M. and Gallet, B. and Cortet, P.-P.},
  journal={Phys. Rev. Lett.},
  volume={125},
  number={25},
  pages={254502},
  year={2020},
  publisher={APS}
}

@article{yarom2014experimental,
  title={Experimental observation of steady inertial wave turbulence in deep rotating flows},
  author={Yarom, E. and Sharon, E.},
  journal={Nat. Phys.},
  volume={10},
  number={7},
  pages={510--514},
  year={2014},
  publisher={Nature Publishing Group},
}

@article{malkus1989experimental,
  title={An experimental study of global instabilities due to the tidal (elliptical) distortion of a rotating elastic cylinder},
  author={Malkus, W. V. R.},
  journal={Geophys. Astrophys. Fluid Dyn.},
  volume={48},
  number={1-3},
  pages={123--134},
  year={1989},
  publisher={Taylor \& Francis}
}

@article{mcewan1970inertial,
  title={Inertial oscillations in a rotating fluid cylinder},
  author={McEwan, A. D.},
  journal={J. Fluid Mech.},
  volume={40},
  number={3},
  pages={603--640},
  year={1970},
  publisher={Cambridge University Press},
}

@article{barker2016turbulence,
  title={On turbulence driven by axial precession and tidal evolution of the spin--orbit angle of close-in giant planets},
  author={Barker, A. J.},
  journal={Mon. Not. R. Astron. Soc.},
  volume={460},
  number={3},
  pages={2339--2350},
  year={2016},
  publisher={Oxford University Press},
}

@article{noir2001numerical,
  title={Numerical study of the motions within a slowly precessing sphere at low {Ekman} number},
  author={Noir, J. and Jault, D. and Cardin, P.},
  journal={J. Fluid Mech.},
  volume={437},
  pages={283--299},
  year={2001},
  publisher={Cambridge University Press}
}

@article{lin2015shear,
  title={Shear-driven parametric instability in a precessing sphere},
  author={Lin, Y. and Marti, P. and Noir, J.},
  journal={Phys. Fluids},
  volume={27},
  number={4},
  year={2015},
  pages={046601},
  publisher={AIP Publishing}
}

@article{lin2016precession,
  title={Precession-driven dynamos in a full sphere and the role of large scale cyclonic vortices},
  author={Lin, Y. and Marti, P. and Noir, J. and Jackson, A.},
  journal={Phys. Fluids},
  volume={28},
  number={6},
  year={2016},
  pages={066601},
  publisher={AIP Publishing},
}

@article{mineev2004viscosity,
  title={Viscosity measurements on metal melts at high pressure and viscosity calculations for the {Earth's} core},
  author={Mineev, V. N. and Funtikov, A. I.},
  journal={Phys.-Usp.},
  volume={47},
  number={7},
  pages={671},
  year={2004},
  publisher={IOP Publishing}
}

@article{labrosse2007crystallizing,
  title={A crystallizing dense magma ocean at the base of the Earth’s mantle},
  author={Labrosse, S. and Hernlund, J.W. and Coltice, N.},
  journal={Nature},
  volume={450},
  number={7171},
  pages={866--869},
  year={2007},
  publisher={Nature Publishing Group},
}

@article{nakajima2015melting,
  title={Melting and mixing states of the {Earth's} mantle after the Moon-forming impact},
  author={Nakajima, M. and Stevenson, D. J.},
  journal={Earth Planet. Sci. Lett.},
  volume={427},
  pages={286--295},
  year={2015},
  publisher={Elsevier},
}

@article{boukare2025solidification,
  title={Solidification of {Earth's} mantle led inevitably to a basal magma ocean},
  author={Boukar{\'e}, C.-{\'E}. and Badro, J. and Samuel, H.},
  journal={Nature},
  volume={640},
  pages={1--6},
  year={2025},
  publisher={Nature Publishing Group},
}

@article{o1992spin,
  title={Spin-up of a two-layer fluid in a rotating cylinder},
  author={O'Donnell, J. and Linden, P. F.},
  journal={Geophys. Astrophys. Fluid Dyn.},
  volume={66},
  number={1-4},
  pages={47--66},
  year={1992},
  publisher={Taylor \& Francis},
}

@article{pedlosky1967spin,
  title={The spin up of a stratified fluid},
  author={Pedlosky, J.},
  journal={J. Fluid Mech.},
  volume={28},
  number={3},
  pages={463--479},
  year={1967},
  publisher={Cambridge University Press},
}

@article{de2023tidal,
  title={Tidal dissipation due to the elliptical instability and turbulent viscosity in convection zones in rotating giant planets and stars},
  author={de Vries, N. B. and Barker, A. J. and Hollerbach, R.},
  journal={Mon. Not. R. Astron. Soc.},
  volume={524},
  number={2},
  pages={2661--2683},
  year={2023},
  publisher={Oxford University Press},
}

@article{de2023interactions,
  title={The interactions of the elliptical instability and convection},
  author={de Vries, N. B. and Barker, A. J. and Hollerbach, R.},
  journal={Phys. Fluids},
  volume={35},
  number={2},
  year={2023},
  pages={024116},
  publisher={AIP Publishing}
}

@article{vidal2020efficiency,
  title={Efficiency of tidal dissipation in slowly rotating fully convective stars or planets},
  author={Vidal, J. and Barker, A. J.},
  journal={Mon. Not. R. Astron. Soc.},
  volume={497},
  number={4},
  pages={4472--4485},
  year={2020},
  publisher={Oxford University Press},
}

@article{duguid2020convective,
  title={Convective turbulent viscosity acting on equilibrium tidal flows: new frequency scaling of the effective viscosity},
  author={Duguid, C. D. and Barker, A. J. and Jones, C. A.},
  journal={Mon. Not. R. Astron. Soc.},
  volume={497},
  number={3},
  pages={3400--3417},
  year={2020},
  publisher={Oxford University Press},
}

@article{vidal2020turbulent,
  title={Turbulent viscosity acting on the equilibrium tidal flow in convective stars},
  author={Vidal, J. and Barker, A. J.},
  journal={Astrophys. J. Lett.},
  volume={888},
  number={2},
  pages={L31},
  year={2020},
  publisher={IOP Publishing}
}

@article{duguid2020tidal,
  title={Tidal flows with convection: frequency dependence of the effective viscosity and evidence for antidissipation},
  author={Duguid, C. D. and Barker, A. J. and Jones, C. A.},
  journal={Mon. Not. R. Astron. Soc.},
  volume={491},
  number={1},
  pages={923--943},
  year={2020},
  publisher={Oxford University Press},
}

@article{luo2022waves1,
  title={Waves in the Earth’s core. {I}. {Mildly} diffusive torsional oscillations},
  author={Luo, J. and Jackson, A.},
  journal={Proc. R. Soc. A},
  volume={478},
  number={2259},
  pages={20210982},
  year={2022},
  publisher={The Royal Society}
}

@article{luo2022waves2,
  title={Waves in the {Earth's} core. {II}. {Magneto}--{Coriolis} modes},
  author={Luo, J. and Marti, P. and Jackson, A.},
  journal={Proc. R. Soc. A},
  volume={478},
  number={2261},
  pages={20220108},
  year={2022},
  publisher={The Royal Society}
}

@article{gerick2024interannual,
  title={Interannual {Magneto}--{Coriolis} modes and their sensitivity on the magnetic field within the {Earth's} core},
  author={Gerick, F. and Livermore, P. W.},
  journal={Proc. R. Soc. A},
  volume={480},
  number={2299},
  pages={20240184},
  year={2024},
  publisher={The Royal Society},
}

@article{friedlander1989asymptotic,
  title={Asymptotic behaviour of decay rates of internal waves in a rotating stratified spherical shell},
  author={Friedlander, S.},
  journal={Geophys. J. Int.},
  volume={96},
  number={2},
  pages={245--252},
  year={1989},
  publisher={Oxford University Press},
}

@article{landeau2016core,
  title={Core merging and stratification following giant impact},
  author={Landeau, M. and Olson, P. and Deguen, R. and Hirsh, B. H.},
  journal={Nat. Geosci.},
  volume={9},
  number={10},
  pages={786--789},
  year={2016},
  publisher={Nature Publishing Group UK London}
}

@article{braginsky1995equations,
  title={Equations governing convection in {Earth's} core and the geodynamo},
  author={Braginsky, Stanislav I. and Roberts, P. H.},
  journal={Geophys. Astrophys. Fluid Dyn.},
  volume={79},
  number={1-4},
  pages={1--97},
  year={1995},
  publisher={Taylor \& Francis}
}

@misc{renaud2023tidalpy,
  title={Tidal{P}y: Moon and exoplanet tidal heating and dynamics estimator},
  author={Renaud, J. P.},
  howpublished={Astrophysics Source Code Library, record ascl:2307},
  pages={ascl:2307},
  year={2023},
}

@article{cebron2021mean,
  title={Mean zonal flows induced by weak mechanical forcings in rotating spheroids},
  author={C{\'e}bron, D. and Vidal, J. and Schaeffer, N. and Borderies, A. and Sauret, A.},
  journal={J. Fluid Mech.},
  volume={916},
  pages={A39},
  year={2021},
  publisher={Cambridge University Press},
}

@article{busse1968steady,
  title={Steady fluid flow in a precessing spheroidal shell},
  author={Busse, F. H.},
  journal={J. Fluid Mech.},
  volume={33},
  number={4},
  pages={739--751},
  year={1968},
  publisher={Cambridge University Press},
}

@article{christensen2009energy,
  title={Energy flux determines magnetic field strength of planets and stars},
  author={Christensen, U. R. and Holzwarth, V. and Reiners, A.},
  journal={Nature},
  volume={457},
  number={7226},
  pages={167--169},
  year={2009},
  publisher={Nature Publishing Group},
}

@article{ivers2017kinematic,
  title={Kinematic dynamos in spheroidal geometries},
  author={Ivers, D. J.},
  journal={Proc. R. Soc. A},
  volume={473},
  number={2206},
  pages={20170432},
  year={2017},
  publisher={The Royal Society},
}

@article{CdV2025spectrum,
  title={The spectrum of the {Poincar{\'e}} operator in an ellipsoid},
  author={Colin de Verdi\`ere, Y. and Vidal, J.},
  journal={J. Spectr. Theory},
  year={2025},
  volume={15},
  number={3},
  pages={1--26},
}

@article{schaeffer2025BMO,
author = {Schaeffer, N. and Labrosse, S. and Aurnou, J. M.},
title = {Energetically expensive dynamo action in {Earth's} basal magma ocean},
journal = { Proc. Natl. Acad. Sci. USA},
volume = {122},
number = {45},
pages = {e2507575122},
year = {2025},
}

@article{sauret2014tide,
  title={Tide-driven shear instability in planetary liquid cores},
  author={Sauret, A. and Le Bars, M. and Le Gal, P.},
  journal={Geophys. Res. Lett.},
  volume={41},
  number={17},
  pages={6078--6083},
  year={2014},
  publisher={Wiley Online Library},
}

@article{lin2025invariance,
  title={Invariance of dynamo action in an early-{Earth} model},
  author={Lin, Y. and Marti, P. and Jackson, A.},
  journal={Nature},
  pages={1--6},
  year={2025},
  volume={644},
  publisher={Nature Publishing Group,}
}

@article{schwaiger2019force,
  title={Force balance in numerical geodynamo simulations: a systematic study},
  author={Schwaiger, T. and Gastine, T. and Aubert, J.},
  journal={Geophys. J. Int.},
  volume={219},
  number={Supplement 1},
  pages={S101--S114},
  year={2019},
  publisher={Oxford University Press},
}

@article{hsieh2025moderate,
  title={Moderate thermal conductivity of Fe-Ni-Si alloy at {Earth's} core conditions: {Implications} for core thermal evolution and geodynamo},
  author={Hsieh, W.-P. and Chiang, Y.-T. and Deschamps, F. and Chen, C.-C. and Chang, J.-W. and Ikuta, D. and Ohtani, E.},
  journal={Geophys. Res. Lett.},
  volume={52},
  number={21},
  pages={e2025GL117576},
  year={2025},
  publisher={Wiley Online Library},
}

@article{andrault2025long,
  title={Long-lived magnetic field in earth-like terrestrial planets},
  author={Andrault, D. and Pacynski, L. Pison and Monteux, J. and Gard{\'e}s, E. and Mathieu, A.},
  journal={Phys. Earth Planet. Int.},
  volume={360},
  pages={107315},
  year={2025},
  publisher={Elsevier}
}

@article{dragulet2025electrical,
  title={Electrical and thermal conductivity of Earth’s iron-enriched basal magma ocean},
  author={Dragulet, F. and Stixrude, L.},
  journal={Proc. Natl. Acad. Sci.},
  volume={122},
  number={42},
  pages={e2509771122},
  year={2025},
  publisher={National Academy of Sciences},
}

@article{de1998viscosity,
  title={The viscosity of liquid iron at the physical conditions of the {Earth's} core},
  author={de Wijs, G. A. and Kresse, G. and Vo{\v{c}}adlo, L. and Dobson, D. and Alfe, D. and Gillan, M. J. and Price, G. D.},
  journal={Nature},
  volume={392},
  number={6678},
  pages={805--807},
  year={1998},
  publisher={Nature Publishing Group},
}

@article{tyler2021tidal,
  title={On the tidal history and future of the {Earth}--{Moon} orbital system},
  author={Tyler, R. H.},
  journal={Planet. Sci. J.},
  volume={2},
  number={2},
  pages={70},
  year={2021},
  publisher={IOP Publishing},
}

@article{daher2021long,
  title={Long-term {Earth}--{Moon} evolution with high-level orbit and ocean tide models},
  author = {Daher, H. and Arbic, B. K. and Williams, J. G. and Ansong, J. K. and Boggs, D. H. and Müller, M. and Schindelegger, M. and Austermann, J. and Cornuelle, B. D. and Crawford, E. B. and Fringer, O. B. and Lau, H. C. P. and Lock, S. J. and Maloof, A. C. and Menemenlis, D. and Mitrovica, J. X. and Green, J. A. Mattias and Huber, M.},
  journal={J. Geophys. Res. Planets},
  volume={126},
  number={12},
  pages={e2021JE006875},
  year={2021},
  publisher={Wiley Online Library},
}

@article{touma1994evolution,
  title={Evolution of the {Earth}--{Moon} system},
  author={Touma, J. and Wisdom, J.},
  journal={Astron. J.},
  volume={108},
  number={5},
  pages={1943--1961},
  year={1994},
}

@article{fuchs1999self,
  title={On self--killing and self--creating dynamos},
  author={Fuchs, H. and R{\"a}dler, K.-H. and Rheinhard, M.},
  journal={Astron. Nachr.},
  volume={320},
  number={3},
  pages={129--133},
  year={1999},
  publisher={Wiley Online Library},
}

@article{reuter2009wave,
  title={Wave-driven dynamo action in spherical magnetohydrodynamic systems},
  author={Reuter, K. and Jenko, F. and Tilgner, A. and Forest, C. B.},
  journal={Phys. Rev. E},
  volume={80},
  number={5},
  pages={056304},
  year={2009},
  publisher={APS},
}

@article{koper2003constraints,
  title={Constraints on aspherical core structure from {PKiKP}--{PcP} differential travel times},
  author={Koper, K. D. and Pyle, M. L. and Franks, J. M.},
  journal={J. Geophys. Res.: Solid Earth},
  volume={108},
  number={B3},
  year={2003},
  pages={2168},
  publisher={Wiley Online Library},
}

@article{sze2003core,
  title={Core mantle boundary topography from short period {PcP}, {PKP}, and {PKKP} data},
  author={Sze, E. K. M. and van der Hilst, R. D.},
  journal={Phys. Earth Planet. Int.},
  volume={135},
  number={1},
  pages={27--46},
  year={2003},
  publisher={Elsevier},
}

@article{green2017explicitly,
  title={Explicitly modelled deep-time tidal dissipation and its implication for {Lunar} history},
  author={Green, J. A.M. and Huber, M. and Waltham, D. and Buzan, J. and Wells, M.},
  journal={Earth Planet. Sci. Lett.},
  volume={461},
  pages={46--53},
  year={2017},
  publisher={Elsevier}
}

@article{zeeden2023earth,
  title={{Earth's} rotation and {Earth}--{Moon} distance in the {Devonian} derived from multiple geological records},
  author={Zeeden, C. and Laskar, J. and De Vleeschouwer, D. and Pas, D. and Da Silva, A.-C.},
  journal={Earth Planet. Sci. Lett.},
  volume={621},
  pages={118348},
  year={2023},
  publisher={Elsevier}
}

@article{korenaga2025tidala,
  title={Tidal dissipation within {Earth's} solidifying magma ocean: {I}. {Effects} of inertia and lunar orbital eccentricity},
  author={Korenaga, J.},
  journal={Icarus},
  pages={116756},
  year={2025},
  volume={442},
  publisher={Elsevier},
}

@article{korenaga2025tidalb,
  title={Tidal dissipation within {Earth's} solidifying magma ocean: {II}. {Atmospheric} blanketing and its constraint on tidal heating},
  author={Korenaga, J.},
  journal={Icarus},
  pages={116743},
  year={2025},
  volume={442},
  publisher={Elsevier},
}

@article{korenaga2025tidalc,
  title={Tidal dissipation within {Earth's} solidifying magma ocean: {III}. {Effects} of matrix compaction},
  author={Korenaga, J.},
  journal={Icarus},
  pages={116759},
  year={2025},
  volume={442},
  publisher={Elsevier},
}

@article{zeman1994note,
  title={A note on the spectra and decay of rotating homogeneous turbulence},
  author={Zeman, O.},
  journal={Phys. Fluids},
  volume={6},
  number={10},
  pages={3221--3223},
  year={1994},
}

@article{li2025energy,
  title={Energy spectrum of two-dimensional isotropic rapidly rotating turbulence},
  author={Li, P.-Y. and Xie, J.-H.},
  journal={Phys. Rev. Fluids},
  volume={10},
  number={5},
  pages={054608},
  year={2025},
  publisher={APS},
}

@article{istas2023transient,
  title={Transient core surface dynamics from ground and satellite geomagnetic data},
  author={Istas, M. and Gillet, N. and Finlay, C. C. and Hammer, M. D. and Huder, L.},
  journal={Geophys. J. Int.},
  volume={233},
  number={3},
  pages={1890--1915},
  year={2023},
  publisher={Oxford University Press},
}

@article{madsen2025modelling,
  title={Modelling geomagnetic jerks with core surface flow derived from satellite gradient tensor elements of secular variation},
  author={Madsen, F. D. and Whaler, K. and Beggan, C. and Brown, W. and Lauridsen, J. and Holme, R.},
  journal={Phys. Earth Planet. Int.},
  pages={107336},
  year={2025},
  volume={366},
  publisher={Elsevier}
}

@article{al2024coupled,
  title={Coupled fates of {Earth's} mantle and core: {Early} sluggish-lid tectonics and a long-lived geodynamo},
  author={Al Asad, M.and Lau, H. C. P.},
  journal={Sci. Adv.},
  volume={10},
  number={31},
  pages={eadp1991},
  year={2024},
  publisher={American Association for the Advancement of Science},
}

@article{moffatt1970turbulent,
  title={Turbulent dynamo action at low magnetic Reynolds number},
  author={Moffatt, H. K.},
  journal={J. Fluid Mech.},
  volume={41},
  number={2},
  pages={435--452},
  year={1970},
  publisher={Cambridge University Press}
}

@article{moffatt1970dynamo,
  title={Dynamo action associated with random inertial waves in a rotating conducting fluid},
  author={Moffatt, H. K.},
  journal={J. Fluid Mech.},
  volume={44},
  number={4},
  pages={705--719},
  year={1970},
  publisher={Cambridge University Press},
}

@article{dehant2017understanding,
  title={Understanding the effects of the core on the nutation of the {Earth}},
  author={Dehant, V. and Laguerre, R. and Rekier, J. and Rivoldini, Attilio and Triana, S. A. and Trinh, A. and Van Hoolst, T. and Zhu, P.},
  journal={Geod. Geodyn.},
  volume={8},
  number={6},
  pages={389--395},
  year={2017},
  publisher={Elsevier}
}

@article{hollerbach1996theory,
  title={On the theory of the geodynamo},
  author={Hollerbach, R.},
  journal={Phys. Earth Planet. Int.},
  volume={98},
  number={3-4},
  pages={163--185},
  year={1996},
  publisher={Elsevier},
}

@article{aubert2017spherical,
  title={Spherical convective dynamos in the rapidly rotating asymptotic regime},
  author={Aubert, J. and Gastine, T. and Fournier, A.},
  journal={J. Fluid Mech.},
  volume={813},
  pages={558--593},
  year={2017},
  publisher={Cambridge University Press},
}
%-----------------------------------------------------------------------

\end{document}